\documentclass[a4paper,10pt,twoside,notitlepage]{article}

\usepackage{latexsym, amsfonts,amsmath,amssymb,graphics,bbm}
\usepackage{feynmp}
\usepackage[cp1250]{inputenc}
\usepackage[dvips]{graphicx}
\DeclareMathAlphabet{\mathpzc}{OT1}{pzc}{m}{it}

\newcommand{\indgd}[3]{{#1}^{#2}_{\phantom{#2}#3}}

\newcommand{\refer}[1]{(\ref{#1})}

\newcommand{\vol}[1]{\!\! \frac{\text{d}^d #1}{\left(2\pi\right)^d}}

\newcommand{\eq}[1]{\begin{equation}#1\end{equation}}
\newcommand{\eqs}[1]{\begin{eqnarray}#1\end{eqnarray}}
\newcommand{\eqsN}[1]{\begin{align}#1\end{align}}
\newcommand{\nawias}[1]{\left(#1\right)}
\newcommand{\pochfunc}[2]{\frac{\delta #1}{\delta #2}}
\newcommand{\poch}[2]{\frac{\partial #1}{\partial #2}}
\newcommand{\anty}[1]{\overline{#1}}

\newcommand{\calka}[3]{\int{\text{d}^{#1} #2 \text{ } #3}}
\newcommand{\calkabp}[3]{\int{\text{d}^{#1} #2 #3}}
\newcommand{\funkcjonal}[2]{#1\left[#2\right]}
\newcommand{\eR}[3]{{e_{{}_{R}}}^{#1}_{\text{ }#2 #3}}
\newcommand{\przerw}[0]{\text{ }}
\newcommand{\przerwpod}[0]{\text{ }\text{ }}
\newcommand{\macierz}[1]{\left[#1\right]}

\author{
Adrian ~Lewandowski\footnote{E-mail: lewandow@fuw.edu.pl}\\
{}\\
\normalsize
\emph{Institute of Theoretical Physics, Faculty of Physics,}\\
\normalsize
\emph{University of Warsaw, Hoża 69, 00-681, Warsaw, Poland}
}

\title{
\begin{flushright}
\normalsize { $\phantom{a}$ }\\
{}
\end{flushright}
\textbf{Renormalization of Nielsen Identities}}
\date{}

\begin{document}

\maketitle

\begin{abstract}
\indent We study renormalization of identities governing the dependence
of 1PI Green's functions on gauge-fixing parameters. For general
dimensionally regularized Yang-Mills theories with gauge groups being
direct products of arbitrary compact simple Lie groups and $U(1)$ groups
coupled to scalar fields, we extend the well known analysis in Fermi
gauges to the class of generalized \mbox{'t Hoot} gauges $R_{\xi,u}$, in
which also symmetry under global gauge transformations is broken by
the gauge-fixing procedure. We also discuss conditions ensuring
homogeneity of the Nielsen identity satisfied by the effective potential.
\end{abstract}

\normalsize

\begin{section}{Introduction}
Renormalizability, unitarity and gauge-independence of Yang-Mills gauge
theories (without the Adler-Bardeen anomaly) were proved by Becchi, Rouet,
Stora \cite{BRS} and Tyutin \cite{Tyutin}. The case of a general algebra
of a compact gauge group was considered in \cite{BBBC}. While unitarity
of the $S$-matrix, owing to the Kugo-Ojima quartet mechanism
\cite{KugoOjima}, is an immediate consequence of the BRST symmetry,
gauge-fixing independence follows only from an \emph{extended} BRST
symmetry \cite{Nielsen, KlubergZuber}, which acts also on gauge-fixing
parameters. Slavnov-Taylor identities of this symmetry are usually called
Nielsen identities \cite{AitchisonFraser,Johnston}. They were originally
used by Nielsen \cite{Nielsen} in the study of gauge-independence of
spontaneous symmetry breaking. Since then, Nielsen identities constitute
an efficient tool for studying virtually all problem related to gauge-independence. Under the extended BRST symmetry, gauge-fixing parameters
are transformed into anticommuting classical fields (`Nielsen sources')
coupled to composite operators. Renormalization of Green's functions with
insertions of these operators requires additional counterterms, which at
the same time control gauge-dependence of ordinary counterterms. For pure
Yang-Mills theories quantized in the so-called Fermi gauges with a single
gauge-fixing parameter $\xi$ this was demonstrated in the work
\cite{PiguetSibold} of Piguet and Sibold in which the problem of
renormalization of the Nielsen identities in such theories was worked out.

If spontaneous gauge symmetry breaking is anticipated, the quantization
scheme should in general be more complex and involve several gauge-fixing
parameters (a very general class of such schemes will be considered in
this paper). Yet, in theories in which gauge symmetries are broken
spontaneously by a vacuum expectation value (VEV) of a scalar field
existing already at the tree level (as in the Standard Model) one commonly
uses the \mbox{'t Hooft} $R_{\xi}$ gauge \cite{tHootfFujikawaLeeSanda} which
is effectively a one parameter scheme. Using it one avoids non-diagonal
scalar-vector propagators and (except for the Landau $\xi=0$ gauge)
infrared divergences which are typical of Fermi gauges. In this class of
theories the perturbative expansion is constructed around the tree level
vacuum and in practical computations of  Green's functions there is no need
to investigate the effective potential. Hence the fact that it is well
defined only in the Landau $\xi=0$ gauge does not preclude the possibility
of checking gauge-independence of the calculated physical quantities.
Nielsen identities applied to this case allow e.g. to simplify the proof
of gauge-independence of the S-matrix \cite{Grassi,Kummer} and to analyze
gauge-independence of masses and widths of unstable particles in the
Standard Model \cite{Grassi}.

Minimization of the effective potential becomes an important ingredient
of the perturbative expansion in theories in which symmetry breaking is
triggered only by radiative corrections \cite{CW} (see \cite{Meiss} for
a recent proposal). In order to obtain a well defined effective potential
for $\xi\neq 0$, another gauge-fixing parameter $u$ has to be introduced
\cite{JackiwDolan}. This leads to the class of generalized \mbox{'t Hooft}
$R_{\xi,u}$ gauges (in theories like the Standard Model, identifying $u$ with
the VEV of the scalar field, one recovers the \mbox{'t Hooft}
$R_\xi$ gauges). In this class of gauges the Nielsen identities were
derived in \cite{Johnston} and checked in various one-loop calculations
mainly in the abelian Higgs model \cite{AitchisonFraser,Baz,BLS} for both
bare (regularized) and renormalized Green's functions (see also
\cite{Met} for the proof of the gauge-independence of the false vacuum
decay rate in abelian theories with radiative
symmetry breaking). General considerations of necessary additional counterterms
based on power-counting arguments for abelian models quantized in non-linear
and background field gauges can be found in \cite{BPP} and \cite{KS},
respectively. Although ordinary counterterms necessary for renormalization
of Green's functions in the $R_{\xi,u}$ gauge in general non-abelian theories
are well known \cite{BBBC, JackiwDolan}, to the best of our knowledge,
based on power-counting arguments determination of all possible additional
counterterms necessary to renormalize the action with operators coupled
to the Nielsen sources has never been presented in the literature.

In this paper we find all these additional counterterms for the
action with Nielsen sources of a general Yang-Mills theory coupled to
bosonic matter fields in the generalized 't Hooft $R_{\xi,u}$ gauges. We
work in the Dimensional Regularization which (in the considered class
of theories) is consistent with the usual
BRST symmetry as well as with the extended one. Thus, the counterterms in
the \mbox{$MS$-scheme} are directly constrained by the symmetry
requirements. Renormalization of
the Nielsen identities is indispensable to obtain equations that govern
gauge-dependence of the renormalized effective potential, which partly
motivated the analysis presented here.

We are particularly interested in the $u$-dependence, which was not
studied in \cite{PiguetSibold}. We will show that the Nielsen identities
governing the $u$-dependence of the renormalized action allow to determine
(in the MS-scheme) the additive VEV counterterm $\delta v$ in terms of a
two-point function of composite operators, which elucidates the origin of
$\delta v$. This method of determination of $\delta v$ turns out to be
very convenient in the one-loop approximation. It reproduces the well
known one-loop results found in the Standard Model \cite{Hollik} and in
the MSSM \cite{ChanPoRo} and can be used in any model irrespectively of
whether the tree-level VEV exists or not.

Furthermore we extend the results of \cite{PiguetSibold} by allowing
for the $\xi$ parameters to be a general matrix in the space of the gauge
group generators. We find that in the most general case, when also the
symmetry of the action with respect to global transformations is broken
by such $\xi$ parameters (as happens e.g. in the Standard Model quantized
with a separate $\xi$ parameter for each mass eigenstate), an additional
superficially divergent three-point function of composite operators
may appear. Its renormalization would therefore require a new counterterm
(which we call `the curvature'). At the one-loop level, as we have
checked by explicit calculation, this three-point function is however
finite owing to a cancellation between two diagrams.

Finally we discuss the Nielsen identity satisfied by the effective
potential. In the generalized 't Hooft $R_{\xi,u}$ gauge this identity
is homogeneous only if
the scalar fields are restricted to the subspace on which the
gauge-fixing function vanishes. The condition of no spontaneous breaking
of the BRST symmetry requires that the scalar fields VEVs belong precisely
to this subspace. We will show that in the class of theories considered
in this paper, a stationary point $\phi_0$ of the effective potential restricted to this subspace is also a stationary point of the full effective potential, provided that $\phi_0$ obeys a certain
condition which does not relay on invariance of the action under
additional discrete transformations (like CP) which may be not exact or
can be broken spontaneously.

The results presented in this paper can immediately be extended to theories
with fermions in nonchiral representations of the gauge group. They have
been omitted for the sake of simplicity - the relevant formulae are
analogous to the ones presented here. Inclusion of chiral fermions (in
nonanomalous representations) is possible but requires a dedicated analysis
of the necessary counterterms.

The paper is organized in the following way. Section 2 contains, for
the reader's convenience, the derivation of the Nielsen identities based
on the method of Piguet and Sibold \cite{PiguetSibold}. In Section 3 the
complete action with all possible counterterms is presented together with
constraints imposed on it by the Nielsen identities; a detailed derivation
is given in appendices. In Section 4 the gauge-independence of bare
coupling constants is proved with the help of the Nielsen identities.
In Section 5 some of the formal results are checked by explicit one-loop
calculations. Section 6 contains explicit computation of the counterterm
$\delta v$ for the effective potential in the $R_{\xi,u}$ gauges and a
discussion of the homogeneity of the Nielsen identity satisfied by the
effective potential. Section 7 is devoted to our conclusions.
\end{section}

\begin{section}{Nielsen Identities}

\noindent We begin with the gauge-fixed action of general Yang-Mills fields
coupled to scalar fields in an arbitrary representation of the gauge group.
In terms of parameters and fields which after inclusion of counterterms
will acquire the interpretation of renormalized field and
parameters\footnote{This is why these parameters carry the
subscript $R$; on the other hand to keep the notation manageable
on renormalized fields this subscript is omitted.}
the Lagrangian density reads
\eqs{\label{LeffModWczesn}
\mathcal{L}^h(x)&=&-\frac{1}{4}
\delta_{\alpha \beta}F^\alpha_{\text{ }\mu\nu}(x)F^{\beta\text{}\mu\nu}(x)+
\frac{1}{2}\delta_{ab}\nawias{D_\mu\phi}^a\nawias{x}
\nawias{D^\mu\phi}^b\nawias{x}-
\mathcal{V}\nawias{\phi\nawias{x}}\nonumber\\
&{}&+s\left(\overline{\omega}_\alpha(x)f^\alpha(x)+
\frac{1}{2}~\!\overline{\omega}_\alpha(x)\xi^{\alpha\beta}h_\beta(x)\right),
}
in which
\eqs{\label{WiekosciKowariantneStart}
F^\alpha_{\text{ }\mu\nu}(x)&=&\partial_\mu A^\alpha_\nu\nawias{x}
-\partial_\nu A^\alpha_\mu\nawias{x}+{e_{{}_R}}^\alpha_{\przerwpod\beta\gamma}
A^\beta_\mu\nawias{x} A^\gamma_\nu\nawias{x},\\
\nawias{D_\mu\phi}^a\nawias{x} &=&\partial_\mu \phi^a\nawias{x}
+ A^\alpha_\mu\nawias{x} \macierz{{T_{{}_R}}_\alpha}^a_{\przerw b}
\nawias{\phi^b\nawias{x}+v^b_{{}_R}},\label{WiekosciKowariantneStart2}
}
and $s$ denotes the BRST operator
(see e.g. \cite{Wein,ZinnJustinQFTCritical})
\begin{eqnarray}
s\left(\phi^a\left(x\right)\right)&=&\omega^{\alpha}\left(x\right)
\left({T_{{}_{R\alpha}}}\left(\phi\left(x\right)+v_{{}_R}\right)\right)^a,\\
s\left(A^\gamma_\mu\left(x\right)\right)&=&
-\partial_\mu\omega^{\gamma}\left(x\right)+
{e_{{}_{R}}}^{\gamma}_{\text{ }\alpha \beta}~\!\omega^{\alpha}\left(x\right)
A^\beta_\mu\left(x\right),\\
s\left(\omega^\alpha\left(x\right)\right)&=&\frac{1}{2}~\!
{e_{{}_{R}}}^{\alpha}_{\text{ }\beta \gamma}~\!\omega^\beta(x)\omega^\gamma(x),\\
s\left(\overline{\omega}_\alpha\left(x\right)\right)&=&
h_\alpha\left(x\right),\\
s\left(h_\alpha\left(x\right)\right)&=&0.
\end{eqnarray}

The matrices $\macierz{{T_{{}_R}}_\alpha}^a_{\przerw c}$ are antisymmetric and
span a representation of the gauge Lie algebra (which is the direct sum of simple compact Lie algebras and $\mathfrak{u}(1)$ algebras) with
totally antisymmetric structure constants ${e_{{}_R}}^\alpha_{\przerwpod\beta\gamma}$
(we prefer nevertheless to distinguish the upper and lower indices).
We assume that $\macierz{{T_{{}_R}}_\alpha}^a_{\przerw c}$ and
${e_{{}_R}}^\alpha_{\przerwpod\beta\gamma}$ have been brought into the usual
block-diagonal form. The scalar potential has the
form $\mathcal{V}\nawias{\phi}\equiv\mathcal{V}_{sym}\nawias{\phi+v_{{}_R}}$
and satisfies the following symmetry conditions
\eq{\label{Vinvstart}
\macierz{{T_{{}_R}}_\alpha}^a_{\przerw b}\nawias{\phi^b+v^b_{{}_R}}
\poch{\mathcal{V}\nawias{\phi}}{\phi^a}=0.
}
We work in the class of linear $R_{\xi,u}$ gauges specified by the functions
\eq{\label{falphastart}
f^\alpha\nawias{x}=-\partial_\mu A^{\alpha\mu}\nawias{x}
-\xi^{\alpha\beta}\delta_{a c} u^a  \macierz{{T_{{}_R}}_\beta}^c_{\przerw b}
\nawias{\phi^b\nawias{x}+v^b_{{}_R}}.
}
in which $u^a$ are additional gauge-fixing parameters. For greater generality we allow for the parameters $\xi^{\alpha\beta}$ which are arbitrary matrices in the space of the gauge Lie algebra generators. The Lagrangian (\ref{LeffModWczesn}) depends on the constant background $v_{{}_R}$ only through the sum $\phi+v_{{}_R}$ and the effective potential may be calculated by the usual methods~\cite{Jackiw}. In the derivations presented below it is convenient to treat the background $v_{{}_R}$ (similarly as other renormalized parameters) as independent of gauge-fixing parameters.\footnote{
If studying the effective potential is not needed, 
instead of treating $v_{{}_R}$ as the constant background, 
one can determine $v_{{}_R}$ from the condition $\langle\phi\rangle=0$,
that is from the requirement that the loop corrections cancel the tree
level tadpole diagrams. This is equivalent to the
minimization of the effective potential, and the gauge-dependence inherited by $v_{{}_R}$ is then controlled by the appropriate Nielsen identity. For
$u=v_{{}_R}$ and $\xi^{\alpha\beta}\propto\delta^{\alpha\beta}$ this choice of $v_{{}_R}$ reduces the $R_{\xi,u}$
gauges to the ordinary 't Hooft $R_\xi$ gauges.
}\\

The  simplest way to get the Nielsen identities \cite{PiguetSibold} is
to replace the $s$ operator in \refer{LeffModWczesn} with its extended
counterpart $s_{\mathrm{ext}}$ defined so that $s_{\mathrm{ext}}=s$ on quantum
fields and
\eq{
s_{\mathrm{ext}} \nawias{u^{a}\nawias{x}}=q^a\nawias{x},
\quad s_{\mathrm{ext}} \nawias{\xi^{\alpha\beta}\nawias{x}}
=q^{\alpha\beta}\nawias{x},
}
where $q^a\nawias{x}$ and $q^{\alpha\beta}\nawias{x}$ are fermionic external
fields, called `Nielsen sources' in the rest of the paper
(\mbox{$s_{\mathrm{ext}} \nawias{q\nawias{x}}=0$}  to ensure nilpotency of
$s_{\mathrm{ext}}$). Unlike \cite{Nielsen,KlubergZuber,PiguetSibold}, we treat
$\xi$, $u$ and, consequently $q$, as $x$-dependent. (Of course $\xi$ and
$u$ should be eventually restricted to constant configurations). This
approach will allow us to avoid some of the IR divergences in explicit
one-loop calculations presented in section \ref{Sec:WyznPrzec}. We choose
to work without the Nakanishi-Lautrup multipliers $h_\alpha(x)$, what seems
to make the perturbative calculations easier. After elimination of
$h_\alpha(x)$ by using their equations of motion we get the following
Lagrangian density
\eqs{\label{LtildeN}
{\mathcal{L}}^N(x)&=&-\frac{1}{4}\delta_{\alpha \beta}
F^\alpha_{\text{ }\mu\nu}(x)F^{\beta\text{}\mu\nu}(x)+
\frac{1}{2}\delta_{ab}\nawias{D_\mu\phi}^a\nawias{x}
\nawias{D^\mu\phi}^b\nawias{x}-
\mathcal{V}\nawias{\phi\nawias{x}}\nonumber\\
{}&{}&-\int{\text{d}^4 y\int\text{d}^4 z\text{ }
\overline{\omega}_{\alpha}(x)\frac{\delta f^{\alpha}(x)}
{\delta \mathcal{A}^i(z)}{D_{{}_R}}^i_{\text{ }\gamma}(z,y)
\omega^\gamma(y)}\\
{}&{}&-\frac{1}{2}(\xi^{-1}\nawias{x})_{\alpha \beta}
\nawias{f^\alpha(x)+\frac{1}{2}q^{\alpha\delta}\nawias{x}
\anty{\omega}_\delta\nawias{x}}
\nawias{f^\beta(x)+\frac{1}{2}q^{\beta\gamma}\nawias{x}
\anty{\omega}_\gamma\nawias{x}}\nonumber\\
{}&{}&+q^{\alpha\beta}\nawias{x}\anty{\omega}_\alpha
\nawias{x}\nawias{\phi^a\nawias{x}+
v^a_{{}_R}}\delta_{ac}\macierz{{T_{{}_R}}_\beta}^c_{\przerw b}u^b\nawias{x}
-q^{a}\nawias{x}\anty{\omega}_\alpha\nawias{x}\xi^{\alpha\beta}
\nawias{x}\delta_{ac}\macierz{{T_{{}_R}}_\beta}^c_{\przerw b}
\nawias{\phi^b\nawias{x}+v^b_{{}_R}}\nonumber\\
{}&{}&+\frac{1}{2}L_{\alpha}\nawias{x}
{e_{{}_R}}^\alpha_{\phantom{\alpha}\beta\gamma}
\omega^{\beta}\nawias{x}\omega^{\gamma}\nawias{x}
+\mathcal{K}_i\nawias{x}\calka{d}{y}{}\omega^\alpha\nawias{y}
{{D_{{}_R}}^i_{\text{ }\alpha}}\nawias{x,y}.\nonumber
}
In the last line of \refer{LtildeN} we have added the usual BRST sources
(see e.g. \cite{ZinnJustinQFTCritical}). We use the notation
$$
\mathcal{A}^i(z)=\nawias{\phi^a\nawias{z},A^\alpha_\mu\nawias{z}},\quad
\mathcal{K}_i(z)=\nawias{K_a\nawias{z},K_\alpha^\mu\nawias{z}},
$$
and
\eq{
{D_{{}_R}}^i_{\text{ }\gamma}(z,y)\equiv\pochfunc{ }{\omega^\gamma\nawias{y}}~\!
s\nawias{\mathcal{A}^i\nawias{z}}.
}
The action $\mathcal{I}^N$ corresponding to \refer{LtildeN} satisfies
the following Nielsen identity
\eq{\label{NielsenEqLiniowN}
\pochfunc{\mathcal{I}^N}{\mathcal{K}_i}\cdot\pochfunc{\mathcal{I}^N}
{\mathcal{A}^i}+
\pochfunc{\mathcal{I}^N}{L_{\alpha}}\cdot\pochfunc{\mathcal{I}^N}
{\omega^{\alpha}}-(\xi^{-1}(x))_{\beta\alpha}\nawias{f^\beta+
\frac{1}{2}q^{\beta\gamma}\anty{\omega}_\gamma}\cdot
\pochfunc{\mathcal{I}^N}{\anty{\omega}_{\alpha}}
+q^{\alpha\beta}\cdotp\pochfunc{\mathcal{I}^N}{\xi^{\alpha\beta}}
+q^{a}\cdotp\pochfunc{\mathcal{I}^N}{u^a}=0,
}
and the ghost equation
\eqs{\label{DuchEqLinN}
\pochfunc{\mathcal{I}^N}{\anty{\omega}_{\alpha}\nawias{x}}+
\pochfunc{f^\alpha\nawias{x}}{\mathcal{A}^i}\cdot\pochfunc{\mathcal{I}^N}
{\mathcal{K}_i}\!\!\!&=&\!\!\!
\frac{1}{2}q^{\alpha\delta}(x)(\xi^{-1}(x))_{\delta \beta}
\nawias{f^\beta(x)+\frac{1}{2}q^{\beta\gamma}\nawias{x}\anty{\omega}_\gamma
\nawias{x}}
-q^{\alpha\beta}\nawias{x}\nawias{\phi^a\nawias{x}+v^a_{{}_R}}\delta_{ac}
\macierz{{T_{{}_R}}_\beta}^c_{\przerw b}u^b(x)\nonumber\\
&{}&\!\!\!+q^a\nawias{x}\xi^{\alpha\beta}\nawias{x}\delta_{ac}
\macierz{{T_{{}_R}}_\beta}^c_{\przerw b}\nawias{\phi^b\nawias{x}
+v^b_{{}_R}},\phantom{aaaaaaaaaaaaaaaaaaaaaaaaaaaaaaaaaaaaaaaaa}
}
in which
$\mathcal{F}\cdotp\mathcal{G}
\equiv\calka{4}{x}{\mathcal{F}(x)~\!\mathcal{G}(x)}$.
For $q=0$, the formula \refer{NielsenEqLiniowN} reduces to the
Slavnov-Taylor identity of the BRST symmetry, the so-called Zinn-Justin
equation \cite{ZinnJustin1974}.
The right-hand side of \refer{DuchEqLinN} is at most linear in the
quantum fields. The same is true for the coefficients multiplying the
functional derivatives in \refer{NielsenEqLiniowN}. Putting
(\ref{NielsenEqLiniowN}) and (\ref{DuchEqLinN}) under the path
integral and integrating by parts, we obtain the corresponding identities
satisfied by the functional generating all Green's functions.
Converting them into identities for the functional generating connected
Green's  functions and, finally, performing the Legendre transform, we
find that the regularized  effective action also satisfies
(\ref{NielsenEqLiniowN}) and (\ref{DuchEqLinN}).
This is true because the dimensional regularization,
which we implicitly use, preserves the (extended) BRST symmetry
of the Yang-Mills theories coupled to scalars and vector-like fermions.
On the other hand, if there are  chiral fermions additional counterterms
are needed to restore the identities \cite{MartinSan, BarPas}.
\end{section}

\begin{section}{Renormalized Action}

\noindent Since the regularized effective action
respects the extended BRST symmetry, the standard Zinn-Justin arguments
(see e.g. \cite{ZinnJustin1974,ZinnJustinQFTCritical} and Appendix
\ref{app:Stability}) imply that the action 
$\tilde{\mathcal{I}}^N$ which includes counterterms (as well as the
renormalized effective action~$\Gamma^N$) also satisfy the equations
(\ref{NielsenEqLiniowN}) and (\ref{DuchEqLinN}). Therefore finding
the most general form of $\tilde{\mathcal{I}}^N$, which is the purpose of
this paper, reduces to writing down the most general, local dimension four
function of the fields and sources with zero ghost number and to extract the
constraints imposed on the coefficients of $\tilde{\mathcal{I}}^N$ by the
Nielsen identity (\ref{NielsenEqLiniowN}) and the ghost equation of
motion (\ref{DuchEqLinN}). Details of the derivation of these constraints (i.e. to the equations (\ref{NielsenEqLiniowN})
and (\ref{DuchEqLinN})) are given in Appendix \ref{app:TozNielPrzec}.
Here we present only the final result.

\vskip0.2cm

The Lagrangian  with all possible counterterms has the form
\eqs{\label{LtildeNPW}
\tilde{\mathcal{L}}^N(x)&=&-\frac{1}{4}\macierz{Z_{{{}_A}}\!
\nawias{\xi\nawias{x}}}_{\alpha\beta}\tilde{F}^\alpha_{\text{ }\mu\nu}(x)
\tilde{F}^{\beta\text{}\mu\nu}(x)
+\frac{1}{2}\macierz{Z_{{{}_\phi}}\!\nawias{\xi\nawias{x}}}_{ab}
\nawias{\tilde{D}_\mu\check{\phi}}^a\!\nawias{x}
\nawias{\tilde{D}^\mu\check{\phi}}^b\!\nawias{x}-\tilde{\mathcal{V}}
\nawias{\check{\phi}(x),\xi\nawias{x}}\nonumber\\
{}&{}&-\int{\text{d}^d y\int\text{d}^d z\text{ }
\overline{\omega}_{\alpha}(x)~\!\frac{\delta f^{\alpha}(x)}
{\delta \mathcal{A}^i(z)}~\!\funkcjonal{D^i_{\text{ }\gamma}}
{\mathcal{A},u,\xi|z,y}\omega^\gamma(y)}\nonumber\\
{}&{}&-\frac{1}{2}(\xi^{-1}\nawias{x})_{\alpha \beta}
\nawias{f^\alpha(x)+\frac{1}{2}q^{\alpha\delta}
\nawias{x}\anty{\omega}_\delta\nawias{x}}
\nawias{f^\beta(x)+\frac{1}{2}q^{\beta\gamma}
\nawias{x}\anty{\omega}_\gamma\nawias{x}}\nonumber\\
{}&{}&+q^{\alpha\beta}\nawias{x}\anty{\omega}_\alpha\nawias{x}
\nawias{\phi^a\nawias{x}+
v^a_{{}_R}}\delta_{ac}\macierz{{T_{{}_R}}_\beta}^c_{\przerw b}u^b\nawias{x}
-q^{a}\nawias{x}\anty{\omega}_\alpha\nawias{x}\xi^{\alpha\beta}
\nawias{x}\delta_{ac}\macierz{{T_{{}_R}}_\beta}^c_{\przerw b}
\nawias{\phi^b\nawias{x}+v^b_{{}_R}}\nonumber\\
{}&{}&+\frac{1}{2}L_{\alpha}\nawias{x}C^\alpha_{\phantom{\alpha}\beta\gamma}
\nawias{\xi\nawias{x}}\omega^{\beta}\nawias{x}\omega^{\gamma}\nawias{x}
+\mathcal{K}_i\nawias{x}\calka{d}{y}{}\omega^\alpha\nawias{y}
\funkcjonal{D^i_{\text{ }\alpha}}{\mathcal{A},u,\xi|x,y}\nonumber\\
&{}&+\Delta\tilde{\mathcal{L}}^N(x).
}
Each line (except the last one) of the Lagrangian \refer{LtildeNPW} is a
counterpart with counterterms included of the corresponding line of the
Lagrangian \refer{LtildeN}. The gauge-fixing function $f^\alpha(x)$ is the
same in both cases because we restrict ourselves to linear gauges
\refer{falphastart}. For the same reason the fourth line of
\refer{LtildeNPW} does not change after renormalization (see Appendix
\ref{app:TozNielPrzec}).
$\Delta\tilde{\mathcal{L}}^N(x)$ denotes additional counterterms
required in the renormalization of Green's functions with insertions of
composite operators, which are coupled to Nielsen sources $q(x)$:
\eqs{\label{LtotLincechNUXIPW}
\Delta\tilde{\mathcal{L}}^N(x)&=&L_{\alpha}
\nawias{x}g^\alpha_{\przerwpod\beta\gamma\delta}
\nawias{\xi\nawias{x}}q^{\beta\gamma}\nawias{x}\omega^{\delta}\nawias{x}
+L_{\alpha}\nawias{x}H^\alpha_{\przerwpod\beta\gamma\delta\epsilon}
\nawias{\xi\nawias{x}}q^{\beta\gamma}\nawias{x}q^{\delta\epsilon}
\nawias{x}\\
{}&{}&+\mathcal{K}_i\nawias{x}\calka{d}{y}{}
\funkcjonal{k^i_{\text{ }\alpha\beta}}{\mathcal{A},u,\xi|x,y}
q^{\alpha\beta}\nawias{y}
-\calkabp{d}{y}{}\calka{d}{z}{}\anty{\omega}_\alpha\nawias{x}
\frac{\delta f^\alpha\nawias{x}}{\delta\mathcal{A}^i(y)}~\!
\funkcjonal{k^i_{\przerw\beta\gamma}}{\mathcal{A},u,\xi|y,z}q^{\beta\gamma}(z)
\nonumber\\
&{}&+K_{c}\nawias{x}b^{c}_{\phantom{c}a}\!
\nawias{\xi\nawias{x}}q^a\nawias{x}
-\calka{d}{z}{}\anty{\omega}_\alpha\nawias{x}
\pochfunc{f^\alpha\nawias{x}}{\phi^c\nawias{z}}~\!{b^{c}_{\phantom{c}a}\!
\nawias{\xi\nawias{z}}}q^a\nawias{z}.\nonumber
}
The gauge invariant part of \refer{LtildeNPW} depends on the combinations
\eqs{\label{WiekosciKowariantneFullNUXIPW}
\tilde{F}^\alpha_{\text{ }\mu\nu}(x)&=&\partial_\mu A^\alpha_\nu\nawias{x}
-\partial_\nu A^\alpha_\mu\nawias{x}
+e^\alpha_{\przerwpod\beta\gamma}\nawias{\xi\nawias{x}}
A^\beta_\mu\nawias{x}A^\gamma_\nu\nawias{x}+\ldots,\\
\nawias{\tilde{D}_\mu\check{\phi}}\nawias{x}
&=&\partial_\mu \check{\phi}\nawias{x}+A^\alpha_\mu\nawias{x}
\macierz{T_\alpha\nawias{\xi\nawias{x}}\check{\phi}
\nawias{x}+ V_{N\alpha}\nawias{\xi\nawias{x}}}+\ldots,
}
where the ellipses stand for contributions which vanish for
$x$-independent $\xi$ configurations; explicit form of these terms is
given in Appendix \ref{app:TozNielPrzec} (formulae following eq.
\refer{WiekosciKowariantneFullNUXI}). Notice that the gauge invariant part
of the Lagrangian depends on $u$ only through the combination
\eq{\label{Eq:checkPhidef}
\check{\phi}(x)\equiv\phi(x)-b\nawias{\xi(x)}u(x),
}
in which $b\nawias{\xi(x)}$ is a matrix which determines the counterterms
to Green's functions with insertions of the composite operators coupled to
the external sources (the third line of \refer{LtotLincechNUXIPW}).
This follows immediately from the Nielsen identity (see Appendix
\ref{app:TozNielPrzec}). The kernels
$\funkcjonal{D^{i}_{\phantom{i}\alpha}}{\mathcal{A},u,\xi|x,y}$ of the gauge
transformations are now given by
\eqs{
\funkcjonal{D^{\beta}_{\mu\alpha}}{A,u,\xi|x,y}
&=&\mathcal{N}^{\delta}_{\przerw\alpha}\nawias{\xi\nawias{x}}
\left[-\delta^{\beta}_{\przerw\delta}\partial_\mu\delta^{(d)}
\nawias{x-y}+e^{\beta}_{\przerwpod\delta \gamma}
\nawias{\xi\nawias{x}}A^\gamma_{\mu}\nawias{x}
\delta\nawias{x-y}\right]+\ldots,
\label{TransCechFull1RozdzNUXIPW}\\
\funkcjonal{D^{a}_{\phantom{a}\alpha}}{\phi,u,\xi|x,y}
&=&\mathcal{N}^{\delta}_{\przerw\alpha}\nawias{\xi\nawias{x}}
\left\{\left[T_{\delta}\nawias{\xi\nawias{x}}\right]^a_{\przerwpod b}
{\check{\phi}}^b\nawias{x}+V^a_{N\delta}\nawias{\xi\nawias{x}}\right\}
\delta\nawias{x-y}.
\label{TransCechFull2RozdzNUXIPW}
}
Finally, the kernels
$\funkcjonal{{k^i}_{\alpha\gamma}}{\mathcal{A},u,\xi|x,y}$
correspond to the extended gauge transformations
\eq{\label{TransCechFull1RozdzNUXIkPW}
\funkcjonal{k^{\beta}_{\mu\alpha\gamma}}{A,u,\xi|x,y}
=-\Omega^{\beta}_{\przerw\alpha\gamma}\nawias{\xi\nawias{x}}
\partial_\mu\delta\nawias{x-y}
+\theta^{\beta}_{\przerwpod\alpha\gamma \epsilon}
\nawias{\xi\nawias{x}}A^\epsilon_{\mu}\nawias{x}\delta\nawias{x-y}+\ldots,}
\eq{\label{TransCechFull2RozdzNUXIkPW}
\funkcjonal{k^{a}_{\phantom{a}\alpha\gamma}}
{\phi,u,\xi|x,y}=
\funkcjonal{}{\zeta^a_{\przerwpod\alpha\gamma c}
\nawias{\xi\nawias{x}}\check{\phi}^c\nawias{x}
+\poch{b^{a}_{\phantom{a} d}}{\xi^{\alpha\gamma}}
\nawias{\xi\nawias{x}}u^d\nawias{x}
 +\Sigma^a_{\przerw\alpha\gamma}
\nawias{\xi\nawias{x}}}\delta\nawias{x-y},
}
whose dimensionful parameters $V^{a}_{N\delta}\nawias{\xi}$ and
$\Sigma^a_{\przerw\alpha\gamma}\nawias{\xi}$ are independent of $u$.
\vskip0.2cm

The Nielsen identity imposes a number of conditions on the coefficients
of the Lagrangian \refer{LtildeNPW}. Among them one finds of course the
ordinary BRST constraints \cite{Wein}. These imply firstly that the
``bare'' parameters $e^{\gamma}_{\przerwpod\alpha \beta}$ must satisfy the Jacobi
identity whereas the matrices $T_{\alpha}$ must obey the commutation
relations
\eq{\label{RelKomFullRozdzXIPW}
\left[T_{\alpha},~T_{\beta}\right]=e^{\gamma}_{\przerwpod\alpha \beta}
T_{\gamma}.
}
Secondly, that the structure constants $C^{\kappa}_{\phantom{\kappa}\beta\gamma}$
are related to $\indgd{e}{\gamma}{\alpha\beta}$ by the change of the Lie algebra basis
\eq{\label{ZwiazekCieRozdzXIPW}
C^{\kappa}_{\phantom{\kappa}\beta\gamma}=\mathcal{N}^\alpha_{\przerw\beta}
\mathcal{N}^\epsilon_{\przerw\gamma}
\left[\mathcal{N}^{-1}\right]^\kappa_{\przerw\delta}
e^{\delta}_{\przerwpod\alpha \epsilon},
}
and that the usual antisymmetry conditions must hold
\eq{\label{ZinvDfullXIPW}
\macierz{Z_A}_{\epsilon\alpha}{e}^\alpha_{\przerwpod\beta\gamma}=
-\macierz{Z_A}_{\gamma\alpha}{e}^\alpha_{\przerwpod\beta\epsilon},
\przerwpod\przerwpod\przerwpod\przerwpod
\macierz{Z_\phi}_{c a} \macierz{{T}_\alpha}^a_{\przerw b}=
-\macierz{Z_\phi}_{ba} \macierz{{T}_\alpha}^a_{\przerw c},
}
together with the following equation for the function
$\tilde{\mathcal{V}}\nawias{\check{\phi},\xi}$
\eq{\label{VinvfullNUXIPW}
\nawias{\macierz{T_\alpha\nawias{\xi}}^a_{\przerw b}\Phi^b+V^a_{N\alpha}
\nawias{\xi}}\poch{\tilde{\mathcal{V}}\nawias{\Phi,\xi}}{\Phi^a}=0.
}
Finally, that the matrices $V^b_{N\beta}$ must satisfy the standard cocycle
equation \cite{BRS}
\eq{\label{RelKomVRozdzNUXIPW}
\left[T_{\alpha}\right]^a_{\przerw b}V^b_{N\beta}
-\left[T_{\beta}\right]^a_{\przerw b}V^b_{N\alpha}=
e^{\gamma}_{\przerwpod\alpha \beta}V^a_{N\gamma}.
}
Moreover, $V^b_{N\beta}$ are  independent of $u$, because the scalar field
has been shifted as specified in \refer{Eq:checkPhidef}.
\vskip0.2cm

The remaining requirements of the Nielsen identity can
most concisely be expressed in terms of the differential
forms\footnote{Coefficients $H^\alpha_{\przerwpod\beta\gamma\delta\epsilon}$
can be treated as antisymmetric with respect to the interchange
$(\beta\gamma)\leftrightarrow(\delta\epsilon)$.}
\eq{
\hat{g}^\alpha_{\przerwpod\delta}=g^\alpha_{\przerwpod\beta\gamma\delta}
\nawias{\xi}\text{d}\xi^{\beta\gamma},\przerwpod\przerwpod\przerwpod
\hat{H}^\alpha=H^\alpha_{\przerwpod\beta\gamma\delta\epsilon}
\nawias{\xi}\text{d}\xi^{\beta\gamma}\wedge\text{d}\xi^{\delta\epsilon},
}
and
\eq{\label{Eq:forms}
\hat{\theta}^\beta_{\przerw\epsilon}=
\theta^\beta_{\przerwpod\alpha\gamma\epsilon}
\nawias{\xi}\text{d}\xi^{\alpha\gamma},\przerwpod\przerwpod\przerwpod
\hat{\zeta}^a_{\przerw b}=\zeta^a_{\przerw\alpha\gamma b}
\nawias{\xi}\text{d}\xi^{\alpha\gamma},\przerwpod\przerwpod\przerwpod
\hat{\Sigma}^a=\Sigma^a_{\przerw\alpha\gamma}\nawias{\xi}
\text{d}\xi^{\alpha\gamma}.
}
In addition it is convenient to define
\eq{\label{Eq:Psi_defPW}
\hat{\Psi}^\kappa\equiv\mathcal{N}^\kappa_{\przerwpod\epsilon}
\hat{H}^\epsilon.
}
In this language the $\xi$-dependence of the matrix Lie algebra
generators acting on vector and scalar fields, respectively is
governed by the following equations\footnote{In our notation
${e}_{\alpha}=\left[{e}_{\alpha}\right]^{\beta}_{\text{ }\gamma}
\equiv{e}^{\beta}_{\text{ }\alpha \gamma}.$
}
\eqsN{\label{Eq:dePW}
&\text{d}e_\gamma=\macierz{\hat{\theta},~e_\gamma}
-\hat{\theta}^\delta_{\przerwpod\gamma}e_\delta,
}
\eq{\label{Eq:dTPW}
\text{d}T_\gamma=\macierz{\hat{\zeta},~T_\gamma}
-\hat{\theta}^\delta_{\przerwpod\gamma}T_\delta.
}
We also get the relation
\eq{\label{Eq:dVNPW}
\text{d}V^a_{N\gamma}=\hat{\zeta}^a_{\przerw b}V^b_{N\gamma}
-\macierz{T_{\gamma}}^a_{\przerw b}\hat{\Sigma}^b
-V^a_{N\delta}\hat{\theta}^\delta_{\przerw\gamma}.
}
Furthermore, the 1-forms $\hat{\theta}$, $\hat{\zeta}$ and
$\hat{\Sigma}$ satisfy the equations
\eq{\label{Eq:dthetaPW}
\text{d}\hat{\theta}=\hat{\theta}\wedge\hat{\theta}
-\hat{\Psi}^\epsilon e_\epsilon,
}
\eq{\label{Eq:dzetaPW}
\text{d}\hat{\zeta}=\hat{\zeta}\wedge\hat{\zeta}
-\hat{\Psi}^\epsilon{T}_\epsilon,
}
and
\eq{\label{Eq:dSigmaPW}
\text{d}\hat{\Sigma}=\hat{\zeta}\wedge\hat{\Sigma}
-\hat{\Psi}^\epsilon V_{N\epsilon}.
}
In turn, the forms $\hat{\Psi}^{\sigma}$ are constrained by the relation
\eq{\label{Eq:dPsiPW}
\text{d}\hat{\Psi}^{\sigma}=\hat{\theta}^\sigma_{\przerw\alpha}
\wedge\hat{\Psi}^{\alpha}.
}
$\xi$-dependence of the potential $\tilde{\mathcal{V}}\nawias{\Phi,\xi}$
with the counterterms included obeys
\eq{\label{VinvfullNUXIkPW}
\poch{\tilde{\mathcal{V}}\nawias{\Phi,\xi}}{\xi^{\alpha\beta}}+
\nawias{\zeta^a_{\przerw\alpha\beta b}\nawias{\xi}
\Phi^b+\Sigma^a_{\przerw\alpha\beta}\nawias{\xi}}
\poch{\tilde{\mathcal{V}}\nawias{\Phi,\xi}}{\Phi^a}\equiv 0.
}
Finally, we find that the gauge and scalar field renormalization
constants $Z_A$ and $Z_\phi$ satisfy the conditions
\eqsN{
\label{Eq:dZAiphiPW}
&\text{d}\macierz{Z_A}_{\kappa\delta}=
-\macierz{Z_A}_{\kappa\epsilon}\hat{\theta}^\epsilon_{\przerwpod\delta}
-\macierz{Z_A}_{\epsilon\delta}\hat{\theta}^\epsilon_{\przerwpod\kappa},
\!\!\!\!\!\!\!\!\!\!\!\!\!\!\!\!\!\!\!\!\!
&\text{d}\macierz{Z_\phi}_{ab}=-\macierz{Z_\phi}_{a c}
\hat{\zeta}^c_{\przerw b}
-\macierz{Z_\phi}_{cb} \hat{\zeta}^c_{\przerw a},
}
while $\xi$-dependence of the factors $\mathcal{N}$, which in linear
gauges (like (\ref{falphastart})) have the interpretation of the
ghost fields renormalization constants\footnote{This follows from
\refer{LtildeNPW} because \refer{falphastart} is unaffected
by radiative corrections.}
is constrained by the condition
\eq{\label{Eq:dNPW}
\text{d}\mathcal{N}=\hat{\theta}\mathcal{N}-\mathcal{N}\hat{g}.
}
\vskip0.2cm

More information can be obtained by exploiting invariance of the
action $\refer{LtildeN}$ under global gauge transformations which
remain symmetries of the action if $\xi^{\alpha\beta}(x)$ and $u^{a}(x)$ are
treated as external fields, which also undergo transformations. Details
are presented in Appendix \ref{app:WnioskiZsymG}. Introducing the vector
fields $\Lambda_\alpha$ generating transformations of $\xi^{\alpha\beta}$
\eqsN{\label{Eq:DefpolLambdaPW}
&\Lambda_{\alpha}=\nawias{{e_{{}_{R}}}^{\epsilon}_{\przerwpod\alpha\beta}
\xi^{\beta\kappa}
+{e_{{}_{R}}}^{\kappa}_{\przerwpod\alpha\beta}\xi^{\beta\epsilon}}
\poch{}{\xi^{\epsilon\kappa}},
\!\!\!\!\!\!\!\!\!\!\!\!\!\!\!\!\!\!\!\!\!\!\!\!\!\!\!\!\!\!\!\!
&\funkcjonal{}{\Lambda_\alpha,~\Lambda_\beta}=-\Lambda_\gamma
\eR{\gamma}{\alpha}{\beta}.
}
we obtain the relation
\eqs{\label{Eq:LieEPW}
\nawias{\Lambda_\alpha e_\kappa}\nawias{\xi}&=&
\macierz{{e_{{}_R}}_\alpha,~e_\kappa\nawias{\xi}}
-e_\epsilon\nawias{\xi} {e_{{}_{R}}}^{\epsilon}_{\przerwpod\alpha\kappa},\\
\label{Eq:LieTPW}
\nawias{\Lambda_\alpha T_\kappa}\nawias{\xi}&=&\macierz{{T_{{}_R}}_\alpha,~
T_\kappa\nawias{\xi}}-T_\epsilon\nawias{\xi}
{e_{{}_{R}}}^{\epsilon}_{\przerwpod\alpha\kappa},
}
and similar equations for $V^a_{N\kappa}$
\eqs{
\label{Eq:LieVNPW}
\nawias{\Lambda_\alpha V_{N\kappa}}\nawias{\xi}&=&{T_{{}_R}}_\alpha
V_{N\kappa}\nawias{\xi}-T_\kappa\nawias{\xi} {T_{{}_{R}}}_\alpha v_{{}_R}
-V_{N\epsilon}\nawias{\xi}{e_{{}_{R}}}^{\epsilon}_{\przerwpod\alpha\kappa}.
}
The matrix valued field renormalization constants
$Z_A$, $Z_\phi$ and $\mathcal{N}$ obey
\eqs{
\label{Eq:LieZAPW}
\nawias{\Lambda_\alpha\macierz{Z_A}_{\beta\kappa}}\nawias{\xi}&=&
-\macierz{Z_A\nawias{\xi}}_{\beta\delta}
{e_{{}_R}}^\delta_{\przerwpod\alpha\kappa}
-\macierz{Z_A\nawias{\xi}}_{\delta\kappa}
{e_{{}_R}}^\delta_{\przerwpod\alpha\beta},\\
\label{Eq:LieZphiPW}
\nawias{\Lambda_\alpha\macierz{Z_\phi}_{b c}}\nawias{\xi}&=&
-\macierz{Z_\phi\nawias{\xi}}_{bd}\macierz{{T_{{}_R}}_\alpha}^d_{\przerw c}
-\macierz{Z_\phi\nawias{\xi}}_{dc}\macierz{{T_{{}_R}}_\alpha}^d_{\przerw b},\\
\label{Eq:LieNPW}
\nawias{\Lambda_\alpha\mathcal{N}}\nawias{\xi}&=&\macierz{{e_{{}_R}}_\alpha,~
\mathcal{N}\nawias{\xi}},
}
while the matrix $b$ appearing in (\ref{Eq:checkPhidef})
satisfies the condition
\eq{
\label{Eq:LieBPW}
\nawias{\Lambda_\alpha b}\nawias{\xi}=\macierz{{T_{{}_R}}_\alpha,~
b\nawias{\xi}}.
}
The corresponding relations satisfied by the differential forms
\refer{Eq:forms} can be compactly expressed with the help of the Lie
derivatives with respect to vector fields
\refer{Eq:DefpolLambdaPW}:\footnote{
Let's remind that for differential forms the most convenient definition
of Lie derivative is by its properties (1)
$\mathfrak{L}_{\cal{X}}f=\mathcal{X} f $, for an arbitrary function $f$ and
a vector field $\mathcal{X}$, (2)
$\mathfrak{L}_{\cal{X}}\mathrm{d}\hat{\rho}
=\mathrm{d}\mathfrak{L}_{\cal{X}}\hat{\rho}$, (3)
$\mathfrak{L}_{\cal{X}}\nawias{\hat{\rho}+\hat{\sigma}}=
\mathfrak{L}_{\cal{X}}\hat{\rho}+\mathfrak{L}_{\cal{X}}\hat{\sigma}$,
and (4)
$\mathfrak{L}_{\cal{X}}\nawias{\hat{\rho}\wedge\hat{\sigma}}=
\nawias{\mathfrak{L}_{\cal{X}}\hat{\rho}}\wedge\hat{\sigma}
+\hat{\rho}\wedge\mathfrak{L}_{\cal{X}}\hat{\sigma}$,
for any forms $\hat{\rho}$ and $\hat{\sigma}$.
}
\eqs{
\label{Eq:LieThetaPW}
\mathfrak{L}_{{\Lambda_\alpha}}\hat{\theta}&=&
\macierz{{e_{{}_R}}_\alpha,~\hat{\theta}},\\
\label{Eq:LieZetaPW}
\mathfrak{L}_{\Lambda_\alpha}\hat{\zeta}&=&
\macierz{{T_{{}_R}}_\alpha,~\hat{\zeta}},\\
\label{Eq:LieSigmaPW}
\mathfrak{L}_{\Lambda_\alpha}\hat{\Sigma}
&=&{T_{{}_R}}_\alpha\hat{\Sigma}-\hat{\zeta}~\!{T_{{}_{R}}}_\alpha v_{{}_R}.
}
Similarly, the form $\hat\Psi$ defined in \refer{Eq:Psi_defPW} satisfies
\eq{\label{Eq:LiePsiPW}
\mathfrak{L}_{{\Lambda_\alpha}}\hat{\Psi}={e_{{}_R}}_\alpha\hat{\Psi}.
}
Defining $\hat{\Omega}^\alpha\equiv\indgd{\Omega}
{\alpha}{\beta\gamma}\nawias{\xi}\mathrm{d}\xi^{\beta\gamma}$
for the coefficient of $\indgd{k}{i}{\alpha\beta}$ distribution we find
\eqsN{\label{EqR11nowePW}
\mathfrak{L}_{{\Lambda_\alpha}}\hat{\Omega}={e_{{}_R}}_\alpha\hat{\Omega}.
}
The analogous equation for $\hat{g}$ follows now from \refer{Eq:dNPW}
(see Appendix \ref{app:WnioskiZsymG}). Finally, invariance with respect to
global transformations implies the relation
\eq{\label{Eq:SymStartVPW}
\nawias{\Lambda_{\alpha}\tilde{\mathcal{V}}}\nawias{\Phi,\xi}
+\macierz{{T_{{}_R}}_\alpha}^a_{\przerw b}\nawias{\Phi^b
+v^b_{{}_R}}\poch{\tilde{\mathcal{V}}\nawias{\Phi,\xi}}{\Phi^a}\equiv 0.
}

In Appendix \ref{app:AbelPod} we consider the case in which the
gauge Lie algebra contains an abelian ideal. For any abelian gauge
field~$A^{\alpha_0}_\mu$ the Ward-Takahashi identity gives
\eqs{
T_{\alpha_0}\nawias{\xi}&=&{T_{{}_R}}_{\alpha_0},\label{Eq:Z1rZ2PW}\\
V_{N\alpha_0}\nawias{\xi}&=&{T_{{}_R}}_{\alpha_0}v_{{}_R}.
\label{Eq:VNsolAbelPW}
}
Equation \refer{Eq:Z1rZ2PW} is a QED-like \mbox{`$Z_1=Z_2$'} identity.
It is also shown in Appendix \ref{app:AbelPod} that the factors
$e^{\alpha}_{\phantom{\alpha}\beta\gamma}$,
$\hat{\theta}^{\alpha}_{\phantom{\alpha}\beta}$ and $\hat{\Psi}^\alpha$
vanish, when any of their indices corresponds to an abelian field. (In
particular, the first equation \refer{Eq:dZAiphiPW} tells us that the
abelian field renormalization constants $\macierz{Z_A}_{\alpha_0\beta_0}$ are
\mbox{$\xi$-independent}, what is well known.) Similar non-renormalization
theorems hold for any gauge-singlet scalar field $\phi^{a_0}$:
\eq{\label{Eq:SinglPW}
\macierz{T_{\alpha}}^{a_0}_{\phantom{a_0}b}=V^{a_0}_{N\alpha}
=\hat{\zeta}^{a_0}_{\phantom{{a_0}}b}
=\hat{\Sigma}^{a_0}=b^{{a_0}}_{\phantom{{{\alpha_0}}}c}=0.
}

In this way we have exhausted general information coming from the Nielsen
identities as well as from the symmetry under global gauge transformations.
(In specific models other global symmetries can of course provide additional
constraints). The relations \refer{Eq:dePW}, \refer{Eq:dTPW} and
\refer{VinvfullNUXIkPW} show that the 1-forms $\hat{\theta}$ and
$\hat{\zeta}$ control the $\xi$-dependence of the ordinary counterterms
carrying the indices respectively of the vector and scalar fields. On
the other hand, the remaining equations that govern the $\xi$-dependence,
i.e. \refer{Eq:dthetaPW}, \refer{Eq:dzetaPW} and \refer{Eq:dPsiPW}, can
be treated as the consistency conditions which ensure that $\text{d}^2=0$.
In particular, the 2-form $\hat{R}\equiv\hat{\Psi}^\epsilon e_\epsilon$ is the
curvature associated with the extended gauge invariance. Comparing with
the case of a single $\xi$ parameter considered in \cite{PiguetSibold},
$\hat{\Psi}^\epsilon$ is an additional counterterm
a priori necessary to make finite
Green's functions of three composite operators; it is shown in Appendix
\ref{appAlgebryPP}, that this counterterm is required only if global gauge
invariance is broken by $\xi^{\alpha\beta}$. Moreover, $\hat{\Psi}^\epsilon$
vanishes in the one-loop order because divergences of two graphs cancel
each other (see Section \ref{Sec:WyznPrzec}). As was noticed in
\cite{PiguetSibold}, in the case of Fermi gauges with a single $\xi$
parameter the equations controlling the $\xi$-dependence of the
counterterms ensure that the bare gauge coupling constant of bare gauge
fields is $\xi$-independent. We will show in the next section that
occurrence of $\hat{\Psi}^\epsilon$ does not spoil this property
in the case of the general $R_{\xi,u}$ gauges \refer{falphastart} and
that the formula \refer{VinvfullNUXIkPW} leads to the similar conclusion
for all coupling constants of bare scalar fields.
\vskip0.2cm

Since the gauge-fixing parameter $u$ has positive dimension, the dependence
of counterterms on $u$ is even more constrained. The gauge transformations,
the covariant derivative of the scalar fields and the potential with
counterterms depend on $u$ only through the shifted field
\refer{Eq:checkPhidef}. It is therefore natural to check, whether this
shift makes
the entire contribution to the infinite VEV counterterm. Thus, we are
interested in the relation between $V_{N\alpha}$ and the background
$v_{{}_R}$. Comparing  the action $\refer{LtildeN}$ with its
renormalized counterpart \refer{LtildeNPW}, we see that
to the lowest order $V_{N\alpha}=T_{\alpha}v_{{}_R}$. For an abelian index
$\alpha_0$ the equality $V_{N\alpha_0}=T_{\alpha_0}v_{{}_R}$ is exact, as follows
from \refer{Eq:Z1rZ2PW} and \refer{Eq:VNsolAbelPW}. Moreover,
$\Lambda_{\alpha_0}\equiv 0\equiv\mathfrak{L}_{\Lambda_{\alpha_0}}$, hence
the equations \refer{Eq:LieTPW} and
\refer{Eq:LieZetaPW} read
\eq{\label{Eq:KomzAbel}
\macierz{{T_{{}_R}}_{\alpha_0},~T_\kappa\nawias{\xi}}=0,
\qquad\macierz{{T_{{}_R}}_{\alpha_0},~\hat{\zeta}}=0,
}
while \refer{Eq:LieVNPW} and \refer{Eq:LieSigmaPW} reduce to
\eq{
{T_{{}_R}}_{\alpha_0}V_{N\kappa}\nawias{\xi}-T_\kappa
\nawias{\xi} {T_{{}_{R}}}_{\alpha_0} v_{{}_R}=0,
\qquad {T_{{}_R}}_{\alpha_0}\hat{\Sigma}-\hat{\zeta}
{T_{{}_{R}}}_{\alpha_0} v_{{}_R}=0,
}
so that the commutativity \refer{Eq:KomzAbel} leads to the relations
\eq{\label{Eq:RelAbelDlaVN}
{T_{{}_R}}_{\alpha_0}\nawias{V_{N\kappa}\nawias{\xi}
-T_\kappa\nawias{\xi}v_{{}_R}}=0,
\qquad {T_{{}_R}}_{\alpha_0}\nawias{\hat{\Sigma}-\hat{\zeta}v_{{}_R}}=0.
}
Let us first consider the class of theories (containing the Standard
Model), in which the gauge algebra is not semisimple and there are no
scalar singlets with respect to the abelian gauge ideal. In such cases
equations \refer{Eq:RelAbelDlaVN} yield
\eq{\label{Eq:ZwiazekVNivR}
V_{N\kappa}\nawias{\xi}=T_\kappa\nawias{\xi}v_{{}_R}, \qquad
\hat{\Sigma}=\hat{\zeta}v_{{}_R},
}
for all gauge indices $\kappa$, so that the gauge invariant part of
the renormalized action depends only on the sum
\eq{\label{Eq:kombphi}
\check{\phi}+v_{{}_R}\equiv\phi+v_{{}_R}-b\nawias{\xi}u,
}
and the parameter
\eq{\label{KontrczlonVEVXI}
v_{{}_B}\equiv v_{{}_R}+\delta v=v_{{}_R}-b\nawias{\xi}u,
}
can be interpreted as the bare background with
\eq{\label{Eq:deltaV}
\delta v= -b\nawias{\xi}u,
}
being the scalar field VEV counterterm. Considering more general theories,
we know only that $V_{N\alpha}$ are $u$-independent on account of the Nielsen
identities. On the other hand the tree level action \refer{LtildeN} depends
on the background $v_{{}_R}$ only through the sum \mbox{$\phi+v_{{}_R}$}, and
the same has to be true for the renormalized action \refer{LtildeNPW},
because the Dimensional Regularization respects formal invariance of the
path integral under translations. This implies that
\eq{
V_{N\alpha}\nawias{\xi}=T_\alpha\nawias{\xi}v_{{}_R}+W_\alpha(\xi),\qquad
}
with $W_\alpha(\xi)$ independent of $u$ and $v_{{}_R}$.
As argued above, $W_{\alpha_0}(\xi)=0$ for any abelian index $\alpha_0$.
For arbitrary indices the equation
\refer{RelKomVRozdzNUXIPW} yields
\eq{\label{Eq:cohW}
\left[T_{\alpha}\right]^a_{\przerw b}W^b_{\beta}
-\left[T_{\beta}\right]^a_{\przerw b}W^b_{\alpha}=
e^{\gamma}_{\przerwpod\alpha \beta}W^a_{\gamma}.}
As we have seen, the coefficients $e^{\gamma}_{\przerwpod\alpha \beta}$ are
non-vanishing only for non-abelian indices $\gamma_1$, $\alpha_1$ and
$\beta_1$ which means that the matrices $T_{\alpha_1}$ form a representation
of a semisimple Lie algebra. The solution to the equation (\ref{Eq:cohW})
must  therefore have the form
\eq{\label{Eq:Wsol}
W_{\gamma_1}=T_{\gamma_1}w,
}
because the first cohomology space is trivial for any representation of a
semisimple Lie algebra \cite{BRS,BBBC}. In general one should not expect
that $w=0$ because global symmetry breaking by $\xi$ can produce a
scalar-vector mixing even in the symmetric phase, if the Lagrangian
involves trilinear scalar couplings. On the other hand $\Lambda_\alpha=0$,
if the parameters $\xi$ are introduced without spoiling the invariance with
respect to global gauge transformations. Hence, comparing equations
\refer{Eq:LieVNPW} and \refer{Eq:LieTPW} one finds in this case that
\eqs{
\label{Eq:SymW}
{T_{{}_R}}_\alpha W_{\kappa}
=W_{\epsilon}{e_{{}_{R}}}^{\epsilon}_{\przerwpod\alpha\kappa}.
}
Using the solution \refer{Eq:Wsol} in \refer{Eq:SymW} and comparing with
\refer{Eq:LieTPW} one obtains
\eqs{
\label{Eq:Symwupr}
{T}_{\gamma_1} {T_{{}_R}}_\alpha w=0.
}
It is well known that if $\xi$ preserves global invariance, $T_{\gamma_1}$
differ from $T_{{}_R\gamma_1}$ only by the separate renormalizations of
gauge couplings of each simple ideal (see e.g. \cite{Wein}). As a result,
owing to the block-diagonal form of $T_{{}_R\gamma_1}$, the equality
\refer{Eq:Symwupr} implies that ${T}_{{}_{R}\beta_1}w=0$ and, finally,
the formula \refer{Eq:Wsol} yields
\eq{
W_{\gamma_1}=0,
}
so that \refer{Eq:ZwiazekVNivR} remains true. Furthermore, in the
presence of
singlets under the entire gauge group, additional global symmetries can be
used (like in models of spontaneous lepton number violation \cite{Meiss})
to ensure that singlets do not acquire an infinite VEV, while for singlets
neutral with respect to global symmetries any $\delta v$ can be absorbed into
a linear term in the scalar potential. Therefore, if the global gauge symmetry
is broken only by the $u$ parameters, the equation \refer{Eq:deltaV} is
satisfied and the Nielsen identity allows to completely determine the VEV
counterterm $\delta v$ in terms of $b(\xi)$. To our knowledge this relation
has never been presented in the literature. At one-loop the relation
\refer{Eq:deltaV} offers a simple way to compute $\delta v$ (see section
\ref{Sec:WyznPrzec}).
\end{section}

\begin{section}{Gauge Independence of Bare Coupling Constants}

\indent In this section we will show that the Nielsen identities for
counterterms express the $\xi$-independence of bare coupling constants
of properly defined bare fields. Since the Nielsen identity
\refer{NielsenEqLiniowN} does not involve the tree level
representations, $e_{{}_R\alpha}$ and $T_{{}_R\alpha}$, the general solution of
\refer{NielsenEqLiniowN} depends on arbitrary generators $e_\alpha$ and
$T_\alpha$. The relation between these two sets of generators follows from the
linearized Nielsen identity \refer{eq:LinZJoper} and reads
\eq{\label{eq:ZeIZt}
e^{\kappa}_{\przerwpod\beta\gamma}=\macierz{Z_e}^\alpha_{\przerw\beta}
\macierz{Z_e}^\epsilon_{\przerw\gamma}
\macierz{Z_{e}^{-1}}^\kappa_{\przerw\delta}
{{e}_{{}_R}}^{\delta}_{\przerwpod\alpha \epsilon},\quad
}
\eq{\label{eq:ZtSamo}
T_{\beta}=\macierz{{Z_e}}^{\alpha}_{\przerw\beta}
\macierz{{Z_{{}_T}}^{-1}}{{T}_{{}_R}}_{\alpha}\macierz{Z_{{{}_T}}},
}
(see \cite{BRS,BBBC} and discussion in appendix \ref{app:Stability}),
where $Z_e=1+\mathcal{O}(\hbar)$ and $Z_T=1+\mathcal{O}(\hbar)$ are
arbitrary matrices, which eventually have to be determined from Feynman
diagrams and all indices are restricted to the semisimple ideal.

We first focus on the $\xi$-dependence of the structure constants.
Differentiating the formula \refer{eq:ZeIZt} and comparing the outcome
with \refer{Eq:dePW} one finds that a matrix
$\hat{\mathcal{E}}=\hat{\theta}+Z_e^{-1}\mathrm{d}Z_e$ obeys
\eq{\label{eq:RownNaE0}
\macierz{\hat{\mathcal{E}},~e_\gamma}=
e_\beta\hat{\mathcal{E}}^\beta_{\phantom{\beta} \gamma},
}
where all indices (including those hidden in the matrix multiplication)
are effectively restricted to non-abelian ones. Thus $\hat{\mathcal{E}}$
is a linear combination of generators $e_\alpha$:\footnote{After some
manipulations \refer{eq:RownNaE0} yields
$\hat{\mathcal{E}}^\beta_{\phantom{\beta} \gamma}=e^\beta_{\phantom{\beta}
\tau\gamma}
\macierz{\mathfrak{K}^{-1}}^{\tau\delta}\mathrm{tr}
\nawias{e_\delta\hat{\mathcal{E}}}$,
where $\mathfrak{K}_{\alpha\beta}=\mathrm{tr}\nawias{e_\alpha e_\beta}$
is invertible as a formal series (assuming a restriction to non-abelian
indices) since $\indgd{e}{\alpha}{\beta\gamma}
=\indgd{e_{{}_R}}{\alpha}{\beta\gamma}+\mathcal{O}\nawias{\hbar}$.
}
$\hat{\mathcal{E}}=\hat{\eta}^\sigma e_\sigma$ and
\eq{\label{eq:thetaeta}
\hat{\theta}=-Z_e^{-1}\mathrm{d}Z_e+\hat{\eta}^\sigma e_\sigma.
}
Equation \refer{eq:thetaeta} is also true for abelian indices provided that
$Z_e$ is extended to a block-diagonal matrix with the identity matrix in the
abelian sector (1-forms $\hat{\eta}^\sigma$ are nonzero only for non-abelian
indices). Computing $\hat{\theta}\wedge\hat{\theta}-\mathrm{d}\hat{\theta}$
and using \refer{Eq:dthetaPW}, one gets
\eq{\label{eq:Psieta}
\hat{\Psi}^\sigma=
-\mathrm{d}\hat{\eta}^\sigma+\indgd{\hat{\theta}}{\sigma}{\delta}
\wedge\hat{\eta}^\delta
-\frac{1}{2}\indgd{e}{\sigma}{\kappa\lambda}\hat{\eta}^\kappa
\wedge\hat{\eta}^\lambda.
}
Equations \refer{Eq:dZAiphiPW} and \refer{eq:thetaeta} yield
\eq{\label{eq:dZAjaw}
\mathrm{d}Z_A=Z_A Z_e^{-1} \mathrm{d}Z_e+\macierz{\mathrm{d}Z_e}^T
\macierz{Z^{-1}_e}^T Z_A,
}
($\hat{\eta}$-terms cancel each other owing to \refer{ZinvDfullXIPW}).
For a symmetric matrix
\eq{
\mathcal{R}_A=\macierz{Z^{-1}_e}^T Z_A Z^{-1}_e,
}
formula \refer{eq:dZAjaw} gives
\eq{\label{eq:dRA}
\mathrm{d}\mathcal{R}_A=0,
}
thus $\mathcal{R}_A$ is gauge-independent. Furthermore, using the
parametrization \refer{eq:ZeIZt} in \refer{ZinvDfullXIPW} and taking
into account the antisymmetry of generators ${e_{{}_R}}_\alpha$, we find
\eq{\label{eq:KomRA}
\macierz{\mathcal{R}_A,~{e_{{}_R}}_\alpha}=0.
}
The above condition holds also for a (symmetric) matrix
$\sqrt{\mathcal{R}_A}$.  Thus $\sqrt{\mathcal{R}_A}$ is a block-diagonal
matrix with blocks corresponding to the entire abelian ideal and
different simple ideals, moreover blocks corresponding to simple ideals
are proportional to the identity matrix. Defining an orthogonal matrix
\eq{
\mathcal{U}_A=\sqrt{Z_A}~\!Z^{-1}_e\sqrt{\mathcal{R}^{-1}_A},\qquad
\mathcal{U}_A\mathcal{U}_A^T=1,
}
we can introduce bare gauge fields
\eq{
A^\alpha_{{}_B}\equiv\indgd{\macierz{\mathcal{U}^{-1}_A
\sqrt{Z_A}}}{\alpha}{\delta}A^\delta.
}
Rewriting \refer{WiekosciKowariantneFullNUXIPW} in terms of
$A^\alpha_{{}_B}$, we obtain
\eqs{
\tilde{F}^\delta_{\text{ }\mu\nu}
&=&
\indgd{\macierz{\sqrt{Z^{-1}_A}~\!\mathcal{U}_A}}{\delta}{\alpha}
\nawias{
\partial_\mu A^\alpha_{{}_B\nu}
-\partial_\nu A^\alpha_{{}_B\mu}
+{e_{{}_B}}^\alpha_{\phantom{\alpha}\beta\gamma}
A^\beta_{{}_B\mu}A^\gamma_{{}_B\nu}+\ldots
},
}
where the bare structure constants read
\eq{
{e_{{}_B}}^{\alpha '}_{\phantom{ \alpha '}{\beta^\prime\gamma^\prime}}=
\indgd{\macierz{\mathcal{U}^{-1}_A\sqrt{Z_A}}}{\alpha '}{\alpha}
{e}^{\alpha}_{\phantom{ \alpha}{\beta \gamma}}
\indgd{\macierz{\sqrt{Z^{-1}_A}~\!\mathcal{U}_A}}{\beta}{\beta^\prime}
\indgd{\macierz{\sqrt{Z^{-1}_A}~\!\mathcal{U}_A}}{\gamma}{\gamma^\prime}
}
Taking into account the relation $\mathcal{U}^{-1}_A\sqrt{Z_A}
=\sqrt{\mathcal{R}_A}Z_e$ and equation \refer{eq:ZeIZt}, one finds
\eq{
{e_{{}_B}}^{\alpha '}_{\phantom{ \alpha '}{\beta ' \gamma '}}=
\indgd{\macierz{\sqrt{\mathcal{R}_A}}}{\alpha '}{\alpha}
{e_{{}_R}}^{\alpha}_{\phantom{ \alpha}{\beta \gamma}}
\indgd{\macierz{\sqrt{\mathcal{R}^{-1}_A}}}{\beta}{\beta '}
\indgd{\macierz{\sqrt{\mathcal{R}^{-1}_A}}}{\gamma}{\gamma '}
}
Finally, using \refer{eq:KomRA} we get
\eq{\label{eq:eB}
{e_{{}_B}}^{\alpha '}_{\phantom{ \alpha '}{\beta ' \gamma '}}=
{e_{{}_R}}^{\alpha '}_{\phantom{ \alpha '}{\beta \gamma '}}
\indgd{\macierz{\sqrt{\mathcal{R}^{-1}_A}}}{\beta}{\beta '}
}
thus the bare structure constants differ from the renormalized ones only
by the separate renormalization of coupling constants of each simple
ideal. The Nielsen identity \refer{eq:dRA} ensures that these bare
couplings are gauge-independent.

Consider now the $\xi$-dependence of generators $T_{\alpha}$ in the space
of scalar fields. Since the formula \refer{eq:ZtSamo} is correct only for
non-abelian indices, we denote them as $\alpha_1$, $\beta_1$, etc.
Differentiating \refer{eq:ZtSamo} and eliminating $\mathrm{d}T_{\gamma_1}$
with the aid of \refer{Eq:dTPW} one finds
\eq{
\macierz{T_{\gamma_1},~\hat{\rho}}=0,
}
with $\hat{\rho}$ defined by
\eq{\label{eq:zetaeta}
\hat{\zeta}=\hat{\rho}-Z_T^{-1}\mathrm{d}Z_T+\hat{\eta}^\sigma T_\sigma.
}
In terms of parametrization
\eq{\hat{\rho}=Z_T^{-1}\hat{r}Z_T,}
the condition $\macierz{T_{\gamma_1}~,~\hat{\rho}}=0$ reads
\eq{\label{eq:Komr}
\macierz{T_{{}_R\gamma_1},~\hat{r}}=0.
}
Computing $\hat{\zeta}\wedge\hat{\zeta}-\mathrm{d}\hat{\zeta}$, one finds
\eq{
\hat{\zeta}\wedge\hat{\zeta}-\mathrm{d}\hat{\zeta}
=Z^{-1}_T\nawias{\hat{r}\wedge\hat{r}-\mathrm{d}\hat{r}}Z_T
+T_\sigma\nawias{
-\mathrm{d}\hat{\eta}^\sigma+\indgd{\hat{\theta}}{\sigma}{\delta}
\wedge\hat{\eta}^\delta
-\frac{1}{2}\indgd{e}{\sigma}{\kappa\lambda}\hat{\eta}^\kappa
\wedge\hat{\eta}^\lambda
}.
}
Owing to \refer{Eq:dzetaPW} and \refer{eq:Psieta} the above formula yields
the Maurer-Cartan equation
\eq{\label{eq:MCEQ}
\mathrm{d}\hat{r}-\hat{r}\wedge\hat{r}=0.
}
Any 1-form obeying \refer{eq:MCEQ} can be represented as (see e.g.
\cite{Sharpe})
\eq{\label{eq:RprzezM}
\hat{r}=-\mathcal{M}^{-1}\mathrm{d}\mathcal{M},
}
moreover, for a given $\hat{r}$, equation \refer{eq:RprzezM} determines
$\mathcal{M}$ uniquely up to a constant of integration
$\mathcal{M}\to c\mathcal{M}$.  This freedom allows us to choose
$\mathcal{M}\nawias{\xi_0}=1$, which together with \refer{eq:Komr} ensures
\eq{\label{eq:KomM}
\macierz{T_{{}_R\gamma_1}~,~\mathcal{M}}=0.
}
With the aid of \refer{eq:RprzezM} we can rewrite $\hat{\zeta}$ in the form
\eq{\label{eq:zetaetaFin}
\hat{\zeta}=-\tilde{Z}_T^{-1}\mathrm{d}\tilde{Z}_T+\hat{\eta}^\sigma T_\sigma.
}
where
\eq{\tilde{Z}_T\equiv \mathcal{M}Z_T,
}
and \refer{eq:ZtSamo} takes the form (owing to \refer{eq:KomM})
\eq{
T_{\beta_1}=\macierz{{Z_e}}^{\alpha_1}_{\przerw\beta_1}
{{\tilde{Z}_{{}_T}}^{-1}}{{T}_{{}_R}}_{\alpha_1}{\tilde{Z}_{{{}_T}}},
}
Defining
\eq{
\mathcal{R}_\phi=\macierz{\tilde{Z}^{-1}_T}^T Z_\phi \tilde{Z}^{-1}_T,
}
we find (similarly to the case of $\mathcal{R}_A$)
\eq{\label{eq:dRphi}
\mathrm{d}\mathcal{R}_\phi=0,
}
\eq{\label{eq:KomRphi}
\macierz{\mathcal{R}_\phi~,~{T_{{}_R}}_{\alpha_1}}=0.
}
We need also the following matrix
\eq{
\mathcal{U}_\phi=\sqrt{Z_\phi}\tilde{Z}^{-1}_T\sqrt{\mathcal{R}^{-1}_\phi},
\qquad \mathcal{U}_\phi \mathcal{U}_\phi^T=1,
}
which allows us to define the bare scalar field
\eq{\label{eq:GolePhi}
\phi^a_{{}_B}\equiv\indgd{\macierz{\mathcal{U}^{-1}_\phi
\sqrt{Z_\phi}}}{a}{d}\phi^d.
}
In order to rewrite the covariant derivative in terms of bare fields we
have to compute $A^{\alpha_1}_\mu T_{\alpha_1}\phi$ and
$A^{\alpha_0}_\mu T_{\alpha_0}\phi$ separately:
\eqs{
A^{\alpha_1}_\mu T_{\alpha_1}\phi&=&A^{\beta_1}_{{}_B\mu}
\macierz{Z^{-1}_e \sqrt{\mathcal{R}^{-1}_A}}^{\alpha_1}_{~\beta_1}
T_{\alpha_1}\tilde{Z}^{-1}_T\sqrt{\mathcal{R}^{-1}_\phi}\phi_{{}_B}
=A^{\beta_1}_{{}_B\mu}
\macierz{\sqrt{\mathcal{R}^{-1}_A}}^{\alpha_1}_{~\beta_1}
\tilde{Z}^{-1}_T T_{{}_R\alpha_1}\sqrt{\mathcal{R}^{-1}_\phi}\phi_{{}_B}
\nonumber\\
&=&\sqrt{Z^{-1}_\phi}\mathcal{U}_\phi
\nawias{A^{\beta_1}_{{}_B\mu}
\macierz{\sqrt{\mathcal{R}^{-1}_A}}^{\alpha_1}_{~\beta_1}
T_{{}_R\alpha_1}\phi_{{}_B}
},
}
hence bare generators have the form
\eq{
T_{{}_B\beta_1}=\macierz{\sqrt{\mathcal{R}^{-1}_A}}^{\alpha_1}_{~\beta_1}
T_{{}_R\alpha_1},
}
in agreement with \refer{eq:eB}. On the other hand (due to
$T_{\alpha_0}=T_{{{}_R}\alpha_0}$ and $\macierz{T_{{{}_R}\alpha_0}~,~Z_\phi}=0$ )
\eqs{
A^{\alpha_0}_\mu T_{\alpha_0}\phi&=&A^{\beta_0}_{{}_B\mu}
\macierz{Z^{-1}_e \sqrt{\mathcal{R}^{-1}_A}}^{\alpha_0}_{~\beta_0}
T_{{{}_R}\alpha_0}\sqrt{Z_\phi^{-1}}\mathcal{U}_\phi\phi_{{}_B}
=A^{\beta_0}_{{}_B\mu}
\macierz{\sqrt{\mathcal{R}^{-1}_A}}^{\alpha_0}_{~\beta_0}
\sqrt{Z_\phi^{-1}} T_{{}_R\alpha_0}\mathcal{U}_\phi\phi_{{}_B}=\nonumber\\
&=&\sqrt{Z^{-1}_\phi}\mathcal{U}_\phi
\nawias{A^{\beta_0}_{{}_B\mu}
\macierz{\sqrt{\mathcal{R}^{-1}_A}}^{\alpha_0}_{~\beta_0}
\mathcal{U}^T_\phi T_{{}_R\alpha_0}\mathcal{U}_\phi\phi_{{}_B}
},
}
thus
\eq{\label{eq:TB}
T_{{}_B\beta_0}=\macierz{\sqrt{\mathcal{R}^{-1}_A}}^{\alpha_0}_{~\beta_0}
\mathcal{U}^T_\phi T_{{}_R\alpha_0}\mathcal{U}_\phi,
}
with
\eq{\label{eq:UTU}
\mathcal{U}^T_\phi T_{{}_R\alpha_0}\mathcal{U}_\phi=
\sqrt{\mathcal{R}^{-1}_\phi} \macierz{\tilde{Z}^{-1}_T}^T Z_\phi
T_{{}_R\alpha_0}\tilde{Z}^{-1}_T\sqrt{\mathcal{R}^{-1}_\phi}=
\sqrt{\mathcal{R}_\phi} \tilde{Z}_T T_{{}_R\alpha_0}\tilde{Z}^{-1}_T
\sqrt{\mathcal{R}^{-1}_\phi}.
}
Differentiating the above equation and eliminating $\mathrm{d}\tilde{Z}_T$ with the help of \refer{eq:zetaetaFin}, one gets
\eq{\label{eq:dUTU}
\mathrm{d}\macierz{\mathcal{U}^T_\phi T_{{}_R\alpha_0}\mathcal{U}_\phi}
=
\sqrt{\mathcal{R}_\phi} \tilde{Z}_T
\macierz{T_{{}_R\alpha_0}~,~\hat{\zeta}-\hat{\eta}^\sigma T_\sigma}
\tilde{Z}^{-1}_T\sqrt{\mathcal{R}^{-1}_\phi}=0,
}
since the commutator vanishes according to \refer{Eq:KomzAbel}. Hence the
bare generators \refer{eq:TB} are gauge-independent, and one can calculate
them for $\xi$'s preserving the symmetry under global gauge transformations.
In this case $Z_{T}\equiv 1$ (see e.g. \cite{Wein}), so that
\eq{
\macierz{ T_{{}_R\alpha_0} ~,~\hat{r} }=\macierz{ T_{{}_R\alpha_0} ~,~
\hat{\zeta} }=0,
}
and thus $\macierz{ T_{{}_R\alpha_0} ~,~\mathcal{M} }=0$, yielding
$\macierz{ T_{{}_R\alpha_0} ~,~\mathcal{R}_\phi }=0$. Finally, equation
\refer{eq:UTU} gives
\eq{\label{eq:TBfin}
T_{{}_B\beta_0}=\macierz{\sqrt{\mathcal{R}^{-1}_A}}^{\alpha_0}_{~\beta_0}
T_{{}_R\alpha_0}.
}
Since $\macierz{\mathcal{R}_A}_{\alpha_0\beta_0}=\macierz{Z_A}_{\alpha_0\beta_0}$,
the above equation is yet another form of the `$Z_1=Z_2$' identity.

Having verified that \refer{eq:GolePhi} is the correct bare field, one can
show that all other bare coupling constants are also $\xi$-independent. As
we have argued $\tilde{\mathcal{V}}\nawias{\Phi,\xi}
=\tilde{\mathcal{V}}_{sym}\nawias{\Phi+v_{{}_R},\xi}$ with
$\tilde{\mathcal{V}}_{sym}\nawias{\varphi,\xi}$
independent of $v_{{}_R}$. Assuming that equations \refer{Eq:ZwiazekVNivR}
are satisfied, one can rewrite \refer{VinvfullNUXIPW} and
\refer{VinvfullNUXIkPW} as
\eq{\label{VinvfullNUXIPW2}
\macierz{T_\alpha}^a_{\przerw b}\varphi^b\poch{\tilde{\mathcal{V}}_{sym}
\nawias{\varphi,\xi}}{\varphi^a}=0,
}
\eq{\label{VinvfullNUXIkPW2}
\mathrm{d}\tilde{\mathcal{V}}_{sym}\nawias{\varphi,\xi}+
\hat{\zeta}^a_{\phantom{a} b}
\varphi^b
\poch{\tilde{\mathcal{V}}_{sym}\nawias{\varphi,\xi}}{\varphi^a}=0,
}
where $\mathrm{d}\equiv \mathrm{d}
\xi^{\alpha\beta}\partial /\partial\xi^{\alpha\beta}$. The potential of the bare
fields
\eq{
\tilde{\mathcal{V}}^B_{sym}\nawias{\varphi_{{}_B},\xi}\equiv
\tilde{\mathcal{V}}_{sym}\nawias{\sqrt{Z^{-1}_\phi}~\!
\mathcal{U}_\phi\varphi_{{}_B},\xi},
}
obeys
\eq{
\mathrm{d}\tilde{\mathcal{V}}^B_{sym}\nawias{\varphi_{{}_B},\xi}
=\left.
-\indgd{\macierz{\tilde{Z}^{-1}_T\mathrm{d}{\tilde{Z}_T}+\hat{\zeta}}}{a}{b}~\!
\varphi^b~\!
\poch{\tilde{\mathcal{V}}_{sym}\nawias{\varphi,\xi}}{\varphi^a}
\right|_{\varphi=\sqrt{Z^{-1}_\phi}\mathcal{U}_\phi\varphi_{{}_B}}
}
thus, taking into account \refer{eq:zetaetaFin} and \refer{VinvfullNUXIPW2},
$\tilde{\mathcal{V}}^B_{sym}\nawias{\varphi_{{}_B},\xi}$ is
$\xi$-independent.
\end{section}

\begin{section}{Explicit Calculation of Counterterms}\label{Sec:WyznPrzec}
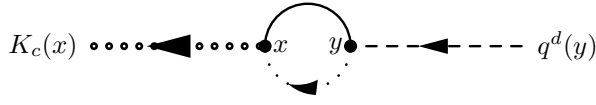
\begin{figure}[t]
\centering
\begin{fmffile}{WkladDob}
\begin{fmfgraph*}(160,100)
\fmfleft{i} \fmfright{o}
\fmfv{label=$K_c(x)$,label.angle=-180}{i}
\fmfv{label=$q^d(y)$,label.angle=0}{o}
\fmfv{label=$x$,label.angle=0,label.dist=3}{v1}
\fmfv{label=$y$,label.angle=-180,label.dist=3}{v2}
\fmf{dbl_dots_arrow,label=$ $}{v1,i}
\fmf{dashes_arrow,label=$ $}{o,v2}
\fmf{dots_arrow,left}{v2,v1}
\fmf{plain,right}{v2,v1}
\fmfdot{v1,v2}
\end{fmfgraph*}
\end{fmffile}
\caption{One-loop contribution to $b^{c}_{\text{  } d}$.}
\label{Rys:WkladDob}
\end{figure}
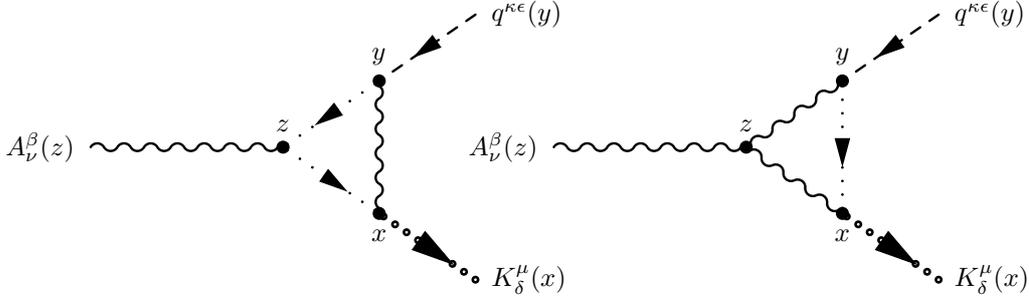
\begin{figure}[t]
\centering
%Diagram1
\begin{fmffile}{WkladDoTheta1}
\begin{fmfgraph*}(160,100)
\fmfleft{i1}
\fmfv{label=$A^\beta_\nu(z)$,label.angle=-180}{i1}
\fmfright{o2,o3}
\fmfv{label=$q^{\kappa\epsilon}(y)$,label.angle=0}{o3}
\fmfv{label=$K^\mu_\delta(x)$,label.angle=0}{o2}
\fmfv{label=$z$,label.angle=90}{v1}
\fmfv{label=$x$,label.angle=-90}{v2}
\fmfv{label=$y$,label.angle=90}{v3}
\fmf{photon}{i1,v1}
\fmf{dots_arrow}{v1,v2}
\fmf{dots_arrow}{v3,v1}
\fmf{dbl_dots_arrow}{v2,o2}
\fmf{dashes_arrow}{o3,v3}
\fmffreeze
\fmf{photon}{v2,v3}
\fmfdot{v1,v2,v3}
\end{fmfgraph*}
\end{fmffile}
%Diagram2
\begin{fmffile}{WkladDoTheta2}
\begin{fmfgraph*}(160,100)
\fmfleft{i1}
\fmfv{label=$A^\beta_\nu(z)$,label.angle=-180}{i1}
\fmfright{o2,o3}
\fmfv{label=$q^{\kappa\epsilon}(y)$,label.angle=0}{o3}
\fmfv{label=$K^\mu_\delta(x)$,label.angle=0}{o2}
\fmfv{label=$z$,label.angle=90}{v1}
\fmfv{label=$x$,label.angle=-90}{v2}
\fmfv{label=$y$,label.angle=90}{v3}
\fmf{photon}{i1,v1}
\fmf{photon}{v1,v2}
\fmf{photon}{v3,v1}
\fmf{dbl_dots_arrow}{v2,o2}
\fmf{dashes_arrow}{o3,v3}
\fmffreeze
\fmf{dots_arrow}{v3,v2}
\fmfdot{v1,v2,v3}
\end{fmfgraph*}
\end{fmffile}
\caption{One-loop contributions to $\theta^\delta_{\text{  }\kappa\epsilon\beta}$.}
\label{Rys:WkladyDoTheta}
\end{figure}
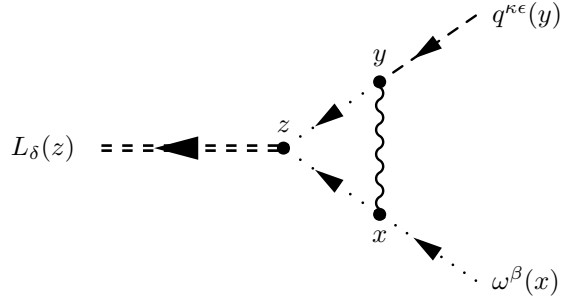
\begin{figure}[t]
\centering
\begin{fmffile}{WkladDoG}
\begin{fmfgraph*}(160,100)
\fmfleft{i1} \fmfright{o2,o3}
\fmfv{label=$L_\delta(z)$,label.angle=-180}{i1}
\fmfv{label=$q^{\kappa\epsilon}(y)$,label.angle=0}{o3}
\fmfv{label=$\omega^\beta(x)$,label.angle=0}{o2}
\fmfv{label=$z$,label.angle=90}{v1}
\fmfv{label=$x$,label.angle=-90}{v2}
\fmfv{label=$y$,label.angle=90}{v3}
\fmf{dbl_dashes_arrow,label=$ $}{v1,i1}
\fmf{dots_arrow}{v2,v1}
\fmf{dots_arrow}{v3,v1}
\fmf{dots_arrow}{o2,v2}
\fmf{dashes_arrow}{o3,v3}
\fmffreeze
\fmf{photon}{v2,v3}
\fmfdot{v1,v2,v3}
\end{fmfgraph*}
\end{fmffile}
\caption{One-loop contribution to $g^\delta_{\text{  }\kappa\epsilon\beta}$.}
\label{Rys:WkladyDoG}
\end{figure}
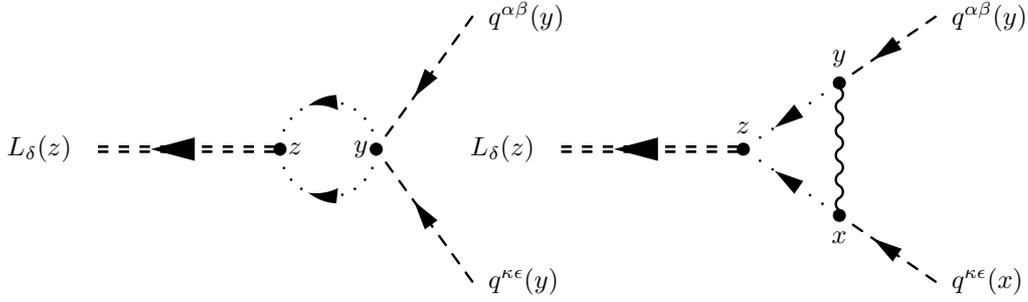
\begin{figure}[t]
\centering
%Diagram1
\begin{fmffile}{WkladDoF1}
\begin{fmfgraph*}(160,100)
\fmfleft{i1} \fmfright{o2,o3}
\fmfv{label=$L_\delta(z)$,label.angle=-180}{i1}
\fmfv{label=$q^{\alpha\beta}(y)$,label.angle=0}{o3}
\fmfv{label=$q^{\kappa\epsilon}(y)$,label.angle=0}{o2}
\fmfv{label=$z$,label.angle=0,label.dist=3}{v1}
\fmfv{label=$y$,label.angle=-180,label.dist=3}{v2}
\fmf{dbl_dashes_arrow}{v1,i1}
\fmf{dots_arrow,left}{v2,v1}
\fmf{dots_arrow,right}{v2,v1}
\fmf{dashes_arrow}{o2,v2}
\fmf{dashes_arrow}{o3,v2}
\fmfdot{v1,v2}
\end{fmfgraph*}
\end{fmffile}
%Diagram2
\begin{fmffile}{WkladDoF2}
\begin{fmfgraph*}(160,100)
\fmfleft{i1} \fmfright{o2,o3}
\fmfv{label=$L_\delta(z)$,label.angle=-180}{i1}
\fmfv{label=$q^{\alpha\beta}(y)$,label.angle=0}{o3}
\fmfv{label=$q^{\kappa\epsilon}(x)$,label.angle=0}{o2}
\fmfv{label=$z$,label.angle=90}{v1}
\fmfv{label=$x$,label.angle=-90}{v2}
\fmfv{label=$y$,label.angle=90}{v3}
\fmf{dbl_dashes_arrow}{v1,i1}
\fmf{dots_arrow}{v2,v1}
\fmf{dots_arrow}{v3,v1}
\fmf{dashes_arrow,label=$ $}{o2,v2}
\fmf{dashes_arrow,label=$ $}{o3,v3}
\fmffreeze 
\fmf{photon}{v2,v3}
\fmfdot{v1,v2,v3}
\end{fmfgraph*}
\end{fmffile}
\caption{One-loop diagrams contributing to $H^\delta_{\text{  }\alpha\beta\kappa\epsilon}$.}
\label{Rys:WkladyDoF}
\end{figure}
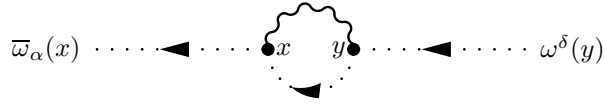
\begin{figure}[t]
\centering
\begin{fmffile}{WkladDoN}
\begin{fmfgraph*}(160,100)
\fmfleft{i} \fmfright{o}
\fmfv{label=$\overline{\omega}_\alpha(x)$,label.angle=-180}{i}
\fmfv{label=$\omega^\delta(y)$,label.angle=0}{o}
\fmfv{label=$x$,label.angle=0,label.dist=3}{v1}
\fmfv{label=$y$,label.angle=-180,label.dist=3}{v2}
\fmf{dots_arrow,label=$ $}{v1,i}
\fmf{dots_arrow,label=$ $}{o,v2}
\fmf{dots_arrow,left}{v2,v1}
\fmf{photon,right}{v2,v1}
\fmfdot{v1,v2}
\end{fmfgraph*}
\end{fmffile}
\caption{One-loop contribution to $\mathcal{N^\alpha_{\text{ }\text{ }\delta}}$.}
\label{Rys:WkladDoN}
\end{figure}
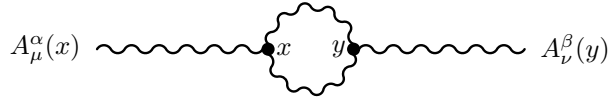
\begin{figure}[t]
\centering
\begin{fmffile}{WkladDodZA}
\begin{fmfgraph*}(160,100)
\fmfleft{i} \fmfright{o}
\fmfv{label=$A^\alpha_\mu(x)$,label.angle=-180}{i}
\fmfv{label=$A^\beta_\nu(y)$,label.angle=0}{o}
\fmfv{label=$x$,label.angle=0,label.dist=3}{v1}
\fmfv{label=$y$,label.angle=-180,label.dist=3}{v2}
\fmf{photon,label=$ $}{i,v1}
\fmf{photon,label=$ $}{v2,o}
\fmf{photon,left}{v1,v2}
\fmf{photon,right}{v1,v2}
\fmfdot{v1,v2}
\end{fmfgraph*}
\end{fmffile}
\caption{One-loop contribution to $\text{d}\left[Z_A\right]_{\alpha\beta}$.}
\label{Rys:WkladDodZA}
\end{figure}
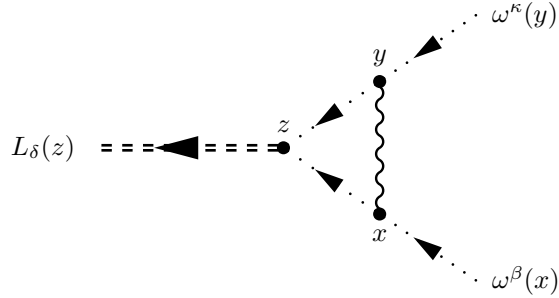
\begin{figure}[t]
\centering
\begin{fmffile}{WkladDoC}
\begin{fmfgraph*}(160,100)
\fmfleft{i1} \fmfright{o2,o3}
\fmfv{label=$L_\delta(z)$,label.angle=-180}{i1}
\fmfv{label=$\omega^{\kappa}(y)$,label.angle=0}{o3}
\fmfv{label=$\omega^\beta(x)$,label.angle=0}{o2}
\fmfv{label=$z$,label.angle=90}{v1}
\fmfv{label=$x$,label.angle=-90}{v2}
\fmfv{label=$y$,label.angle=90}{v3}
\fmf{dbl_dashes_arrow,label=$ $}{v1,i1}
\fmf{dots_arrow}{v2,v1}
\fmf{dots_arrow}{v3,v1}
\fmf{dots_arrow}{o2,v2}
\fmf{dots_arrow}{o3,v3}
\fmffreeze
\fmf{photon}{v2,v3}
\fmfdot{v1,v2,v3}
\end{fmfgraph*}
\end{fmffile}
\caption{One-loop contribution to $C^\delta_{~\kappa\beta}$.}
\label{Rys:WkladyDoC}
\end{figure}
\begin{figure}[t]
\centering
\begin{fmffile}{WkladDoZpsi}
\begin{fmfgraph*}(160,100)
\fmfleft{i1} \fmfright{o2,o3}
\fmfv{label=$\overline{K}_{A}(x)$,label.angle=-180}{i1}
\fmfv{label=$\psi^{B}(z)$,label.angle=0}{o3}
\fmfv{label=$q(y)$,label.angle=0}{o2}
\fmf{dbl_dots_arrow}{v1,i1}
\fmf{dots_arrow}{v2,v1}
\fmf{dbl_plain_arrow}{v3,v1}
\fmf{dashes_arrow,label=$ $}{o2,v2}
\fmf{plain_arrow,label=$ $}{o3,v3}
\fmffreeze
\fmf{dbl_wiggly}{v2,v3}
\fmfdot{v1,v2}
\fmfblob{.10w}{v3}
\end{fmfgraph*}
\end{fmffile}
\caption{`Dressed' diagram contributing to $\zeta$ in QED.}
\label{Rys:WkladDoZpsi}
\end{figure}
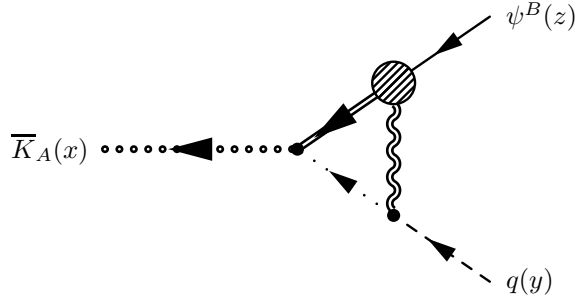
\noindent In this section we compute (at the one-loop order) some of the
counterterms in order to check validity of our results. We are interested
in one-particle-irreducible (1PI) diagrams in the presence of the Nielsen
sources $q(x)$. External lines of the diagrams correspond therefore either
to `quantum' fields ($\phi$, $A_\mu$, $\omega$, $\overline{\omega}$) or to
classical sources ($q$, $\mathcal{K}$, $L$). Solid and dotted lines
represent respectively matter fields and the Faddeev-Popov ghosts.
The necessary Feynman rules can be read off directly from the Lagrangian
(\ref{LtildeNPW}).

The one-loop diagram shown in figure \ref{Rys:WkladDob}, after including
the $b^c_{\phantom{c} d}$ counterterm, gives the renormalized two-point
function $F^c_{\przerw d}(x,y)$ of composite operators coupled to the external
sources $K_c(x)$ and $q^d(y)$:
\eq{
F^c_{\przerw d}(x,y)=b^{c}_{\przerw d}\delta^{(4)}(x-y)
+i\hbar~\!\mathcal{N}^\delta_{\przerw\alpha}
\macierz{T_\delta}^c_{\przerw a}\xi^{\beta\gamma}\delta_{db}
\macierz{{T_{{}_R}}_{\gamma}}^b_{\przerw e}
\left<\phi^a\nawias{x}\phi^e\nawias{y}\right>_0\times
\left<\omega^\alpha\nawias{x}\anty{\omega}_\beta\nawias{y}\right>_0
+\mathcal{O}\nawias{\hbar^2}.
}
In the dimensional regularization ($d=4-2\epsilon$) the products of the
tree level propagators gives
\eqs{
\left<\phi^a\nawias{x}\phi^e\nawias{y}\right>_0\times
\left<\omega^\alpha\nawias{x}\anty{\omega}_\beta\nawias{y}\right>_0
&=&i\int\vol{p}~\!e^{-ip\nawias{x-y}}
\int\vol{k}i\macierz{\frac{1}{k^2-\mathcal{M}^2_\phi+i\varepsilon}
+\ldots}^{ae}\macierz{\frac{1}{\nawias{k-p}^2-\mathcal{M}^2_\omega
+i\varepsilon}}^{\alpha}_{\przerw\beta}\nonumber\\
&=&-i\frac{1}{\nawias{4\pi}^2\epsilon}~\!\delta^{ae}
\delta^{\alpha}_{\przerw\beta}\delta^{(4)}
\nawias{x-y}+\mathcal{O}\nawias{\epsilon^0}.
}
The ellipses in the first square bracket stand for terms arising from the
mixing of the scalar and vector fields. These terms do not change the
leading UV behavior of the propagator and can, therefore, be omitted in
the present calculation. Since to the order we are working
$\mathcal{N}^\delta_{\przerw\alpha}=\delta^\delta_{\przerw\alpha}$
and $T_\delta={T_{{}_R}}_\delta$, we get (in the $MS$~scheme)
\eq{\label{b1loopXI}
b^{c}_{\przerw d}=\frac{\hbar}{\nawias{4\pi}^2\epsilon}~\!\xi^{\alpha\gamma}
\macierz{{T_{{}_R}}_\alpha{T_{{}_R}}_{\gamma}}^c_{\przerw d}+\mathcal{O}
\nawias{\hbar^2}.
}
The above matrix clearly respects the symmetry requirements
\refer{Eq:LieBPW}. If $(\xi^{-1})_{\alpha\gamma}$ is an invariant form on
the gauge Lie algebra, the formula \refer{b1loopXI} tells us that
$b^{c}_{\przerw d}$ is proportional to the Casimir operator of the
representation $T_{{}_R\alpha}$.\\

Other dimensionless counterterms such as $\hat{\theta}$, $\hat{g}$
etc., can be calculated in the restricted 't Hooft gauge ($u=v_{{}_R}$ with
$\left<\phi\right>=0$). This choice removes the tree-level
mixing between scalar and vector fields and leads to the standard form of
the propagator \cite{Wein}:
\eqs{\label{PropAA}
\left<A^\alpha_\mu A^\beta_\nu\right>_0\nawias{k}&=&
-i\macierz{\eta_{\mu\nu}\frac{1}{k^2-\mathcal{M}^2+i\varepsilon}
+k_\mu k_\nu \frac{1}{k^2-\xi\mathcal{M}^2+i\varepsilon}
\nawias{\xi-\mathbb{I}}\frac{1}{k^2-\mathcal{M}^2
+i\varepsilon}}^{\alpha\beta}.
}
As long as we are interested in dimensionless parameters, the non-diagonal
form of the mass matrices is immaterial and calculations with general
$\xi^{\alpha\beta}$ parameters can be easily performed. Computing divergent parts of the diagrams shown in figure
\ref{Rys:WkladyDoTheta}, we find ($\left\{\cdot,\cdot\right\}$ denotes the
anticommutator)
\eq{\label{Eq:ThetaJawn}
\theta^\delta_{\text{  }\kappa\epsilon\beta}=-\frac{1}{(4\pi)^2\epsilon}~\!
\frac{1}{8}\left\{e_{{}_R\epsilon},~e_{{}_R\kappa}
\right\}^\delta_{\przerw\beta},
}
while for $g^\delta_{\text{  }\kappa\epsilon\beta}$ we obtain (see figure
\ref{Rys:WkladyDoG}) the result:
\eq{\label{Eq:GJawn}
g^\delta_{\text{  }\kappa\epsilon\beta}=-\frac{1}{(4\pi)^2\epsilon}~\!
\frac{1}{4}\left\{e_{{}_R\epsilon},~e_{{}_R\kappa}
\right\}^\delta_{\przerw\beta}.
}
The sum of two divergent diagrams shown in figure \ref{Rys:WkladyDoF}
is finite, hence $\hat{H}^\epsilon=\mathcal{O}\nawias{\hbar^2}$.
This result agrees with \refer{Eq:Psi_defPW} and \refer{Eq:dthetaPW}, since
$\text{d}\hat{\theta}=0$ while $\hat{\theta}\wedge\hat{\theta}$ is of the
order of $\hbar^2$.\\

The one-loop correction to the ghost propagator, which is relevant for the
computation of $\text{d}\mathcal{N}$, is shown in figure~\ref{Rys:WkladDoN}.
It gives
\eq{\label{Eq:dNJawn}
\text{d}\mathcal{N}^\delta_{\przerw\beta}=\frac{1}{(4\pi)^2\epsilon}~\!
\frac{1}{8}
\left\{e_{{}_R\epsilon},~e_{{}_R\kappa}\right\}^\delta_{\przerw\beta}
\text{d}\xi^{\kappa\epsilon}.
}
The results \refer{Eq:ThetaJawn}, \refer{Eq:GJawn} and \refer{Eq:dNJawn}
are consistent with  the Nielsen identity requirements \refer{Eq:dNPW}.
Among various corrections to the gauge field propagator, only the diagram
of figure \ref{Rys:WkladDodZA} contributes to $\text{d} Z_A$. A short
calculation gives:
\eq{\label{Eq:dZAJawn}
\text{d}\macierz{{Z}_A}_{\alpha\beta}=\frac{1}{(4\pi)^2\epsilon}~\!\frac{1}{4}
\delta_{\alpha\delta}\left\{e_{{}_R\epsilon},~e_{{}_R\kappa}
\right\}^\delta_{\przerw\beta}
\text{d}\xi^{\kappa\epsilon}.
}
This agrees with \refer{Eq:ThetaJawn} and \refer{Eq:dZAiphiPW}. Finally, the diagram shown in figure \ref{Rys:WkladyDoC} determines renormalization of the structure constants $\indgd{C}{\alpha}{\beta\gamma}$ yielding
\eq{\label{Eq:CJawn}
\indgd{C}{\alpha}{\sigma\delta}=
\indgd{e_{{}_R}}{\alpha}{\sigma\delta}
-\frac{\hbar}{(4\pi)^2\epsilon}~\!
\indgd{e_{{}_R}}{\alpha}{\beta\gamma}
\xi^{\epsilon\tau}
\indgd{e_{{}_R}}{\beta}{\tau\sigma}
\indgd{e_{{}_R}}{\gamma}{\epsilon\delta}
=
\macierz{{Z_C}}^\alpha_{\przerw \alpha'}
\indgd{e_{{}_R}}{\alpha'}{\sigma' \delta'}
\left[Z_C^{-1}\right]^{\sigma'}_{\przerw\sigma}
\left[Z_C^{-1}\right]^{\delta'}_{\przerw\delta}+\mathcal{O}\nawias{\hbar^2},
}
with
\eq{
\macierz{{Z_C}}^\delta_{\przerw \beta}
=
\delta^\delta_{\przerw \beta}-\frac{\hbar}{(4\pi)^2\epsilon}~\!
\frac{1}{4}\left\{e_{{}_R\epsilon},~e_{{}_R\kappa}
\right\}^\delta_{\przerw\beta}\xi^{\epsilon\kappa},
}
so that $\hat{g}=\macierz{dZ_C}Z_C^{-1}+\mathcal{O}\nawias{\hbar^2}$. The $\xi$-dependent part of the structure constants $\indgd{e}{\alpha}{\beta\gamma}$ can be then obtained with the aid of \refer{ZwiazekCieRozdzXIPW} and \refer{Eq:dNJawn} - the resulting expression is consistent with \refer{Eq:dePW} and \refer{Eq:ThetaJawn}.\\

We end this section by rederiving, with the help of the Nielsen
identities, the well known equation for the gauge-dependence of the
electron field renormalization constant in QED (see
e.g. \cite{ZinnJustinQFTCritical}).
(As we have shown in the preceding section, in abelian theories
gauge-independence of the gauge field renormalization constant follows
immediately from the Nielsen identities.) In QED the
renormalized Lagrangian includes the terms
\eq{
\tilde{\mathcal{L}}^N(x)\supset i e_{R} \overline{K}_A(x)\omega(x)\psi^A(x)
+\zeta\overline{K}_A(x)q(x)\psi^A(x),
}
(and analogous couplings of the $\overline{\psi}_A(x)$ field), in which
$e_{R}$ denotes the renormalized charge. Instead of the second equation
\refer{Eq:dZAiphiPW} we now have
\eq{
\frac{1}{Z_\psi}\poch{Z_\psi}{\xi}=-\zeta-\overline{\zeta}.
}
Owing to the  decoupling of ghosts there is only one (`dressed') diagram
contributing to $\zeta$. It is shown in figure \ref{Rys:WkladDoZpsi}
in which double lines represent the full (renormalized) propagators of
photon and electrons. The blob stands for the (renormalized) 1PI vertex
$A_\mu\overline{\psi}_B\psi^C$. The vertex $q~\!\anty{\omega}A_\mu$ comes from
the third line of the Lagrangian \refer{LtildeNPW}, and depends only on
$\partial^\mu A_\mu$. Therefore, to compute this diagram we need only the
transverse part of the photon propagator, which is unaffected
by radiative corrections. Consequently, we can eliminate the
divergence of the $A_\mu\overline{\psi}_B\psi^C$ vertex, by using the
Ward-Takahashi identity. In this way we get the following contribution to
the effective action
\eqs{
\Gamma^N\supset\frac{1}{2}e^2_R \int\vol{p}\int\vol{l}~\!
\anty{K}_A(-p-l)q(l)\psi^B(p)\int\vol{k}\frac{i}{\nawias{k+l}^2 k^2}
\macierz{1+i \tilde{G}_R^{(2)}\nawias{p-k}\tilde{\Gamma}_R^{(2)}
\nawias{-p}}^A_{\phantom{A}B}.
}
The electron propagator $\tilde{G}_R^{(2)}\nawias{p-k}$ makes the integral
of the second component convergent. Hence,
\eq{
\zeta=\frac{1}{\nawias{4\pi}^2\epsilon}~\!\frac{1}{2}e^2_R.
}
Computing an analogous diagram with $K$ and $\overline{\psi}$ external
lines instead of $\overline{K}$ and ${\psi}$ we find
$\overline{\zeta}=\zeta$, and finally
\eq{
\frac{1}{Z_\psi}\poch{Z_\psi}{\xi}=-\frac{1}{\nawias{4\pi}^2\epsilon}~\!
e^2_R.
}
The derivation presented here should be compared with the standard one
based on the Ward-Takahashi identity, which can be found e.g. in
\cite{ZinnJustinQFTCritical}.
\end{section}

\begin{section}{Determination of $\delta v$}

One-loop checks of the Nielsen identity for the effective potential
in the abelian Higgs model in $R_{\xi,u}$-gauges can be found in
\cite{AitchisonFraser,Baz}. In order to verify the relation
\refer{Eq:deltaV}, as well as other requirements of the Nielsen identity,
we have computed the effective potential in a simplified version of the
Standard Model, with $g_t\approx 1$ as the only non-vanishing Yukawa
coupling. In this calculation  known problems with $\gamma_5$ do not
play any role, and one expects that \refer{Eq:deltaV} should hold at the
one-loop order. Compared to its Landau gauge form,\footnote{In the Landau
gauge the effective potential has been computed in a general
renormalizable theory up to two-loops \cite{Ford,Martin}.}
the effective potential in the $R_{\xi,u}$ gauge has some
unusual features which deserve special discussion. In particular,
the vacuum direction depends on the gauge already at tree level.

On the quartet of the real scalar fields $\phi\in\mathbb{R}^4$
the generators of the $\mathfrak{u}(1)_Y\times \mathfrak{su}(2)_L$
algebra are represented by the following four matrices:
\begin{eqnarray}\label{Bazat}
{T_{{}_R}}_0=\frac{g_{{}_Y}}{2} \left(
\begin{array}{cccc}
 0 & 0 & 1 & 0 \\
 0 & 0 & 0 & 1 \\
 -1 & 0 & 0 & 0 \\
 0 & -1 & 0 & 0
\end{array}
\right),
&\przerwpod&
{T_{{}_R}}_1=\frac{g}{2} \left(
\begin{array}{cccc}
 0 & 0 & 0 & 1 \\
 0 & 0 & 1 & 0 \\
 0 & -1 & 0 & 0 \\
 -1 & 0 & 0 & 0
\end{array}
\right),
\nonumber\\
{T_{{}_R}}_2=\frac{g}{2} \left(
\begin{array}{cccc}
 0 & -1 & 0 & 0 \\
 1 & 0 & 0 & 0 \\
 0 & 0 & 0 & -1 \\
 0 & 0 & 1 & 0
\end{array}
\right),
&\przerwpod&
{T_{{}_R}}_3=\frac{g}{2} \left(
\begin{array}{cccc}
 0 & 0 & 1 & 0 \\
 0 & 0 & 0 & -1 \\
 -1 & 0 & 0 & 0 \\
 0 & 1 & 0 & 0
\end{array}
\right).
\end{eqnarray}
The structure constants read
\eq{
{e_{{}_R}}^\alpha_{\przerwpod\beta\gamma}=
\left\{
\begin{array}{ll}
g \epsilon_{\alpha\beta\gamma}\przerwpod&{\rm for}\przerwpod\alpha,
\beta,\gamma\in\{1,2,3\} \\
0&\przerwpod \text{otherwise}
\end{array}
\right.~\!.
}
For simplicity, we take
\eq{
\xi_{\alpha\beta}=\xi~\!\delta_{\alpha\beta},
}
and
\eq{\label{Eq:u}
u=\nawias{0,~\bar{u},~0,~0}^T.
}
In the presence of a constant background $v_{{}_R}$ the scalar potential
with all counterterms allowed by \refer{Eq:SymStartVPW} has the form
\eq{\label{VSMjawnapostac}
\tilde{\mathcal{V}}\nawias{\check{\phi},\xi}=\frac{1}{2}Z_{{}_\phi}\!
\nawias{m^2+\delta m^2}\nawias{\check{\phi}+v_{{}_R}}^T\!\nawias{\check{\phi}
+v_{{}_R}}+
\frac{1}{4!}Z^2_{{}_\phi}\!\nawias{\lambda+\delta\lambda}
\nawias{\nawias{\check{\phi}+v_{{}_R}}^T\!
\nawias{\check{\phi}+v_{{}_R}}}^2,
}
in which (see \refer{Eq:checkPhidef})
\eq{
\check{\phi}=\phi+\delta v.
}
The tree level \emph{effective} potential
written in terms of the background field $v_{\scriptscriptstyle R}$
includes also the contributions of the gauge-fixing term and reads
\eq{\label{Vefftree}
\mathcal{V}^{(0-loop)}_{eff}\nawias{v_{{}_R}}=\frac{1}{2}m^2v^T_{{}_R}
v_{{}_R}+\frac{1}{4!}\lambda\nawias{v^T_{{}_R} v_{{}_R}}^2
+\frac{\xi}{2} \delta^{\alpha\beta}\nawias{v^T_{{}_R}{T_{{}_R}}_\alpha u}
\nawias{v^T_{{}_R}{T_{{}_R}}_\beta u}.
}
For $\xi>0$, $\mathcal{V}^{(0-loop)}_{eff}$ has a minimum at
\eq{\label{Kiervr}
v_{{}_R}=\nawias{0,~\bar{v}_{{}_R},~0,~0}^T,
}
with (assuming $m^2<0$)
\eq{\label{Wartvr}
\bar{v}^2_{{}_R}=-{6m^2\over\lambda}~\!.
}
While the occurrence of spontaneous symmetry breaking (i.e. the existence
of the solution \refer{Wartvr}) depends only on the parameters of the
gauge invariant part of the Lagrangian,\footnote{Is is worth stressing
here again that in the Nielsen identity \refer{NielsenEqLiniowN} we
differentiate with respect to $\xi$ and $u$ keeping the background
$v_{{}_R}$ fixed. From the Nielsen identity satisfied by $\Gamma^N$ it then
follows that the gauge-dependence of the VEV of the scalar field (i.e of
the minimum of $\mathcal{V}_{eff}$) cancels with the explicit
gauge-dependence of 1PI functions, ensuring that physical masses and
couplings expressed as functions of parameters of the tree level action
\refer{LtildeN} do not depend on $\xi$ and $u$ (see \cite{Nielsen}).}
the form \refer{Kiervr} of $v_{{}_R}$ indicates that the vacuum alignment
depends on the gauge (i.e. on the direction of $u$) already at the tree
level. This is reminiscent of the well-known Dashen vacuum alignment
condition \cite{Dashen}. In the $R_{\xi,u}$ gauge with the choice \refer{Eq:u},
the vacuum degeneracy is entirely removed - we have to choose the solution
to \refer{Wartvr} which has the same sign as $\bar{u}$. Otherwise mass
squares of unphysical `particles' would become negative and the usual
interpretation of the Cutkosky rules in terms of the (pseudo)unitarity
would be destroyed. The solution \refer{Kiervr} implies the following
identification of the electromagnetic $\mathfrak{u}(1)_{\rm EM}$ generator
\eq{\label{eq:Qdef}
Q=\frac{e}{g_{{}_Y}}{T_{{}_R}}_0+\frac{e}{g}{T_{{}_R}}_3, \quad
e=\frac{g g_{{}_Y}}{g_{{}_Z}},\quad g_{{}_Z}=\sqrt{g^2+g_{{}_Y}^2},
}
and leads to the usual parametrization of the scalar field
\eq{
\phi=\nawias{G^1,~h,~G^2,~G^0}^T.
}
Computing the one-point function of $h(x)$, we obtain\footnote{In
\refer{Veff1loopdiv} the background $v_{{}_R}$ is restricted to the
$h(x)$-direction. Under our assumptions, the vanishing of other tadpoles
is then ensured by the $CP$ symmetry and the $\mathfrak{u}(1)_{\rm EM}$ symmetry.}
\eqs{\label{Veff1loopdiv}
-\poch{\mathcal{V}_{eff}^{(1-loop)}\nawias{v_{{}_R}}}{v_{{}_R}}&=&
\frac{\hbar}{\nawias{4\pi}^2\epsilon}\left[-3 g_{{}_t}^4 v^3_{{}_R}
+\lambda v_{{}_R}\nawias{m^2+\frac{\lambda}{3} v^2_{{}_R}}
+\frac{3}{16}\nawias{2 g^4+g^4_{{}_Z}}v^3_{{}_R}+{}\right.\nonumber\\
&{}&\left.-\frac{1}{4}\xi v_{{}_R}\nawias{2 g^2+g^2_{{}_Z}}
\nawias{m^2+\frac{\lambda}{3}v^2_{{}_R}}+
\frac{1}{4}\xi u\nawias{2 g^2+g^2_{{}_Z}}
\nawias{m^2+\frac{\lambda}{2}v^2_{{}_R}}\right]
+\mathcal{O}\nawias{\epsilon^0}+{}\nonumber\\
&{}&-Z_{{}_\phi}\nawias{m^2+\delta m^2}\nawias{v_{{}_R}+\delta v}
-\frac{1}{3!}Z^2_{{}_\phi}\nawias{\lambda+\delta\lambda}
\nawias{v_{{}_R}+\delta v}^3.
}
The above function is indeed finite (at $\mathcal{O}\nawias{\hbar}$ order)
for any value of $v_{{}_R}$, provided that
\eq{\label{deltavjaw}
\delta v=\frac{\hbar}{\nawias{4\pi}^2\epsilon}\frac{1}{4}\xi
\nawias{2 g^2+g^2_{{}_Z}} u+\mathcal{O}\nawias{\hbar^2},
}
\eq{\label{deltamkw}
\delta m^2=\frac{\hbar}{\nawias{4\pi}^2\epsilon}~\!m^2
\macierz{\lambda-\frac{1}{4}\xi\nawias{2 g^2+g^2_{{}_Z}}}
-m^2\delta Z_{{}_\phi}+\mathcal{O}\nawias{\hbar^2},
}
\eq{\label{deltalam}
\delta \lambda=\frac{\hbar}{\nawias{4\pi}^2\epsilon}
\macierz{2\lambda^2-18 g^4_{{}_t}+\frac{9}{8}\nawias{2 g^4+g^4_{{}_Z}}
-\frac{1}{2}\lambda\xi\nawias{2 g^2+g^2_{{}_Z}}}
-2\lambda \delta Z_{{}_\phi} +\mathcal{O}\nawias{\hbar^2},
}
where $\delta Z_{{}_\phi}=Z_{{}_\phi}-1$. For the generators (\ref{Bazat}), the
formula (\ref{b1loopXI}) yields
\eq{
b^{a}_{\przerw c}=-\frac{\hbar}{\nawias{4\pi}^2\epsilon}
\frac{1}{4}\xi\nawias{2 g^2+g^2_{{}_Z}}\delta^{a}_{\przerw c}
+\mathcal{O}\nawias{\hbar^2},
}
so that \refer{Eq:deltaV} holds true as required by the
Nielsen identity. Furthermore, the mass counterterm (\ref{deltamkw}) is
independent of $u$, as it should be - the scalar potential in
\refer{LtildeNPW} can depend on $u$ only through $\check{\phi}$.
Since $Z_{{}_\phi}$ cannot depend on $u$ and $v_{{}_R}$, we have computed the
two-point function of $h(x)$ in the restricted 't Hooft gauge (i.e. setting
$u=v_{{}_R}$ with the background $v_{{}_R}$ chosen so that
$\left<\phi\right>=0$). This gives
\eq{\label{deltalaZ}
\delta Z_{{}_\phi}=\frac{\hbar}{\nawias{4\pi}^2\epsilon}
\macierz{-3 g^2_{{}_t}-\frac{1}{4}\nawias{\xi-3}
\nawias{2 g^2+g^2_{{}_Z}}} +\mathcal{O}\nawias{\hbar^2}.
}
The explicit $\xi$-dependence of $\delta m^2$ and $\delta\lambda$ given by
(\ref{deltalam}) and (\ref{deltamkw}), respectively is canceled by
that of $Z_{{}_\phi}$. This also agrees with the Nielsen identity, since
the relations \refer{VinvfullNUXIkPW} and \refer{Eq:dZAiphiPW} allow the
scalar potential \refer{VSMjawnapostac} to depend on $\xi$ only through
$Z_{{}_\phi}$, as we have shown in Section 4.
\vskip0.2cm

We end this section by deriving a condition which ensures homogeneity
of the Nielsen identity satisfied by the (renormalized) effective
potential. To this end we set $v_{{}_R}=0$, so that now the field $\phi$
implicitly includes its VEV and differentiate the Nielsen identity
(\ref{NielsenEqLiniowN}) for the effective action $\Gamma^N$ with
respect to $q^{\kappa\epsilon}\nawias{x}$. In this way we obtain the equation
originally derived by Nielsen \cite{Nielsen}
\eq{\label{NielsenEqLiniowNGammaGH0}
\left.\macierz{\nawias{\pochfunc{}{q^{\kappa\epsilon}\nawias{x}}
\pochfunc{\Gamma^N}{\mathcal{K}_i}}\cdot\pochfunc{\Gamma^N}{\mathcal{A}^i}
-(\xi^{-1})_{\beta\alpha}f^\beta\cdot\pochfunc{}{q^{\kappa\epsilon}\nawias{x}}
\pochfunc{\Gamma^N}{\anty{\omega}_{\alpha}}
+\pochfunc{\Gamma^N}{\xi^{\kappa\epsilon}\nawias{x}}}\right|_{gh.n.=0}=0,
}
where $gh.n.=0$ indicates restriction to
terms having zero the ghost number. For spacetime-independent configurations of $\phi$, $\xi$,
$u$ and vanishing vector fields (\ref{NielsenEqLiniowNGammaGH0})
reduces to
\eq{\label{NielsenIDniejed}
\poch{\mathcal{V}_{eff}\nawias{\phi,\xi}}{\xi^{\kappa\epsilon}}
+\poch{\mathcal{V}_{eff}\nawias{\phi,\xi}}{\phi^a}~\!C^a_{\kappa\epsilon}
\nawias{\phi,\xi}=\nawias{u^T {T_{{}_R}}_\alpha\phi}
B^\alpha_{\kappa\epsilon}\nawias{\phi,\xi},
}
with
\eq{\label{Eq:BiC}
C^a_{\kappa\epsilon}\nawias{\phi,\xi}\equiv
\left.\macierz{\calka{4}{y}{}\pochfunc{}{q^{\kappa\epsilon}\nawias{x}}
\pochfunc{\Gamma^N}{K_a\nawias{y}}}\right|_{ {\phi=const \atop rest=0}},\qquad
B^\alpha_{\kappa\epsilon}\nawias{\phi,\xi}\equiv\left.\macierz{\calka{4}{y}{}
\pochfunc{}{q^{\kappa\epsilon}\nawias{x}}\pochfunc{\Gamma^N}
{\anty{\omega}_{\alpha}\nawias{y}}}\right|_{\phi=const\atop rest=0}.
}
Nielsen worked in the $u=0$ gauge, for which the right-hand side of
\refer{NielsenIDniejed} is zero and the resulting identity has the form
analogous to the renormalization group equation satisfied by
$\mathcal{V}_{eff}$. In this case $\xi$-independence of the occurrence
of spontaneous symmetry breaking is ensured (see \cite{Nielsen} for
details). As pointed out in \cite{AitchisonFraser}, to reach the same
conclusion for $u\neq 0$, the effective potential should 
be restricted to field configurations obeying
$u^T {T_{{}_R}}_\alpha\phi=0$. The physical minimum of the full effective potential must belong to this subspace anyway - for the gauge-fixing function
\refer{falphastart} this condition is equivalent to the requirement that
\eq{\label{Eq:vevcond}
\left<f^\alpha(x)\right>=0.
}
Violation of \refer{Eq:vevcond} would mean spontaneous BRST symmetry
breaking \cite{deWit,FukudaKugo} which would spoil the Kugo-Ojima
quartet mechanism and, consequently, unitarity of the physical
$S$-matrix. Still, one should
check whether the minima of the potential restricted to this
subspace are indeed stationary points of the full effective
action. Usually this is guaranteed by discrete symmetries
\cite{FukudaKugo} what in most cases requires invariance of the theory
under $CP$ which however need not always be an exact symmetry of the
theory of interest. As we now show, even in the absence of requisite
discrete symmetries, the Slavnov-Taylor identity of the ordinary BRST
symmetry itself (which always holds true if there are no anomalies)
ensures that at the minimum of the restricted potential the remaining tadpoles vanish automatically provided a certain condition is satisfied at the tree level. This generalizes the observation made in \cite{Grassi} for the Standard Model quantized in the ordinary 't Hooft $R_\xi$ gauge.

Differentiating the identity \refer{NielsenEqLiniowN} written for
$\Gamma^N$ with respect to $\omega^\gamma(x)$ one finds that only one term on the left-hand side contributes when $\phi$ is spacetime-independent and satisfies $u^T  {T_{{}_R}}_\alpha\phi=0$ while all other fields are taken to zero. This gives rise to the relation
\eq{\label{eq:Vtad}
\poch{\mathcal{V}_{eff}\nawias{\phi}}{\phi^a}~\!\chi^a_{\phantom{a}\gamma}
\nawias{\phi}=0,
}
in which 
\eq{\label{Eq:chi}
\chi^a_{\phantom{a}\gamma}\nawias{\phi}\equiv
\left.\macierz{\calka{4}{y}{}\pochfunc{}{\omega^\gamma\nawias{x}}
\pochfunc{\Gamma^N}{K_a\nawias{y}}}\right|_{ {\phi=const \atop rest=0}}=\macierz{T_{{}_R\gamma}}^a_{\phantom{a}b}\phi^b+\mathcal{O}(\hbar).
}
Defining a matrix
\eq{
\Omega_{\beta a}\equiv(u^{{}_T} T_{{{}_{R}\beta}})_a,
}
one gets
\eq{\label{eq:VminLag}
\left.\poch{\mathcal{V}_{eff}\nawias{\phi}}{\phi^a}\right|_{\phi=\phi_0}=\lambda^B \Omega_{Ba},
}
where $\phi_0$ is a stationary point of the effective potential restricted to $\mathrm{ker}~\!\Omega$ , $\lambda^B$ are Lagrange multipliers and the index $B$ runs over a set $\mathrm{R}$ of linearly independent rows of the matrix $\Omega_{\beta a}$. Comparing \refer{eq:Vtad} and \refer{eq:VminLag} gives
\eq{\label{eq:lambdCon}
\lambda^B \Omega_{Ba}~\!\chi^a_{\phantom{a}\gamma}\nawias{\phi_0}=0.
}
Finally, taking into account the expansion \refer{Eq:chi}, one finds
\eq{
\Omega_{Ba}~\!\chi^a_{\phantom{a}\gamma}\nawias{\phi_0}=
\frac{1}{2}
u^{T}
\left\{
T_{{{}_R}B},~
T_{{{}_R}\gamma}
\right\}
\phi_0+\mathcal{O}(\hbar).
}
Thus, if the quadratic form
\eq{\label{eq:muDef}
\mu_{BG}\equiv u^{T}
\left\{
T_{{{}_R}B},~
T_{{{}_R}G}
\right\}
\phi_0,\qquad B,G\in\mathrm{R},
}
is nondegenerate, then from the relation \refer{eq:lambdCon} one recursively infers
that $\lambda^B$ vanishes to all orders. Hence, according to \refer{eq:VminLag}, $\phi_0$ is the stationary point of the full effective potential.
\footnote{It is easy to
see that the same conclusion readily follows if $\phi_0$ in the definition \refer{eq:muDef} is replaced by its tree approximation. However, in 
theories in which a nonzero VEV is generated only radiatively, it is better to treat $\phi_0$ as the minimum of the potential calculated to a given order in the loop expansion.
}

For example, in the case of two Higgs doublets with hypercharges +1/2,
one can take $u$ as an arbitrary vector which preserves the
$\mathfrak{u}(1)_{EM}$ generator defined in analogy with \refer{eq:Qdef}.
The quadratic form \refer{eq:muDef} is then always nondegenerate with the single
exception of $\phi_0$ being orthogonal to the gauge-fixing vector~$u$. Similarly, in the simplified Standard Model considered in the first
part of this section, $\mu_{BG}$ is nondegenerate in the gauge specified by \refer{Eq:u}, if $\phi_0$ is identified with the solution \refer{Kiervr}. However even the tree level effective potential \refer{Vefftree} has
also another stationary point (a minimum for some values of gauge-fixing
parameters) which is reminiscent of the solution found by Jackiw and
Dolan \cite{JackiwDolan} in the abelian Higgs model.
This stationary point does not satisfy the condition \refer{Eq:vevcond} and should be rejected, because one cannot built a physically acceptable theory around such a solution.
\footnote{We disagree with the suggestions made in \cite{PiguetiInni},
that no relation between $\left<\phi\right>$ and $u$ is required, and that
one can construct a quantum theory around any minimum of the modified
effective potential $\mathcal{V}^{mod}_{eff}\nawias{\phi}$, which in our
notation reads $$\mathcal{V}^{mod}_{eff}\nawias{\phi}\equiv
\mathcal{V}_{eff}\nawias{\phi}-\frac{\xi^{\alpha\beta}}{2}
\nawias{u^T {T_{{}_R}}_\alpha\phi}\nawias{u^T {T_{{}_R}}_\beta\phi}.$$
$\mathcal{V}^{mod}_{eff}$ defined in this way is gauge-independent at the
tree level and satisfies the homogeneous Nielsen identity (this follows
immediately from \refer{NielsenIDniejed}, since the functions
defined in \refer{Eq:BiC} are related to each other by the ghost equation
(\ref{DuchEqLinN}) for $\Gamma^N$). The modified potential was obtained
in \cite{PiguetiInni} from the effective action by setting to zero the
Nakanishi-Lautrup multipliers. This is however an \emph{off-shell
configuration of fields} if the scalar fields do not satisfy the condition
$u^T {T_{{}_R}}_\alpha\phi=0$. Thus, the potential
$\mathcal{V}_{eff}$, in which the Nakanishi-Lautrup fields are always
on-shell (with respect to a given configuration of the scalar fields)
seems more physical despite its
gauge-dependence. Of course, if one restricts the space of scalar fields
to configurations obeying $u^T {T_{{}_R}}_\alpha\phi=0$, then stationary
points of $\mathcal{V}_{eff}$ are the same as those of
$\mathcal{V}^{mod}_{eff}$, however - contrary to the conclusions of
\cite{PiguetiInni} - by replacing $\mathcal{V}_{eff}$ with
$\mathcal{V}^{mod}_{eff}$ one cannot avoid the condition
$u^T {T_{{}_R}}_\alpha\phi=0$, since it is necessary on physical grounds.
On the other hand, the usual potential $\mathcal{V}_{eff}$ naturally `feels'
this conditions, owing to the Dashen mechanism mentioned above.
}

The above reasoning shows that in theories without chiral fermions
in which the requisite discrete symmetries are not exact (e.g. CP can be
explicitly and/or spontaneously broken in the extended Higgs sector) 
the effective potential can be restricted to configurations satisfying the
condition $u^T  {T_{{}_R}}_\alpha\phi=0$ under relatively mild
requirements. On the other hand, in theories with chiral fermions in which (as in the Standard Model) CP is not an exact symmetry, our reasoning
formally can still be applied, since it is based only on the Slavnov-Taylor
identity satisfied by the effective action $\Gamma^N$. However, in this
case the renormalized action $\tilde{\cal I}^N$ must include counterterms
explicitly violating its BRST invariance in order to restore the invariance
of $\Gamma^N$ \cite{BarPas}. It appears that as such BRST-noninvariant
counterterms appropriate terms linear in the scalar fields (linear in the
would-be Goldstone bosons) can be indispensable to ensure restoration of the
identity \refer{eq:Vtad} which has been the starting point of our
arguments.

\end{section}

\begin{section}{Conclusions}
In this paper we have found all counterterms required to render finite
the effective action with Nielsen sources included
of a general Yang-Mills theory  coupled to arbitrary scalar
fields and quantized in linear $R_{\xi,u}$ gauges. We have shown that the
dependence
of all counterterms on $u$ is controlled by a single matrix $b(\xi)$ that
acts as a counterterm for a two-point function of certain composite
operators. In particular, the gauge-invariant part of the action with
counterterms depends on $u$ only through the shifted scalar field
$\check{\phi}=\phi-b(\xi)u$. Assuming that the parameters $\xi$ are
consistent with the symmetry under global gauge transformations, we
have proved that this shift constitutes the only contribution to the
well known \mbox{VEV counterterm $\delta v$}. This is our main new result
that allows a simple calculation of $\delta v$ at one-loop and clarifies
its origin. The resulting expression for $\delta v$ agrees with explicit
computations in Section 5 as well as those of
\cite{Hollik, ChanPoRo, LoiWill}.

We have also considered the case of multiple parameters $\xi$. We have
shown that an additional counterterm (`the curvature') $\hat{\Psi}$,
which trivially vanishes in the situation studied in \cite{PiguetSibold},
can be generated only if the matrix $\xi^{-1}$ breaks the symmetry under
global gauge transformations. We have also shown that the interpretation of the Nielsen identities for counterterms in terms of \mbox{$\xi$-independence} of coupling constants of bare fields is unaffected by $\hat{\Psi}$.

Finally, we have considered the problem of homogeneity of the Nielsen
identities that control the gauge-dependence of the effective potential.
There has been much discussion in the literature of this issue, e.g.
\cite{FukudaKugo,AitchisonFraser,PiguetiInni}. The effective potential restricted to configurations which preserve the BRST symmetry 
satisfies the homogeneous Nielsen identities. We have introduced a condition 
which allows to check whether a minimum of such a restricted potential is a stationary point of the full effective action.

\vskip0.5cm

\noindent\textbf{Acknowledgments.} I am indebted to Professors P. H. Chankowski
and K. A. Meissner for enlightening conversations and comments on an early version of this paper.
\end{section}

\appendix

\begin{section}{Determination of $\tilde{\mathcal{I}}^N$}

\begin{subsection}{Nielsen Identities for
$\tilde{\mathcal{I}}^N$}\label{app:TozNielPrzec}

The renormalized action
$\tilde{\mathcal{I}}^N$ which includes all possible counterterms is a
general local solution to the Nielsen identities and the ghost equation,
constrained by the power-counting and the ghost number conservation. Scalar
and vector fields have dimension 1. We treat parameters $u^a(x)$ and
$\xi(x)$ as external fields of vanishing ghost number and dimension 1 and 0,
respectively. If, as is customary, we ascribe dimension 1 and ghost number
$+1$ to the ghost field $\omega^\alpha$, then for the other fields dimension
4 and zero ghost number of the Lagrangian \refer{LtildeN} implies:
\eq{
\anty{\omega}_\alpha\sim (1,-1), \quad q^a\sim (2, +1),
\quad q^{\alpha\beta}\sim (1,+1),
\quad \mathcal{K}_i\sim (2, -1),\quad L_\alpha(x) \sim (2, -2).
}
The renormalized action functional $\tilde{\mathcal{I}}^N$ must satisfy
the Nielsen identity
\eq{\label{NielsenEqLiniowNUXI}
\pochfunc{\tilde{\mathcal{I}}^N}{\mathcal{K}_i}\cdot
\pochfunc{\tilde{\mathcal{I}}^N}{\mathcal{A}^i}+
\pochfunc{\tilde{\mathcal{I}}^N}{L_{\alpha}}\cdot
\pochfunc{\tilde{\mathcal{I}}^N}{\omega^{\alpha}}
-(\xi^{-1})_{\beta\alpha}\nawias{f^\beta+
\frac{1}{2}q^{\beta\gamma}~\!\anty{\omega}_\gamma}\cdot
\pochfunc{\tilde{\mathcal{I}}^N}{\anty{\omega}_{\alpha}}
+q^{\alpha\beta}\cdotp\pochfunc{\tilde{\mathcal{I}}^N}{\xi^{\alpha\beta}}+
q^{a}\cdotp\pochfunc{\tilde{\mathcal{I}}^N}{u^a}=0,
}
and the ghost equation
\eqs{\label{DuchEqLinNUXI}
\pochfunc{\tilde{\mathcal{I}}^N}{\anty{\omega}_{\alpha}\nawias{x}}+
\pochfunc{f^\alpha\!\nawias{x}}{\mathcal{A}^i}\cdot
\pochfunc{\tilde{\mathcal{I}}^N}{\mathcal{K}_i}&=&
\frac{1}{2}q^{\alpha\delta}(x)(\xi(x)^{-1})_{\delta \beta}
\nawias{f^\beta(x)+\frac{1}{2}q^{\beta\gamma}
\nawias{x}\anty{\omega}_\gamma\nawias{x}}
-q^{\alpha\beta}\nawias{x}\nawias{\phi^a\!\nawias{x}+v^a_{{}_R}}
\delta_{ac}\macierz{{T_{{}_R}}_\beta}^c_{\przerw b}u^b(x)\nonumber\\
&{}&+q^a\nawias{x}\xi^{\alpha\beta}\nawias{x}\delta_{ac}
\macierz{{T_{{}_R}}_\beta}^c_{\przerw b}\nawias{\phi^b\!\nawias{x}+v^b_{{}_R}}.
}
The most general functional of~dimension~4 allowed by the ghost
number conservation has the following dependence on the external sources
$q^a(x)$, $q^{\alpha\beta}(x)$, $\mathcal{K}_i(x)$ and $L_\alpha(x)$
\eq{\label{DzialanieZkontrzlonamiNUXI}
\tilde{\mathcal{I}}^N
\equiv\calka{4}{x}{\funkcjonal{\tilde{\mathcal{L}}^N}
{{\mathcal{A},\anty{\omega},\omega;\mathcal{K},L;q,u,\xi}|x}},
}
where
\begin{eqnarray}\label{LtotLincechNUXI}
\tilde{\mathcal{L}}^N(x)&=&\frac{1}{2}L_{\alpha}
\nawias{x}C^\alpha_{\przerwpod\beta\gamma}\nawias{\xi\nawias{x}}
\omega^{\beta}\nawias{x}\omega^{\gamma}\nawias{x}
+L_{\alpha}\nawias{x}g^\alpha_{\przerwpod\beta\gamma\delta}
\nawias{\xi\nawias{x}}q^{\beta\gamma}\nawias{x}\omega^{\delta}\nawias{x}
+L_{\alpha}\nawias{x}H^\alpha_{\przerwpod\beta\gamma\delta\epsilon}
\nawias{\xi\nawias{x}}q^{\beta\gamma}\nawias{x}q^{\delta\epsilon}\nawias{x}
\nonumber\\
{}&{}&+\mathcal{K}_i\nawias{x}\calka{d}{y}{}\omega^\alpha\nawias{y}
\funkcjonal{D^i_{\text{ }\alpha}}{\mathcal{A},u,\xi|x,y}+
\mathcal{K}_i\nawias{x}\calka{d}{y}{}\funkcjonal{k^i_{\text{ }\alpha\beta}}
{\mathcal{A},u,\xi|x,y}q^{\alpha\beta}\nawias{y}+\mathcal{K}_i
\nawias{x}b^i_{\przerw a}\nawias{\xi\nawias{x}}q^a\nawias{x}\nonumber\\
&{}&+q^a\nawias{x}\calka{d}{y}{}\funkcjonal{d_{a}^{\przerw \alpha}}
{\mathcal{A},u,\xi|x,y}\anty{\omega}_\alpha\nawias{y}+
q^{\alpha\beta}\nawias{x}\calka{d}{y}{}
\funkcjonal{r_{\alpha\beta}^{\przerw\przerwpod \kappa}}
{\mathcal{A},u,\xi|x,y}\anty{\omega}_\kappa\nawias{y}\nonumber\\
{}&{}&+\anty{\omega}_\kappa\nawias{x}\anty{\omega}_\sigma
\nawias{x}{\mathit{l}}^{\kappa\sigma}_{\przerw\przerwpod\beta\gamma\delta\epsilon}
\nawias{\xi\nawias{x}}q^{\beta\gamma}\nawias{x}q^{\delta\epsilon}\nawias{x}+
\anty{\omega}_\kappa\nawias{x}\anty{\omega}_\sigma
\nawias{x}{\mathit{m}}^{\kappa\sigma}_{\przerw\przerwpod\beta\gamma\delta}
\nawias{\xi\nawias{x}}q^{\beta\gamma}\nawias{x}\omega^{\delta}\nawias{x}
\nonumber\\
{}&{}&+\funkcjonal{\tilde{\mathcal{L}}^N_{FP}}
{\mathcal{A},\anty{\omega},\omega;u,\xi|x}.
\end{eqnarray}
The functions $C^\alpha_{\przerw\beta\gamma}\nawias{\xi}$,
$g^\alpha_{\przerwpod\beta\gamma\delta}\nawias{\xi}$,
$H^\alpha_{\przerwpod\beta\gamma\delta\epsilon}\nawias{\xi}$,
$b^i_{\przerw a}\nawias{\xi}$,
${\mathit{l}}^{\kappa\sigma}_{\przerw\przerwpod\beta\gamma\delta\epsilon}
\nawias{\xi}$ and
${\mathit{m}}^{\kappa\sigma}_{\przerw\przerwpod\beta\gamma\delta}\nawias{\xi}$
have all dimension 0. $H^\alpha_{\przerwpod\beta\gamma\delta\epsilon}$
and ${\mathit{l}}^{\kappa\sigma}_{\przerw\przerwpod\beta\gamma\delta\epsilon}$ are
antisymmetric under the interchange
$(\beta\gamma)\leftrightarrow(\delta\epsilon)$ and symmetric with respect
to interchanges $\beta\leftrightarrow\gamma$ or
$\delta\leftrightarrow\epsilon$.
The kernels $\funkcjonal{D^i_{\text{ }\alpha}}{\mathcal{A},u,\xi|x,y}$,
$\funkcjonal{k^i_{\text{ }\alpha\beta}}{\mathcal{A},u,\xi|x,y}$ and
$\funkcjonal{d_{a}^{\przerw \alpha}}{\mathcal{A},u,\xi|x,y}$ have dimension 1,
while $\funkcjonal{r_{\beta\gamma}^{\przerw\przerwpod\kappa}}
{\mathcal{A},u,\xi|x,y}$
has dimension 2.\footnote{Assigning dimensions to these kernels we can
treat $\delta^{(4)}\nawias{x-y}$ as dimensionless since in the lagrangian
it is accompanied by the measure $\calkabp{4}{}{x}$.}

Constraints imposed by the Nielsen identity can be
conveniently expressed in terms of the differential forms
\eq{
\hat{g}^\alpha_{\przerwpod\delta}=g^\alpha_{\przerwpod\beta\gamma\delta}
\nawias{\xi}\text{d}\xi^{\beta\gamma},\przerwpod\przerwpod\przerwpod
\hat{H}^\alpha=H^\alpha_{\przerwpod\beta\gamma\delta\epsilon}
\nawias{\xi}\text{d}\xi^{\beta\gamma}\wedge\text{d}\xi^{\delta\epsilon},
}
and
\eq{
\hat{\mathit{l}}^{\kappa\sigma}
=\mathit{l}^{\kappa\sigma}_{\przerwpod\przerw\beta\gamma\delta\epsilon}
\nawias{\xi}\text{d}\xi^{\beta\gamma}\wedge\text{d}\xi^{\delta\epsilon}.
}
Terms of the identity (\ref{NielsenEqLiniowNUXI}) which are
linear in $L_\alpha$ and proportional to different powers of
$q^{\alpha\beta}$ impose respectively the relations
\eq{\label{TozJacNielXI}
\left[C_\beta,~C_\gamma\right]=C^\alpha_{\przerwpod\beta\gamma} C_\alpha,
}
and
\eqs{
&\text{d}C^{\sigma}_{\przerw\beta\gamma}&
=\hat{g}^\sigma_{\przerw\alpha}C^{\alpha}_{\przerw\beta\gamma}
-C^{\sigma}_{\przerw\beta\alpha}\hat{g}^\alpha_{\przerw\gamma}
-C^{\sigma}_{\przerw\alpha\gamma}\hat{g}^\alpha_{\przerw\beta},\label{EqOA}\\
&\text{d}\hat{g}^{\sigma}_{\przerw\gamma}&
=\hat{g}^\sigma_{\przerw\alpha}\wedge\hat{g}^{\alpha}_{\przerw\gamma}
-C^{\sigma}_{\przerw\alpha\gamma}\hat{H}^\alpha,\label{EqOB}\\
&\text{d}\hat{H}^{\sigma}&
=\hat{g}^\sigma_{\przerw\alpha}\wedge\hat{H}^{\alpha}.\label{EqOC}
}
In turn, terms involving products $q^a \times\mathcal{K}_i\times\omega^\beta$
and $q^a\times\mathcal{K}_j\times q^{\alpha\beta}$ give
respectively\footnote{The coefficients ${b^i_{\przerw a}}$ are
treated here as vanishing if the index $i$ corresponds to
vector fields.}
\eq{
\nawias{{b^i_{\przerw a}\!\nawias{\xi\nawias{z}}}
\pochfunc{}{\mathcal{A}^i\nawias{z}}+\pochfunc{}{u^a\nawias{z}}}
\funkcjonal{D^j_{\text{ }\alpha}}{\mathcal{A},u,\xi|x,y}=0,
}
\eq{
\nawias{{b^i_{\przerw a}\!\nawias{\xi\nawias{z}}}\pochfunc{}
{\mathcal{A}^i\nawias{z}}+\pochfunc{}{u^a\nawias{z}}}
\funkcjonal{k^j_{\text{ }\alpha\beta}}{\mathcal{A},u,\xi|x,y}
\funkcjonal{D^i_{\text{ }\alpha}}{\mathcal{A},u,\xi|x,y}
=\funkcjonal{D^i_{\text{ }\alpha}}{\mathcal{A}-b\nawias{\xi} u,0,\xi|x,y}.
}
Hence
\eq{\label{DNUXIroz}
\funkcjonal{D^i_{\text{ }\alpha}}{\mathcal{A},u,\xi|x,y}
=\funkcjonal{D^i_{\text{ }\alpha}}{\mathcal{A}-b\nawias{\xi} u,0,\xi|x,y},
}
and
\eq{\label{kNUXIroz}
\funkcjonal{k^i_{\text{ }\alpha\beta}}{\mathcal{A},u,\xi|x,y}
=\funkcjonal{k^i_{\text{ }\alpha\beta}}{\mathcal{A}-b\nawias{\xi} u,0,\xi|x,y}
+\poch{b^i_{\przerw a}}{\xi^{\alpha\beta}}
\nawias{\xi\nawias{x}}u^a\nawias{x}\delta^{(4)}\nawias{x-y}.
}
Terms of the Nielsen identity involving products
$q^{\alpha\beta} \times\mathcal{K}_i\times\omega^\gamma$
and $q^{\alpha\beta} \times q^{\delta\gamma}\times \mathcal{K}_i$
yield the relations
\eqs{\label{KomDzkNUXI}
\calkabp{d}{x}{}\nawias{
\funkcjonal{k^i_{\przerw\alpha\beta}}{\mathcal{A},u,\xi|x,y}
\pochfunc{\funkcjonal{D^j_{\przerw\gamma}}
{\mathcal{A},u,\xi|z,w}}{\mathcal{A}^i(x)}
-\funkcjonal{D^i_{\przerw\gamma}}{\mathcal{A},u,\xi|x,w}
\pochfunc{\funkcjonal{k^j_{\przerw\alpha\beta}}{\mathcal{A},u,\xi|z,y}}
{\mathcal{A}^i(x)}}\phantom{aaaaaaa}\nonumber\\
=-g^{\delta}_{\przerw\alpha\beta \gamma}\nawias{\xi\nawias{w}}
\delta^{(d)}(y-w)\funkcjonal{D^j_{\przerw\delta}}{\mathcal{A},u,\xi|z,w}-
\pochfunc{\funkcjonal{D^j_{\przerw\gamma}}
{\mathcal{A},u,\xi|z,w}}{\xi^{\alpha\beta}\nawias{y}},
}
and
\eqs{\label{KomkzkNUXI}
\calkabp{d}{x}{}\nawias{
\funkcjonal{k^i_{\przerw\alpha\beta}}{\mathcal{A},u,\xi|x,y}
\pochfunc{\funkcjonal{k^j_{\przerw\delta\gamma}}
{\mathcal{A},u,\xi|z,w}}{\mathcal{A}^i(x)}
-\funkcjonal{k^i_{\przerw\delta\gamma}}{\mathcal{A},u,\xi|x,w}
\pochfunc{\funkcjonal{k^j_{\przerw\alpha\beta}}
{\mathcal{A},u,\xi|z,y}}{\mathcal{A}^i(x)}}\phantom{aaaaaaa}\nonumber\\
=-2 F^{\epsilon}_{\przerw\alpha\beta\delta \gamma}\nawias{\xi\nawias{y}}
\delta^{(d)}(y-w)\funkcjonal{D^j_{\przerw\epsilon}}{\mathcal{A},u,\xi|z,y}-
\pochfunc{\funkcjonal{k^j_{\przerw\delta\gamma}}
{\mathcal{A},u,\xi|z,w}}{\xi^{\alpha\beta}\nawias{y}}
+\pochfunc{\funkcjonal{k^j_{\przerw\alpha\beta}}
{\mathcal{A},u,\xi|z,y}}{\xi^{\delta\gamma}\nawias{w}}.
}
Finally, terms linear in $\mathcal{K}_i$ impose the commutation relations
\eqs{\label{KomCechFullNUXI}
\calkabp{d}{z}{}\nawias{
\pochfunc{\funkcjonal{D^i_{\przerw\beta}}{\mathcal{A},u,\xi|x,y}}
{\mathcal{A}^j(z)}\funkcjonal{D^j_{\przerw\gamma}}{\mathcal{A},u,\xi|z,w}
-\pochfunc{\funkcjonal{D^i_{\przerw\gamma}}{\mathcal{A},u,\xi|x,w}}
{\mathcal{A}^j(z)}\funkcjonal{D^j_{\przerw\beta}}{\mathcal{A},u,\xi|z,y}}
\phantom{aaaa}\nonumber\\
=C^{\alpha}_{\text{ }\beta \gamma}\nawias{\xi\nawias{y}}
\delta^{(d)}(y-w)\funkcjonal{D^i_{\przerw\alpha}}{\mathcal{A},u,\xi|x,y}.
}

The equations \refer{TozJacNielXI} and \refer{KomCechFullNUXI} are the
ordinary conditions imposed by the BRST symmetry (see e.g.
\cite{ZinnJustinQFTCritical}). Before considering terms of the Nielsen
identity which are independent of $L_\alpha$ and $\mathcal{K}_i$, it will
be convenient to extract information from the ghost equation
(\ref{DuchEqLinNUXI}). Its terms linear in $q^a$ give the relation
\eq{\label{WiezyAntyNUnadXI}
\funkcjonal{d^{\przerw \alpha}_{a}}{\mathcal{A},u,\xi|z,x}=
-\xi^{\alpha\beta}\nawias{z}\delta_{ac}
\macierz{{T_{{}_R}}_\beta}^c_{\przerw b}
\nawias{\phi^b\nawias{x}+v^b_{{}_R}}\delta^{(d)}\nawias{z-x}+
\pochfunc{f^\alpha\nawias{x}}{\mathcal{A}^i\nawias{z}}~\!
{b^i_{\przerw a}\nawias{\xi\nawias{z}}}.
}
Furthermore, (\ref{DuchEqLinNUXI}) implies also that
\eqs{
&{}&{\mathit{m}}^{\kappa\sigma}_{\przerw\przerwpod\beta\gamma\delta}
\nawias{\xi}=0,\\
&{}&{\hat{\mathit{l}}}^{\rho\sigma}=\frac{1}{8}(\xi^{-1})_{\alpha\delta}~\!
\text{d}\xi^{\rho\alpha}\wedge\text{d}\xi^{\delta\sigma},\label{PostacWspl}
}
and that
\eqs{\label{PostacDytrR}
\calka{d}{y}{}q^{\beta\gamma}\nawias{y}
\funkcjonal{r_{\beta\gamma}^{\przerw\przerwpod \alpha}}
{\mathcal{A},u,\xi|y,x}&=&\calkabp{d}{y}~\!
\frac{\delta f^\alpha\nawias{x}}{\delta \mathcal{A}^i(y)}\calka{d}{z}{}
\funkcjonal{k^i_{\przerw\beta\gamma}}{\mathcal{A},u,\xi|y,z}q^{\beta\gamma}(z)
\nonumber\\
&{}&-\frac{1}{2}q^{\alpha\delta}(x)(\xi(x)^{-1})_{\delta \beta}
f^\beta(x)
+q^{\alpha\beta}\nawias{x}\nawias{\phi^a\!\nawias{x}+v^a_{{}_R}}
\delta_{ac}\macierz{{T_{{}_R}}_\beta}^c_{\przerw b}u^b(x),
}
Finally, those terms of (\ref{DuchEqLinNUXI}) which are
independent of the Nielsen sources lead to an equation for
$\delta\tilde{\mathcal{I}}^N_{FP}/\delta\anty{\omega}_\alpha(x)$
(i.e. for the derivative of the last term in \refer{LtotLincechNUXI})
whose solution has the general form
\eq{\label{PostacINeffXI}
\funkcjonal{\tilde{\mathcal{I}}^N_{FP}}
{\mathcal{A},\anty{\omega},\omega;u,\xi}=
-\calkabp{d}{x}{}\calkabp{d}{y}{}
\calka{d}{z}{}\overline{\omega}_{\alpha}(x)
\frac{\delta f^\alpha\nawias{x}}
{\delta \mathcal{A}^i(z)}\funkcjonal{D^i_{\text{ }\gamma}}
{\mathcal{A},u|z,y}\omega^\gamma(y)
+\funkcjonal{\tilde{\mathcal{I}}^N_{rest}}{\mathcal{A};u,\xi}.
}
Returning to the implications of the identity
(\ref{NielsenEqLiniowNUXI}), we find that its
terms cubic in $q^{\alpha\beta}$ vanish automatically after taking into
account the form (\ref{PostacWspl}) of ${\hat{\mathit{l}}}^{\rho\sigma}$.
Similarly, terms quadratic in
$q^{\alpha\beta}$ vanish due to form of ${\hat{\mathit{l}}}^{\rho\sigma}$
and the equations (\ref{PostacDytrR}), (\ref{KomkzkNUXI}).
Terms involving products
$q^a\times q^b$ and $q^a\times q^{\alpha\beta}$ do not give any new
information either, being automatically satisfied by virtue of the
equations (\ref{WiezyAntyNUnadXI}) and (\ref{PostacDytrR}) respectively.
Using (\ref{PostacINeffXI}) and (\ref{WiezyAntyNUnadXI}), we find that the
terms of (\ref{NielsenEqLiniowNUXI}) which are linear in $q^a$ impose
\eq{\label{WarLinqaOstXI}
{b^i_{\przerw a}\!\nawias{\xi\nawias{x}}}~\!
\pochfunc{\tilde{\mathcal{I}}^N_{rest}}{\mathcal{A}^i\nawias{x}}+
\pochfunc{\tilde{\mathcal{I}}^N_{rest}}{u^a\nawias{x}}
=-\calka{d}{y}{}(\xi^{-1})_{\alpha\beta}~\!f^\beta\nawias{y}
\funkcjonal{d^{\przerw \alpha}_{a}}{\mathcal{A},u|x,y}.
}
The above equation can be solved with the help of (\ref{WiezyAntyNUnadXI}):
\eq{\label{DzialanieGF}
\tilde{\mathcal{I}}^N_{rest}
=-\frac{1}{2}(\xi^{-1})_{\alpha\beta}~\!f^\alpha\cdotp f^\beta
+ \funkcjonal{\tilde{\mathcal{I}}^N_{GI}}
{A^\alpha_\mu,~\!\phi^a-b^{a}_{\przerw c} u^c,~\!\xi^{\delta\epsilon}},
}
where $\tilde{\mathcal{I}}^N_{GI}$ is an arbitrary functional of its
arguments. After taking (\ref{PostacDytrR}) and (\ref{KomDzkNUXI})
into account, the terms of the identity \refer{NielsenEqLiniowNUXI}
linear in $q^{\alpha\beta}$ yield
\eq{\label{WarGINUXIk}
\nawias{\pochfunc{}{\xi^{\beta\gamma}(x)}+\calka{d}{z}~\!
\funkcjonal{k^j_{\przerw\beta\gamma}}{\mathcal{A},~\!u,~\!\xi|z,x}
\pochfunc{}{\mathcal{A}^j(z)}}{\funkcjonal{\tilde{\mathcal{I}}^N_{GI}}
{{A^\alpha_\mu,~\!\phi^a-b^{a}_{\przerw c}
u^c,~\!\xi^{\delta\epsilon}}}}=0.
}
Finally, vanishing of the terms of \refer{NielsenEqLiniowNUXI}
independent of external sources gives leads to the constraint
\eq{\label{WarGINUXI}
\calka{d}{z}{}\funkcjonal{D^j_{\przerw\gamma}}{\mathcal{A},u,\xi|z,x}
\pochfunc{\funkcjonal{\tilde{\mathcal{I}}^N_{GI}}
{{A^\alpha_\mu,~\!\phi^a-b^{a}_{\przerw c} u^c,~\!
\xi^{\delta\epsilon}}}}{\mathcal{A}^j(z)}=0,
}
which is the ordinary condition of gauge invariance, whereas
the condition (\ref{WarGINUXIk}) expresses an `additional
gauge invariance' connected with the extended BRST symmetry.
\vskip0.2cm

The most general form $\funkcjonal{D^{i}_{\phantom{a}\alpha}}
{\mathcal{A},u,\xi|x,y}$ of the gauge transformations allowed by
power-counting, Lorentz invariance and their general structure (\ref{DNUXIroz})
is
\eq{\label{TransCechFull1RozdzNUXI}
\funkcjonal{D^{\beta}_{\mu\alpha}}{A,u,\xi|x,y}
=\left\{-\mathcal{N}^{\beta}_{\przerw\alpha}\nawias{\xi\nawias{x}}
\partial^{(x)}_\mu+\mathcal{Q}^{\beta}_{\przerw\alpha\kappa\epsilon}
\nawias{\xi\nawias{x}}\partial_\mu\xi^{\kappa\epsilon}\nawias{x}
+\tilde{e}^{\beta}_{\przerwpod\alpha \epsilon}\nawias{\xi\nawias{x}}
A^\epsilon_{\mu}\nawias{x}\right\}\delta\nawias{x-y},
}
\eq{\label{TransCechFull2RozdzNUXI}
\funkcjonal{D^{a}_{\phantom{a}\alpha}}{\phi,u,\xi|x,y}=
\left\{\left[\tilde{T}_{\alpha}
\nawias{\xi\nawias{x}}\right]^a_{\phantom{a}b}\check{\phi}^b\nawias{x}
+\tilde{V}^a_{N\alpha}\nawias{\xi\nawias{x}}\right\}\delta\nawias{x-y},
}
where $\mathcal{Q}^{\beta}_{\przerw\alpha\kappa\epsilon}\nawias{\xi}$ are
dimensionless functions. Similarly, taking into account
the general form (\ref{kNUXIroz}) one obtains the following
formula for the extended gauge transformations
$\funkcjonal{k^{i}_{\phantom{a}\alpha\gamma}}{\mathcal{A},u,\xi|x,y}$:
\eq{\label{TransCechFull1RozdzNUXIk}
\funkcjonal{k^{\beta}_{\mu\alpha\gamma}}{A,u,\xi|x,y}
=\left\{-\Omega^{\beta}_{\przerw\alpha\gamma}
\nawias{\xi\nawias{x}}\partial^{(x)}_\mu
+\mathcal{P}^{\beta}_{\przerw\alpha\gamma\kappa\epsilon}
\nawias{\xi\nawias{x}}\partial_\mu\xi^{\kappa\epsilon}\nawias{x}
+\theta^{\beta}_{\przerwpod\alpha\gamma \epsilon}
\nawias{\xi\nawias{x}}A^\epsilon_{\mu}\nawias{x}\right\}\delta\nawias{x-y},}
\eq{\label{TransCechFull2RozdzNUXIk}
\funkcjonal{k^{a}_{\phantom{a}\alpha\gamma}}
{\phi,u,\xi|x,y}=
\funkcjonal{}{\zeta^a_{\przerwpod\alpha\gamma b}
\nawias{\xi\nawias{x}}\check{\phi}^b\nawias{x}
+\poch{b^{a}_{\przerw d}}{\xi^{\alpha\gamma}}\nawias{\xi\nawias{x}}
u^d\nawias{x}
+\Sigma^a_{\przerw\alpha\gamma}\nawias{\xi\nawias{x}}}\delta\nawias{x-y}.
}
We have introduced here the notation
\eq{\label{Eq:checkPhidefApp}
\check{\phi}^b\nawias{x}=\phi^b\nawias{x}
-b^{b}_{\przerw d}\nawias{\xi\nawias{x}}u^d\nawias{x}.
}
For the functions of $\xi$ appearing in the formulae
(\ref{TransCechFull1RozdzNUXI}) and (\ref{TransCechFull2RozdzNUXI})
it is convenient to introduce the following parametrization:
\eq{\label{ZwizekWielkosciFalkowINieFalkow}
\tilde{e}^{\beta}_{\przerwpod\alpha \epsilon}\nawias{\xi}
=\mathcal{N}^{\delta}_{\przerw\alpha}
\nawias{\xi}e^{\beta}_{\przerwpod\delta \epsilon}\nawias{\xi},
\przerwpod\przerwpod\przerwpod
\tilde{T}_{\alpha}\nawias{\xi}
=\mathcal{N}^{\delta}_{\przerw\alpha}
\nawias{\xi}T_{\delta}\nawias{\xi},\przerwpod\przerwpod\przerwpod
\tilde{V}^a_{N\alpha}\nawias{\xi}
=\mathcal{N}^{\delta}_{\przerw\alpha}\nawias{\xi}V^a_{N\delta}\nawias{\xi}.
}
This  can be done without loss of generality because in the perturbation
theory $\mathcal{N}^{\delta}_{\przerw\alpha}=\delta^{\delta}_{\przerw\alpha}
+\mathcal{O}\nawias{\hbar}.$ The commutation relations
(\ref{KomCechFullNUXI}) yield
\eq{\label{ZwiazekCieRozdzXI}
C^{\kappa}_{\przerwpod\beta\gamma}
=\mathcal{N}^\alpha_{\przerw\beta}\mathcal{N}^\epsilon_{\przerw\gamma}
\left[\mathcal{N}^{-1}\right]^\kappa_{\przerw\delta}
e^{\delta}_{\przerwpod\alpha \epsilon},
}
and
\eq{\label{RelKomFullRozdzXI}
\left[T_{\alpha},~T_{\beta}\right]
=e^{\gamma}_{\przerwpod\alpha \beta}T_{\gamma},\quad\quad
\left[e_{\alpha},~e_{\beta}\right]
=e^{\gamma}_{\przerwpod\alpha\beta}e_{\gamma}.
}
Moreover (\ref{KomCechFullNUXI}) implies also that
\eq{\label{RelKomVRozdzNUXI}
\left[T_{\alpha}\right]^a_{\przerw b}V^b_{N\beta}
-\left[T_{\beta}\right]^a_{\przerw b}V^b_{N\alpha}
=e^{\gamma}_{\przerwpod\alpha \beta}V^a_{N\gamma},}
and imposes the following constraints on the functions
$\mathcal{Q}^{\beta}_{\przerw\alpha\kappa\epsilon}\nawias{\xi}$:
\eq{\label{RelKomQRozdzNUXI}
\tilde{e}^{\alpha}_{\przerwpod\beta\delta}
\mathcal{Q}^{\delta}_{\przerw\gamma\kappa\epsilon}
-\tilde{e}^{\alpha}_{\przerwpod\gamma\delta}
\mathcal{Q}^{\delta}_{\przerw\beta\kappa\epsilon}
=C^{\delta}_{\przerwpod\beta \gamma}
\mathcal{Q}^{\alpha}_{\przerw\delta\kappa\epsilon}
-\mathcal{N}^\alpha_{\przerw\delta}~\!
\poch{C^{\delta}_{\przerwpod\beta \gamma}}{\xi^{\kappa\epsilon}}.}
Equations \refer{ZwiazekCieRozdzXI}-\refer{RelKomVRozdzNUXI} are the
standard requirements of the BRST symmetry. However, the information about
$u$-independence of $V^a_{N\gamma}$ follows only from the Nielsen identity.
Constraints imposed by
(\ref{KomDzkNUXI}) can be compactly expressed by the 1-forms
\eq{
\hat{\theta}^\beta_{\przerw\epsilon}
=\theta^\beta_{\przerwpod\alpha\gamma\epsilon}\nawias{\xi}
\text{d}\xi^{\alpha\gamma},\przerwpod\przerwpod\przerwpod
\hat{\zeta}^a_{\przerw b}=\zeta^a_{\przerw\alpha\gamma b}
\nawias{\xi}\text{d}\xi^{\alpha\gamma},\przerwpod\przerwpod\przerwpod
\hat{\Sigma}^a=\Sigma^a_{\przerw\alpha\gamma}
\nawias{\xi}\text{d}\xi^{\alpha\gamma},
}
and read
\eq{\label{EqA}
\text{d}\tilde{e}_\gamma=\macierz{\hat{\theta},~\tilde{e}_\gamma}-
\hat{g}^\delta_{\przerwpod\gamma}\tilde{e}_\delta,
\przerwpod\przerwpod\przerwpod
\text{d}\tilde{T}_\gamma=\macierz{\hat{\zeta},~\tilde{T}_\gamma}
-\hat{g}^\delta_{\przerwpod\gamma}\tilde{T}_\delta,
}
\eqs{\label{EqB}\label{Eq:dN}
\text{d}\mathcal{N}=\hat{\theta}\mathcal{N}-\mathcal{N}\hat{g},
}
\eq{\label{EqCeu}
\mathcal{Q}^{\delta}_{\przerw\gamma\alpha\beta}=
\tilde{e}^\delta_{\przerwpod\gamma\epsilon}
\Omega^{\epsilon}_{\przerwpod\alpha\beta}+
\mathcal{N}^{\delta}_{\przerw\epsilon}
g^\epsilon_{\przerwpod\alpha\beta\gamma},
}
\eq{\label{EqD}
\text{d}\tilde{V}^a_{N\gamma}=\hat{\zeta}^a_{\przerw b}\tilde{V}^b_{N\gamma}
-\macierz{\tilde{T}_{\gamma}}^a_{\przerw b}\hat{\Sigma}^b
-\tilde{V}^a_{N\delta}\hat{g}^\delta_{\przerw\gamma},
}
\eq{\label{EqE}
\theta^{\rho}_{\przerwpod\alpha\beta\delta}
\mathcal{Q}^{\delta}_{\przerwpod\gamma\kappa\epsilon}
-\tilde{e}^{\rho}_{\przerwpod\gamma\delta}
\mathcal{P}^{\delta}_{\przerwpod\alpha\beta\kappa\epsilon}
=g^\delta_{\przerwpod\alpha\beta\gamma}
\mathcal{Q}^{\rho}_{\przerwpod\delta\kappa\epsilon}
-\mathcal{N}^{\rho}_{\przerw\delta}
\poch{g^\delta_{\przerwpod\alpha\beta\gamma}}{\xi^{\kappa\epsilon}}
+\poch{\mathcal{Q}^\rho_{\przerwpod\gamma\kappa\epsilon}}
{\xi^{\kappa\epsilon}}.
}
In turn, the relation (\ref{KomkzkNUXI}) leads to
\eq{\label{EqFiJ}
\text{d}\hat{\theta}=\hat{\theta}\wedge\hat{\theta}
-\hat{H}^\epsilon\tilde{e}_\epsilon,\przerwpod\przerwpod\przerwpod
\text{d}\hat{\zeta}=\hat{\zeta}\wedge\hat{\zeta}
-\hat{H}^\epsilon\tilde{T}_\epsilon,\przerwpod\przerwpod\przerwpod
\text{d}\hat{\Sigma}=\hat{\zeta}\wedge\hat{\Sigma}
-\hat{H}^\epsilon\tilde{V}_{N\epsilon},
}
and
\eq{\label{EqLH}
\mathcal{P}^{\rho}_{\przerwpod\delta\gamma\alpha\beta}=
-\poch{\Omega^\rho_{\przerw\alpha\beta}}{\xi^{\delta\gamma}}
+\theta^{\rho}_{\przerwpod\delta\gamma\epsilon}
\Omega^\epsilon_{\przerw\alpha\beta}
+2\mathcal{N}^{\rho}_{\przerw\epsilon}
H^\epsilon_{\przerwpod\alpha\beta\delta\gamma},
}
\eq{\label{EqK}
\frac{1}{2}\nawias{
\theta^\rho_{\przerwpod\delta\gamma\sigma}
\mathcal{P}^{\sigma}_{\przerwpod\alpha\beta\kappa\epsilon}
-\theta^\rho_{\przerwpod\alpha\beta\sigma}
\mathcal{P}^{\sigma}_{\przerwpod\delta\gamma\kappa\epsilon}
}
=
\mathcal{N}^{\rho}_{\przerw\sigma}
\poch{H^\sigma_{\przerwpod\alpha\beta\delta\gamma}}{\xi^{\kappa\epsilon}}
+\mathcal{Q}^\rho_{\przerwpod\sigma\kappa\epsilon}
H^\sigma_{\przerwpod\delta\gamma\alpha\beta}
-\frac{1}{2}\nawias{
\poch{\mathcal{P}^{\rho}_{\przerwpod\delta\gamma\kappa\epsilon}}
{\xi^{\alpha\beta}}
-\poch{\mathcal{P}^{\rho}_{\przerwpod\alpha\beta\kappa\epsilon}}
{\xi^{\delta\gamma}}
}.
}
There is also one additional condition on the antisymmetric component of
$\mathcal{P}^{\rho}_{\przerwpod\delta\gamma\alpha\beta}$ which holds automatically
due~to~(\ref{EqLH}). The coefficients $\mathcal{Q}$ and $\mathcal{P}$
describing
the dependence of gauge transformations on derivatives of $\xi$ are
unambiguously determined by the remaining parameters according to equations
(\ref{EqCeu}) and (\ref{EqLH}). Constraints on them (represented by the eqs.
\refer{RelKomQRozdzNUXI}, \refer{EqE} and \refer{EqK}) are also
automatically ensured by \refer{EqOA}, \refer{EqOB}, \refer{EqB} and
\refer{EqOC}.

Finally, we need the general solution to gauge invariance conditions
(\ref{WarGINUXI}) and (\ref{WarGINUXIk}). The `ordinary' BRST symmetry
(e.g. \cite{Wein,ZinnJustinQFTCritical,BBBC}) suggests that the functional
$\tilde{\mathcal{I}}^N_{GI}$ can be obtained from the Lagrangian
\eqs{\label{LGIfullNUXI}
\funkcjonal{\tilde{\mathcal{L}}^N_{GI}}{\mathcal{A},u,\xi|x}
&=&-\frac{1}{4}\macierz{Z_{{{}_A}}\!\nawias{\xi\nawias{x}}}_{\alpha\beta}
\tilde{F}^\alpha_{\text{ }\mu\nu}(x)\tilde{F}^{\beta\text{}\mu\nu}(x)
+\frac{1}{2}\macierz{Z_{{{}_\phi}}\!\nawias{\xi\nawias{x}}}_{ab}
\nawias{\tilde{D}_\mu\check{\phi}}^a\!
\nawias{x}\nawias{\tilde{D}^\mu\check{\phi}}^b\!\nawias{x}\nonumber\\
{}&{}&-\tilde{\mathcal{V}}\nawias{\check{\phi}\nawias{x},\xi\nawias{x}},
}
where, in terms of parameters introduced in
(\ref{ZwizekWielkosciFalkowINieFalkow}),
\eqs{\label{WiekosciKowariantneFullNUXI}
\tilde{F}^\alpha_{\text{ }\mu\nu}(x)
&=&\partial_\mu A^\alpha_\nu\nawias{x}-\partial_\nu A^\alpha_\mu\nawias{x}
+e^\alpha_{\przerwpod\beta\gamma}\nawias{\xi\nawias{x}}
A^\beta_\mu\nawias{x}A^\gamma_\nu\nawias{x}\\
&{}&+\mathcal{X}^\alpha_{\przerwpod\beta\gamma\kappa}
\nawias{\xi\nawias{x}}\nawias{A^\beta_\mu\nawias{x}
\partial_{\nu}\xi^{\gamma\kappa}\nawias{x}
-A^\beta_\nu\nawias{x}\partial_{\mu}\xi^{\gamma\kappa}\nawias{x}}
+\mathcal{Y}^\alpha_{\przerwpod\beta\epsilon\gamma\kappa}
\nawias{\xi\nawias{x}}
\partial_\mu\xi^{\beta\epsilon}\partial_{\nu}\xi^{\gamma\kappa}
\nawias{x},\nonumber\\
\nawias{\tilde{D}_\mu\check{\phi}}\nawias{x}
&=&\partial_\mu \check{\phi}\nawias{x}
+A^\alpha_\mu\nawias{x}\macierz{T_\alpha\nawias{\xi\nawias{x}}}
\check{\phi}\nawias{x}
+A^\alpha_\mu\nawias{x} V_{N\alpha}\nawias{\xi\nawias{x}}\nonumber\\
&{}&{}+
\partial_\mu\xi^{\kappa\gamma}\nawias{x}\macierz{\mathcal{R}_{\kappa\gamma}
\nawias{\xi\nawias{x}}}\check{\phi}\nawias{x}
+\partial_\mu\xi^{\kappa\gamma}\nawias{x}\mathcal{S}_{\kappa\gamma}
\nawias{\xi\nawias{x}}.
}
The additional terms with derivatives of $\xi$ have been included in these
formulae in agreement with the power-counting and the Lorentz invariance.
The constraint (\ref{WarGINUXI}) yields the usual conditions of the BRST
symmetry
\eq{\label{ZinvDfullXI}
\macierz{Z_A}_{\epsilon\alpha}{e}^\alpha_{\przerwpod\beta\gamma}=
-\macierz{Z_A}_{\gamma\alpha}{e}^\alpha_{\przerwpod\beta\epsilon},\quad\quad
\macierz{Z_\phi}_{c a} \macierz{{T}_\alpha}^a_{\przerw b}=
-\macierz{Z_\phi}_{ba} \macierz{{T}_\alpha}^a_{\przerw c},
}
and leads to the following conditions on
the scalar potential $\tilde{\mathcal{V}}\nawias{\Phi,\xi}$:
\eq{\label{VinvfullNUXI}
\nawias{\macierz{T_\alpha\nawias{\xi}}^a_{\przerw b}\Phi^b
+V^a_{N\alpha}\nawias{\xi}}\poch{\tilde{\mathcal{V}}
\nawias{\Phi,\xi}}{\Phi^a}\equiv 0.
}
Similarly, (\ref{WarGINUXIk}) gives
\eqs{\label{ZinvkfullXI}\label{Eq:dZAiphi}
\text{d}\macierz{Z_A}_{\kappa\delta}=-\macierz{Z_A}_{\kappa\epsilon}
\hat{\theta}^\epsilon_{\przerwpod\delta}
-\macierz{Z_A}_{\epsilon\delta}\hat{\theta}^\epsilon_{\przerwpod\kappa},
\przerwpod\przerwpod\przerwpod
\text{d}\macierz{Z_\phi}_{ab}=-\macierz{Z_\phi}_{a c}
\hat{\zeta}^c_{\przerw b}
-\macierz{Z_\phi}_{cb} \hat{\zeta}^c_{\przerw a,}
}
and
\eq{\label{VinvfullNUXIk}
\poch{\tilde{\mathcal{V}}\nawias{\Phi,\xi}}{\xi^{\alpha\beta}}+
\nawias{\zeta^a_{\przerw\alpha\beta b}\nawias{\xi}
\Phi^b+\Sigma^a_{\przerw\alpha\beta}\nawias{\xi}}
\poch{\tilde{\mathcal{V}}\nawias{\Phi,\xi}}{\Phi^a}\equiv 0.
}
Once again we find that the terms dependent on derivatives of $\xi$
are entirely determined by the remaining ones. Defining:
\eq{
\hat{\mathcal{X}}^\alpha_{\przerwpod\epsilon}
=\mathcal{X}^\alpha_{\przerwpod\epsilon\gamma\kappa}\nawias{\xi}
\text{d}\xi^{\gamma\kappa},\qquad
\hat{\mathcal{Y}}^\epsilon
=\mathcal{Y}^\epsilon_{\przerwpod\kappa\alpha\gamma\beta}\nawias{\xi}
\text{d}\xi^{\kappa\alpha}\wedge\text{d}\xi^{\gamma\beta},
}
and
\eq{
\hat{\mathcal{R}}^a_{\przerw e}
=\mathcal{R}^a_{\przerw e\kappa\gamma}\nawias{\xi}\text{d}\xi^{\kappa\gamma},
\qquad
\hat{\mathcal{S}}^a=\mathcal{S}^a_{\przerw\kappa\gamma}\nawias{\xi}
\text{d}\xi^{\kappa\gamma},\qquad \hat{\Omega}^\epsilon
=\Omega^\epsilon_{\przerwpod\kappa\alpha}\nawias{\xi}
\text{d}\xi^{\kappa\alpha},
}
we conclude that the relations \refer{WarGINUXIk}-\refer{WarGINUXI} require
\eq{
\hat{\mathcal{X}}=\hat{\theta}-\hat{\Omega}^{\epsilon}e_{\epsilon},
\przerwpod\przerwpod\przerwpod
\frac{1}{2}\hat{\mathcal{Y}}^\alpha=\text{d}\hat{\Omega}^{\alpha}
-\hat{\theta}^\alpha_{\przerw\epsilon}
\wedge\hat{\Omega}^{\epsilon}
+\frac{1}{2}e^\alpha_{\przerwpod\delta\epsilon}\hat{\Omega}^{\delta}
\wedge\hat{\Omega}^{\epsilon}
+\mathcal{N}^\alpha_{\przerwpod\epsilon}\hat{H}^\epsilon,
}
and
\eq{
\hat{\mathcal{R}}=\hat{\Omega}^{\epsilon}T_{\epsilon}-\hat{\zeta},
\przerwpod\przerwpod\przerwpod
\hat{\mathcal{S}}=\hat{\Omega}^{\epsilon}V_{N\epsilon}-\hat{\Sigma}.
\przerwpod\przerwpod\przerwpod
}

It will be convenient to rewrite the constraint (\ref{EqA}) in terms of the
structure constants $e^{\sigma}_{\przerw\beta\gamma}$ and the generators $T_\alpha$.
Using (\ref{EqB}) and \refer{ZwizekWielkosciFalkowINieFalkow} one obtains
\eqs{\label{Eq:de}
\text{d}e_\gamma=\macierz{\hat{\theta},~e_\gamma}
-\hat{\theta}^\delta_{\przerwpod\gamma}e_\delta,
}
\eq{\label{Eq:dT}
\text{d}T_\gamma=\macierz{\hat{\zeta},~T_\gamma}
-\hat{\theta}^\delta_{\przerwpod\gamma}T_\delta.\phantom{aaa}
}
Moreover, the condition (\ref{EqOA}) follows immediately from \refer{Eq:de} and
(\ref{EqB}). Similarly, the condition (\ref{EqOB}) is equivalent to the first
relation in (\ref{EqFiJ}) and can be rewritten as
\eq{\label{Eq:dtheta}
\text{d}\hat{\theta}=\hat{\theta}\wedge\hat{\theta}-\hat{\Psi}^\epsilon
e_\epsilon,
}
where
\eq{\label{Eq:Psi_def}
\hat{\Psi}^\kappa\equiv
\mathcal{N}^\kappa_{\przerwpod\epsilon}\hat{H}^\epsilon.
}
Instead of (\ref{EqOC}), we have
\eq{\label{Eq:dPsi}
\text{d}\hat{\Psi}^{\sigma}=
\hat{\theta}^\sigma_{\przerw\alpha}\wedge\hat{\Psi}^{\alpha},
}
while the last two equations in (\ref{EqFiJ}) read respectively
\eq{\label{Eq:dzeta}
\text{d}\hat{\zeta}=\hat{\zeta}\wedge\hat{\zeta}
-\hat{\Psi}^\epsilon{T}_\epsilon,
}
\eq{\label{Eq:dSigma}
\text{d}\hat{\Sigma}=\hat{\zeta}\wedge\hat{\Sigma}
-\hat{\Psi}^\epsilon V_{N\epsilon}.
}
Finally, (\ref{EqD}) takes the form
\eq{\label{Eq:dVN}
\text{d}V^a_{N\gamma}=\hat{\zeta}^a_{\przerw b}V^b_{N\gamma}
-\macierz{T_{\gamma}}^a_{\przerw b}\hat{\Sigma}^b
-V^a_{N\delta}\hat{\theta}^\delta_{\przerw\gamma}.
}
The relations (\ref{Eq:de})-(\ref{Eq:dVN}), supplemented with (\ref{EqB}),
(\ref{ZinvkfullXI}) and (\ref{VinvfullNUXIk}) govern the $\xi$-dependence
of the counterterms. Since $u$ has dimension 1, the $u$-dependence is even
more restricted: the form (\ref{DzialanieGF}) of the solution to the
constraint \refer{WarLinqaOstXI} shows that the gauge-invariant part
of the action depends on $u$ only through the shifted field $\check\phi$
defined in \refer{Eq:checkPhidefApp}. Putting all  this together, we obtain
the renormalized action presented in Section~3.
\end{subsection}

\begin{subsection}{Global gauge invariance}\label{app:WnioskiZsymG}
Aside from the Nielsen identity and the ghost equation, the
action (\ref{LtildeN}) also satisfies the Ward-Takahashi identity
\eq{\label{WTglobrendlaItotNUXI}
\mathcal{W}^N_\alpha\mathcal{I}^N\equiv 0,
}
where $\mathcal{W}^N_\alpha$ are the following differential operators
\eqs{\label{WTglobrenNUXI}
\mathcal{W}^N_\alpha&=&
\macierz{{T_{{}_R}}_\alpha}^a_{\przerw b}u^b\cdotp\pochfunc{}{u^a}+
\macierz{{T_{{}_R}}_\alpha}^a_{\przerw b}q^b\cdotp\pochfunc{}{q^a}
+\macierz{{T_{{}_R}}_\alpha}^a_{\przerw b}\nawias{\phi^b
+v^b_{{}_R}}\cdotp\pochfunc{}{\phi^a}
+{e_{{}_{R}}}^{\epsilon}_{\przerwpod\alpha \kappa}
A^\kappa_{\mu}\cdotp\pochfunc{}{A^\epsilon_{\mu}}
+{e_{{}_{R}}}^{\epsilon}_{\przerwpod\alpha \kappa}
\omega^\kappa\cdotp\pochfunc{}{\omega^\epsilon}\nonumber\\
{}&{}&+\delta_{\epsilon\beta}{e_{{}_{R}}}^{\beta}_{\przerwpod\alpha\gamma}
\delta^{\gamma\kappa}\anty{\omega}_\kappa\cdotp
\pochfunc{}{\anty{\omega}_\epsilon}
+\delta_{\epsilon\beta}{e_{{}_{R}}}^{\beta}_{\przerwpod\alpha \gamma}
\delta^{\gamma\kappa}L_\kappa\cdotp\pochfunc{}{L_\epsilon}
+\delta_{\epsilon\beta}{e_{{}_{R}}}^{\beta}_{\przerwpod\alpha \gamma}
\delta^{\gamma\kappa}K_\kappa^\mu\cdotp\pochfunc{}{K^\mu_\epsilon}
+\delta_{ad}\macierz{{T_{{}_R}}_\alpha}^d_{\przerw c}
\delta^{cb}K_b\cdotp\pochfunc{}{K_a}\nonumber\\
{}&{}&+\nawias{{e_{{}_{R}}}^{\epsilon}_{\przerwpod\alpha\beta}
\xi^{\beta\kappa}
+{e_{{}_{R}}}^{\kappa}_{\przerwpod\alpha\beta}\xi^{\beta\epsilon}}
\cdotp\pochfunc{}{\xi^{\epsilon\kappa}}
+\nawias{{e_{{}_{R}}}^{\epsilon}_{\przerwpod\alpha\beta}q^{\beta\kappa}
+{e_{{}_{R}}}^{\kappa}_{\przerwpod\alpha\beta}q^{\beta\epsilon}}
\cdotp\pochfunc{}{q^{\epsilon\kappa}}.
}
The operators $-\mathcal{W}^N_\alpha$ form a representation of the Lie
algebra with the structure constants ${e_{{}_R}}^\delta_{\przerwpod\beta\gamma}$.
Standard arguments (e.g. \cite{ZinnJustinQFTCritical}) are unaffected by
the presence of Nielsen sources, and one can conclude that the
renormalized action $\tilde{\mathcal{I}}^N$ satisfies the same identity
\eq{\label{WTglobrendlaItotNUfullXI}
\mathcal{W}^N_\alpha\tilde{\mathcal{I}}^N\equiv 0.
}
It is convenient to introduce the vector fields
\eqs{\label{Eq:DefpolLambda}
\Lambda_{\alpha}=\nawias{{e_{{}_{R}}}^{\epsilon}_{\przerwpod\alpha\beta}
\xi^{\beta\kappa}
+{e_{{}_{R}}}^{\kappa}_{\przerwpod\alpha\beta}\xi^{\beta\epsilon}}
\poch{}{\xi^{\epsilon\kappa}},\qquad
\funkcjonal{}{\Lambda_\alpha,~\Lambda_\beta}=-\Lambda_\gamma
\eR{\gamma}{\alpha}{\beta}.
}
In the identity (\ref{WTglobrendlaItotNUfullXI}) vanishing of the
terms proportional to the products $\mathcal{K}_i\times q^a$ and $L_\alpha \times\omega^\beta \omega^\gamma$
implies  respectively the equalities
\eqs{
\label{EqR6}\label{Eq:LieB}
\nawias{\Lambda_\alpha b}\nawias{\xi}&=&\macierz{{T_{{}_R}}_\alpha,~b
\nawias{\xi}},\\
\label{EqR2dlaC}
\nawias{\Lambda_\alpha C_\kappa}\nawias{\xi}&=&\macierz{{e_{{}_R}}_\alpha,~
C_\kappa\nawias{\xi}}-C_\epsilon\nawias{\xi}
{e_{{}_{R}}}^{\epsilon}_{\przerwpod\alpha\kappa},
}
while vanishing of the coefficients of the products
$L_\alpha\times q^{\beta\gamma} \omega^\delta$ and
$L_\alpha \times q^{\beta\gamma} q^{\delta\epsilon}$ give the following equations
for the Lie derivatives
\eqs{
\label{EqR16}
\mathfrak{L}_{{\Lambda_\alpha}}\hat{g}&=&
\macierz{{e_{{}_R}}_\alpha,~\hat{g}},\\
\label{EqR17}
\mathfrak{L}_{{\Lambda_\alpha}}\hat{H}&=&{e_{{}_R}}_\alpha\hat{H},
}
Vanishing of the terms involving the product
$K^{\mu}_{\delta}\times \omega^\alpha$ gives
\eq{\label{EqR0}
{e_{{}_{R}}}^{\beta}_{\przerwpod\alpha\epsilon}
\funkcjonal{D^{\epsilon}_{\mu\kappa}}{\mathcal{A},u,\xi|x,y}
-\funkcjonal{D^{\beta}_{\mu\epsilon}}
{\mathcal{A},u,\xi|x,y}{e_{{}_{R}}}^{\epsilon}_{\przerwpod\alpha\kappa}
=\mathcal{W}^N_\alpha
\funkcjonal{D^{\beta}_{\mu\kappa}}{\mathcal{A},u,\xi|x,y}.
}
Hence
\eqs{
\label{Eq:LieN}
\nawias{\Lambda_\alpha\mathcal{N}}\nawias{\xi}&=&
\macierz{{e_{{}_R}}_\alpha,~\mathcal{N}\nawias{\xi}}.\\
\label{Eq:LieE}
\nawias{\Lambda_\alpha e_\kappa}\nawias{\xi}&=&
\macierz{{e_{{}_R}}_\alpha,~e_\kappa\nawias{\xi}}-e_\epsilon\nawias{\xi}
{e_{{}_{R}}}^{\epsilon}_{\przerwpod\alpha\kappa},
}
and ($\mathcal{Q}_{\rho\epsilon}=
\macierz{\mathcal{Q}_{\rho\epsilon}}^\beta_{\przerw\kappa}\equiv
\mathcal{Q}^\beta_{\przerw\kappa{\rho\epsilon}}$)
\eq{\label{EqR3}
\nawias{\Lambda_\alpha\mathcal{Q}_{\rho\omega}}\nawias{\xi}=
\macierz{{e_{{}_R}}_\alpha,~\mathcal{Q}_{\rho\omega}\nawias{\xi}}
-\nawias{{e_{{}_{R}}}^{\epsilon}_{\przerwpod\alpha\rho}
\mathcal{Q}_{\epsilon\omega}\nawias{\xi}
+{e_{{}_{R}}}^{\epsilon}_{\przerwpod\alpha\omega}
\mathcal{Q}_{\rho\epsilon}\nawias{\xi}}.
}
The condition (\ref{EqR2dlaC}) is automatically satisfied, provided
\refer{Eq:LieN} and \refer{Eq:LieE} hold. The equation for gauge
transformation of scalars which is analogous to (\ref{EqR0}) yields
\eqs{
\label{Eq:LieT}
\nawias{\Lambda_\alpha T_\kappa}\nawias{\xi}&=&\macierz{{T_{{}_R}}_\alpha,~
T_\kappa\nawias{\xi}}-T_\epsilon\nawias{\xi}
{e_{{}_{R}}}^{\epsilon}_{\przerwpod\alpha\kappa},\\
\label{Eq:LieVN}
\nawias{\Lambda_\alpha V_{N\kappa}}\nawias{\xi}&=&{T_{{}_R}}_\alpha
V_{N\kappa}\nawias{\xi}-T_\kappa\nawias{\xi} {T_{{}_{R}}}_\alpha v_{{}_R}
-V_{N\epsilon}\nawias{\xi}{e_{{}_{R}}}^{\epsilon}_{\przerwpod\alpha\kappa}.
}
The terms of the form $K^{\mu}_{\beta}\times q^{\kappa\lambda}$ in
(\ref{WTglobrendlaItotNUfullXI}) yield
\eq{\label{EqR0dlak}
{e_{{}_{R}}}^{\beta}_{\przerwpod\alpha\epsilon}
\funkcjonal{k^{\epsilon}_{\mu\kappa\lambda}}{\mathcal{A},u,\xi|x,y}
-\nawias{\funkcjonal{k^{\beta}_{\mu\epsilon\lambda}}{\mathcal{A},
u,\xi|x,y}{e_{{}_{R}}}^{\epsilon}_{\przerwpod\alpha\kappa}
+\funkcjonal{k^{\beta}_{\mu\epsilon\kappa}}{\mathcal{A},
u,\xi|x,y}{e_{{}_{R}}}^{\epsilon}_{\przerwpod\alpha\lambda}}
=\mathcal{W}^N_\alpha\funkcjonal{k^{\beta}_{\mu\kappa\lambda}}
{\mathcal{A},u,\xi|x,y},
}
whence
\eqs{
\label{EqR10}\label{Eq:LieTheta}
\mathfrak{L}_{{\Lambda_\alpha}}\hat{\theta}&=&\macierz{{e_{{}_R}}_\alpha,~
\hat{\theta}},\\
\label{EqR11}\label{EqR11nowe}
\mathfrak{L}_{{\Lambda_\alpha}}\hat{\Omega}&=&{e_{{}_R}}_\alpha\hat{\Omega}.
}
The constraint \refer{EqR0dlak} gives also a condition on
$\mathcal{P}\nawias{\xi}$, which is however automatically satisfied by
$\mathcal{P}\nawias{\xi}$ determined by (\ref{EqLH}) owing to the
relations (\ref{EqR10}), (\ref{EqR11}), (\ref{EqR17}) and
(\ref{Eq:LieN}). The scalar counterpart of (\ref{EqR0dlak}) leads to
\eqs{
\label{Eq:LieZetaSt}\label{Eq:LieZeta}
\mathfrak{L}_{\Lambda_\alpha}\hat{\zeta}&=&\macierz{{T_{{}_R}}_\alpha,~
\hat{\zeta}},\\
\label{Eq:LieSigmaSt}\label{Eq:LieSigma}
\mathfrak{L}_{\Lambda_\alpha}\hat{\Sigma}
&=&{T_{{}_R}}_\alpha\hat{\Sigma}-\hat{\zeta}~\!{T_{{}_{R}}}_\alpha v_{{}_R}.
}
Taking into account the above equations one can check that the operators
$\tilde{F}^\alpha_{\text{ }\mu\nu}$ and $\tilde{D}_\mu\check{\phi}$
transform covariantly
\eqs{
\mathcal{W}^N_\alpha\tilde{F}^\beta_{\text{ }\mu\nu}\nawias{x}
&=&{e_{{}_R}}^\beta_{\przerwpod\alpha\gamma}
\tilde{F}^\gamma_{\text{ }\mu\nu}\nawias{x},\\
\mathcal{W}^N_\alpha\nawias{\tilde{D}_\mu\check{\phi}}^a\nawias{x}
&=&\macierz{{T_{{}_R}}_\alpha}^a_{\przerw b}
\nawias{\tilde{D}_\mu\check{\phi}}^b\nawias{x}.
}
Thus one obtains the relations
\eqs{
\label{Eq:SymStartZA}\label{Eq:LieZA}
\nawias{\Lambda_\alpha\macierz{Z_A}_{\beta\kappa}}\nawias{\xi}&=&
-\macierz{Z_A\nawias{\xi}}_{\beta\delta}
{e_{{}_R}}^\delta_{\przerwpod\alpha\kappa}
-\macierz{Z_A\nawias{\xi}}_{\delta\kappa}
{e_{{}_R}}^\delta_{\przerwpod\alpha\beta},\\
\label{Eq:SymStartZphi}\label{Eq:LieZphi}
\nawias{\Lambda_\alpha\macierz{Z_\phi}_{b c}}\nawias{\xi}&=&
-\macierz{Z_\phi\nawias{\xi}}_{bd}\macierz{{T_{{}_R}}_\alpha}^d_{\przerw c}
-\macierz{Z_\phi\nawias{\xi}}_{dc}\macierz{{T_{{}_R}}_\alpha}^d_{\przerw b}.
}
Similarly, using the formula
\eqs{
\mathcal{W}^N_\alpha\check{\phi}^a\nawias{x}
=\macierz{{T_{{}_R}}_\alpha}^a_{\przerw b}
\nawias{\check{\phi}+v_{{}_R}}^b\nawias{x},
}
we obtain the following constraint on the counterterms for the potential of the scalar fields:
\eq{\label{Eq:SymStartV}
\nawias{\Lambda_{\alpha}\tilde{\mathcal{V}}}\nawias{\Phi,\xi}
+\macierz{{T_{{}_R}}_\alpha}^a_{\przerw b}\nawias{\Phi^b
+v^b_{{}_R}}\poch{\tilde{\mathcal{V}}\nawias{\Phi,\xi}}{\Phi^a}\equiv 0.
}
The form (\ref{DzialanieGF}) of the solution to the constraint
\refer{WarLinqaOstXI} shows that the gauge-fixing function $f^\beta\nawias{x}$
is not altered by renormalization. From its explicit form \refer{falphastart}
it is easy to find that
\eq{
\mathcal{W}^N_\alpha f^\beta\nawias{x}
={e_{{}_R}}^\beta_{\przerwpod\alpha\gamma} f^\gamma\nawias{x},
}
and the gauge-fixing part of the renormalized action (i.e. the first
term of (\ref{DzialanieGF})) automatically satisfies the identity
(\ref{WTglobrendlaItotNUfullXI}). All other terms in the renormalized
action are also consistent with the global
invariance (\ref{WTglobrendlaItotNUfullXI}), provided the conditions
listed in this appendix are fulfilled. Furthermore, the relations
(\ref{EqR16}) and (\ref{EqR3})
do not give any new information: (\ref{EqR16}) follows from (\ref{EqR10}),
(\ref{Eq:LieN}) and (\ref{Eq:dN}), while (\ref{EqR3}) is ensured by
(\ref{EqCeu}), (\ref{Eq:LieN}), (\ref{Eq:LieE}), (\ref{EqR16}) and
(\ref{EqR11}). It will be convenient to rewrite (\ref{EqR17}) in terms of
the 2-form $\hat{\Psi}$ defined in \refer{Eq:Psi_def}. With the help of
\refer{Eq:LieN} one finds
\eq{\label{Eq:LiePsi}
\mathfrak{L}_{{\Lambda_\alpha}}\hat{\Psi}={e_{{}_R}}_\alpha\hat{\Psi}.
}
Finally, let us notice that if $\xi^{-1}$ is an invariant form on the
gauge Lie algebra, then the vector fields $\Lambda_\alpha$ vanish and the
counterterms are subject to the ordinary algebraic constraints.
\end{subsection}

\begin{subsection}{Singlets}\label{app:AbelPod}
If the gauge Lie algebra
is a direct sum of a semisimple Lie
algebra (of a compact group) $\mathfrak{g}$ and an abelian Lie algebra
$\mathfrak{h}$, then additional constraints are available.
The structure constants ${e_{{}_R}}^\gamma_{\przerw\alpha\beta}$ corresponding
to the basis \mbox{$\{\tau_\alpha\}=\{\tau_{\alpha_1}\}\cup\{\tau_{\alpha_0}\}$}
whose generators $\tau_{\alpha_1}$ span $\mathfrak{g}$ and
$\tau_{\alpha_0}$ span $\mathfrak{h}$ are such that
${e_{{}_R}}^\gamma_{\przerw\alpha\beta}=0$, if any of the
the indices $\alpha$, $\beta$ or $\gamma$ correspond to a generator of
$\mathfrak{h}$. Moreover, as follows from \refer{Eq:DefpolLambda},
$\Lambda_{\alpha_0}\equiv 0$ and the relations
found in Appendix \ref{app:WnioskiZsymG} become algebraic. For instance one has
\eqs{\label{Eq:GlobAbelWiezyAlgebr}
\macierz{{T_{{}_R}}_{\alpha_0},~T_\kappa}=0,
&\macierz{{T_{{}_R}}_{\alpha_0},~b}=0,
&\macierz{{T_{{}_R}}_{\alpha_0},~\hat{\zeta}}=0.
}
For any abelian generator $\tau_{\alpha_0}$ we define
\eqsN{\label{WTlokNoper}
\mathfrak{W}^N_{\alpha_0}\nawias{x}=
&\poch{}{x^\mu}\pochfunc{}{A^{\alpha_0}_\mu\nawias{x}}
+\macierz{{T_{{}_R}}_{\alpha_0}}^a_{\przerw b}
\nawias{\phi^b\nawias{x}+v^b_{{}_R}}\pochfunc{}{\phi^a\nawias{x}}
+\delta_{ad}\macierz{{T_{{}_R}}_{\alpha_0}}^d_{\przerw c}\delta^{cb}K_b
\nawias{x}\pochfunc{}{K_a\nawias{x}}\nonumber\\
&+\macierz{{T_{{}_R}}_{\alpha_0}}^a_{\przerw b}u^b\nawias{x}
\pochfunc{}{u^a\nawias{x}}
+\macierz{{T_{{}_R}}_{\alpha_0}}^a_{\przerw b}q^b\nawias{x}
\pochfunc{}{q^a\nawias{x}}.
}
The operator $\mathfrak{W}^N_{\alpha_0}\nawias{x}$ is obviously connected
with the one introduced in \refer{WTglobrenNUXI}: $\mathcal{W}^N_{\alpha_0}
=\calka{}{x}{\mathfrak{W}^N_{\alpha_0}\nawias{x}}$. The tree level action
(\ref{LtildeN}) satisfies the Ward-Takahashi identity
\eq{\label{WTlokNtoz}
\mathfrak{W}^N_{\alpha_0}\nawias{x}\mathcal{I}^N
=-\partial_x^2
\left\{(\xi^{-1}\nawias{x})_{\alpha_0\beta}
\nawias{f^\beta\nawias{x}+\frac{1}{2}q^{\beta\kappa}\nawias{x}
\anty{\omega}_\kappa\nawias{x}}
\right\}.
}
The right-hand side of \refer{WTlokNtoz} is linear in quantum fields and
therefore the renormalized effective action $\Gamma^N$ as well as the action
with counterterms $\tilde{\mathcal{I}}^N$ also obeys \refer{WTlokNtoz}.
For this reason we have
\eqsN{
&T_{\alpha_0}\nawias{\xi}={T_{{}_R}}_{\alpha_0},\\
&V_{N\alpha_0}\nawias{\xi}={T_{{}_R}}_{\alpha_0}v_{{}_R},
}
\eq{
{e}^\gamma_{\przerwpod{\alpha_0}\beta}\nawias{\xi}
={\text{d}e}^\gamma_{\przerwpod{\alpha_0}\beta}=
{\theta}^\gamma_{\przerwpod\kappa\epsilon{\alpha_0}}\nawias{\xi}=
\poch{{\theta}^\gamma_{\przerwpod\kappa\epsilon{\alpha_0}}}
{\xi^{\rho\sigma}}=0,
}
in addition to the global invariance conditions of the form
\refer{Eq:GlobAbelWiezyAlgebr}. Moreover, since
${e_{{}_R}}^{\alpha_0}_{\phantom{\alpha_0}\beta\gamma}=0$, the following
identities are also satisfied
\eq{\label{eq:SingIden}
\pochfunc{\mathcal{I}^N}{L_{\alpha_0}(x)}=0, \quad\quad
\pochfunc{\mathcal{I}^N}{K^\mu_{\alpha_0}(x)}
=-\partial_\mu\omega^{\alpha_0}\nawias{x}.
}
The corresponding equations for the renormalized action
functional $\tilde{\mathcal{I}}^N$ yield
\eq{\label{Eq:Fvan}
C^{\alpha_0}_{\phantom{\alpha_0}\beta\gamma}
=\hat{g}^{\alpha_0}_{\phantom{\alpha_0}\gamma}=\hat{H}^{\alpha_0}=0,
}
and
\eq{\label{Nab}
\mathcal{N}^{\alpha_0}_{\phantom{\alpha_0}\beta}
=\delta^{\alpha_0}_{\phantom{\alpha_0}\beta},
}
\eq{\label{Eq:Eabup}
{e}^{\alpha_0}_{\phantom{\alpha_0}\beta\gamma}=
\hat{\theta}^{\alpha_0}_{\phantom{\alpha_0}\gamma}=\hat{\Omega}^{\alpha_0}=0.
}
It is worth noting that these equations agree with \refer{ZwiazekCieRozdzXI}
and \refer{Eq:dN}. Expressed in terms of the form $\hat{\Psi}^\epsilon$ the
relations \refer{Nab} and \refer{Eq:Fvan} read
\eq{
\hat{\Psi}^{\alpha_0}=0.
}
Thus we see that coefficients $e^{\alpha}_{\phantom{\alpha}\beta\gamma}$,
$\hat{\theta}^{\alpha}_{\phantom{\alpha}\beta}$ and $\hat{\Psi}^{\alpha}$ vanish,
if any of their indices corresponds to an abelian generator. The summations
in the formulae \refer{Eq:de}, \refer{Eq:dtheta}, \refer{Eq:dPsi} etc. can
be, therefore, restricted to semisimple indices only.

If $\phi^{a_0}$ is a gauge singlet, then
$\macierz{T_{{}_{R\alpha}}}^{a_0}_{\phantom{a_0}b}=0$ since $T_{{}_{R\alpha}}$
is completely reducible. This gives another identity
\eq{
\pochfunc{\mathcal{I}^N}{K_{a_0}(x)}=0.
}
Applied to the functional $\tilde{\mathcal{I}}^N$, this gives
\eq{
\macierz{T_{{\alpha}}}^{a_0}_{\phantom{a_0}b}=V^{a_0}_{N\alpha}
=\hat{\zeta}^{a_0}_{\phantom{{a_0}}b}
=\hat{\Sigma}^{a_0}=b^{{a_0}}_{\phantom{{{\alpha_0}}}c}=0.
}
\end{subsection}
\end{section}

\begin{section}{The case of $\xi$ being an invariant form}\label{appAlgebryPP}
Here we show that $\hat{\Psi}^\epsilon=0$ if $(\xi^{-1})_{\alpha\beta}$ is an
invariant form. As we have seen in Appendix \ref{app:AbelPod}, the
coefficients $e^{\alpha}_{\phantom{\alpha}\beta\gamma}$,
$\hat{\theta}^{\alpha}_{\phantom{\alpha}\beta}$ and $\hat{\Psi}^{\alpha}$ are
non-vanishing only if all their indices correspond to non-abelian
generators. Limiting
ourselves to these indices we can assume that the gauge Lie algebra
is semisimple. Contracting the 1-forms in \refer{Eq:de} with a vector field
$\Lambda_\alpha$ we get
\eq{
\Lambda_\alpha e_\omega=\macierz{\hat{\theta}\nawias{\Lambda_\alpha},~
e_\omega}-e_\rho~\!\hat{\theta}\nawias{\Lambda_\alpha}^\rho_{\przerw\omega}.
}
Comparing this with
\refer{Eq:LieE} we find that the matrices
$\mathcal{E}_\alpha={e_{{}_R}}_\alpha-\hat{\theta}\nawias{\Lambda_\alpha}$
satisfy the rule
\eq{\label{eq:RownNaE}
\macierz{\mathcal{E}_\alpha,~e_\gamma}
=e_\beta~\!\mathcal{E}^\beta_{\phantom{\beta} \alpha\gamma}
}
In the case of semisimple Lie algebras the rule
\refer{eq:RownNaE} can be satisfied
only if $\mathcal{E}_\alpha$ are  linear
combinations of the generators~$e_\alpha$ (see section 4) 
\eq{\label{Eq:KombLinE}
{e_{{}_R}}_\alpha-\hat{\theta}\nawias{\Lambda_\alpha}
=\mathfrak{M}^\beta_{\przerwpod\alpha}e_\beta.
}
Differentiating both sides of \refer{Eq:KombLinE} we get the relation
\eq{\label{Eq:LieDerPoKon}
\mathfrak{L}_{\Lambda_\alpha}\hat{\theta}
-\Lambda_\alpha\lrcorner\text{d}\hat{\theta}
=-\nawias{\text{d}\mathfrak{M}^\beta_{\przerwpod\alpha}} e_\beta
-\mathfrak{M}^\beta_{\przerwpod\alpha}\text{d}e_\beta.
}
which rewritten with the help of
\refer{Eq:de} and \refer{Eq:dtheta} takes the form
\eq{\label{Eq:LieDerPoKonMod}
\mathfrak{L}_{\Lambda_\alpha}\hat{\theta}
=\macierz{\Lambda_\alpha\lrcorner\hat{\theta}
+\mathfrak{M}^\beta_{\przerwpod\alpha}e_\beta,~\hat{\theta}}
-\Lambda_\alpha\lrcorner\hat{\Psi}^\beta e_\beta
-\nawias{\mathrm{d}\mathfrak{M}^\beta_{\przerwpod\alpha}}e_\beta
+\indgd{\hat{\theta}}{\beta}{\gamma}
\mathfrak{M}^\gamma_{\przerwpod\alpha}e_\beta.
}
Using \refer{Eq:KombLinE} once more, one obtains finally
\eq{\label{Eq:LieDerPoKonMod2}
\mathfrak{L}_{\Lambda_\alpha}\hat{\theta}
-\macierz{{e_{{}_R}}_\alpha,~\hat{\theta}}
=
-\Lambda_\alpha\lrcorner\hat{\Psi}^\beta e_\beta
-\nawias{\mathrm{d}\mathfrak{M}^\beta_{\przerwpod\alpha}}e_\beta
+\indgd{\hat{\theta}}{\beta}{\gamma}
\mathfrak{M}^\gamma_{\przerwpod\alpha}e_\beta.
}
The left-hand side of \refer{Eq:LieDerPoKonMod2} vanishes on account of
\refer{Eq:LieTheta}, so the linear independence of $\{e_\beta\}$ leads to
\eq{\label{Eq:N17APP}
\mathrm{d}\indgd{\mathfrak{M}}{\epsilon}{\alpha}=
\indgd{\hat{\theta}}{\epsilon}{\gamma}\indgd{\mathfrak{M}}{\gamma}{\alpha}
-\Lambda_\alpha\lrcorner\hat{\Psi}^\epsilon.
}
Since $\Lambda_\alpha\equiv 0$ on the submanifold specified by the
invariance of $\xi^{-1}$, this means that
\eq{\label{Eq:N17APPmodmodmod}
\indgd{\hat{\theta}}{\epsilon}{\gamma}
=\nawias{\mathrm{d}\indgd{\mathfrak{M}}{\epsilon}{\alpha}}
\indgd{\nawias{{\mathfrak{M}^{-1}}}}{\alpha}{\gamma},
}
and the equation \refer{Eq:dtheta} shows then that on this
submanifold `the curvature' $\hat{\Psi}^\epsilon$ vanishes.
\end{section}

\begin{section}{Stability of the Action $\tilde{\mathcal{I}}^N$ }
\label{app:Stability}
In this appendix we present more detailed arguments
that the renormalized action $\tilde{\mathcal{I}}^N$ obeys the Nielsen
identity~\refer{NielsenEqLiniowN}. We begin by repeating
the standard Zinn-Justin arguments
\cite{ZinnJustin1974,ZinnJustinQFTCritical}.
For simplicity we omit the superscript $N$ on functionals $\mathcal{I}^N$,
$\tilde{\mathcal{I}}^N$, $\Gamma^N$. Let $\mathcal{S}\nawias{\mathcal{I}}$
be the left-hand side of the Nielsen identity \refer{NielsenEqLiniowN}.
Since $\mathcal{S}\nawias{\cdot}$
is a nonlinear differential operator, one needs also its linearized
counterpart $\mathcal{S}_\mathcal{F}$ defined
(for arbitrary functionals $\mathcal{F}$ and $\mathcal{G}$) by
\eq{\nonumber
\mathcal{S}\nawias{\mathcal{F}+\varepsilon~\!\mathcal{G}}
=\mathcal{S}\nawias{\mathcal{F}}+\varepsilon~\!\mathcal{S}_\mathcal{F}
\mathcal{G}+\mathcal{O}(\varepsilon^2).
}
The renormalized action $\tilde{\mathcal{I}}_{(n)}$ generates the 1PI
effective action $\Gamma_{(n)}=\sum_{k}\hbar^k\Gamma^{(k)}_{(n)}$, which is finite up to the order $\hbar^n$.
If $\tilde{\mathcal{I}}_{(n)}$ satisfies \refer{NielsenEqLiniowN}, then
(assuming the Dimensional Regularization is used) so does $\Gamma_{(n)}$,
hence the divergent part of $\Gamma^{(n+1)}_{(n)}$ obeys
\eq{\label{eq:LinZJoper}
\mathcal{S}_{\mathcal{I}}\Gamma^{(n+1,\rm{div})}_{(n)}=0,
}
with $\mathcal{I}$ being the tree level action (and
$\tilde{\mathcal{I}}_{(0)}\equiv \mathcal{I}$). This equation ensures
that in the $MS$-scheme the renormalized action at the order $\hbar^{n+1}$,
i.e.
\eq{\label{eq:Inplusjed}
\tilde{\mathcal{I}}_{(n+1)}=\tilde{\mathcal{I}}_{(n)}
-\hbar^{n+1}\Gamma^{(n+1,\rm{div})}_{(n)}+\mathcal{O}\nawias{\hbar^{n+2}},
}
satisfies
\eq{\nonumber
\mathcal{S}\nawias{\tilde{\mathcal{I}}_{(n+1)}}
=\mathcal{O}\nawias{\hbar^{n+2}}.
}
To extend the Nielsen identity to the next order, the terms denoted
$\mathcal{O}\nawias{\hbar^{n+2}}$ in \refer{eq:Inplusjed} have to be
chosen so that $\mathcal{S}\nawias{\tilde{\mathcal{I}}_{(n+1)}}\equiv 0$.
For $q=0$ (i.e. for the ordinary Zinn-Justin equation) a proof
that this can be done was given in~\cite{BBBC}; it
provides additional constraints on possible counterterms,
giving rise to the equations \refer{eq:ZeIZt} and \refer{eq:ZtSamo}. Since the Nielsen sources $q$ enter
$\mathcal{S}\nawias{\cdot}$ linearly and multiply WT-like differential
operators, extension of this analysis to $q\neq 0$ does not give any new
information beyond those already contained in the relations
\refer{eq:ZeIZt}-\refer{eq:ZtSamo}. We will demonstrate this for equations
\refer{TozJacNielXI}-\refer{EqOC} (Appendix \ref{app:TozNielPrzec}).
To this end we focus on the following part of the Lagrangian
renormalized through the $n$-th order
\eq{\nonumber
\tilde{\mathcal{L}}_{(n)}\nawias{x}\supset
\frac{1}{2}L_{\alpha}\nawias{x}
\stackrel{(n)}{C}\!\!{}^\alpha_{\przerwpod\beta\gamma}\omega^{\beta}
\nawias{x}\omega^{\gamma}\nawias{x}+
L_{\alpha}
\nawias{x}\stackrel{(n)}{g}\!\!{}^\alpha_{\przerwpod\beta\gamma\delta}
q^{\beta\gamma}\nawias{x}\omega^{\delta}\nawias{x}
+L_{\alpha}\nawias{x}\stackrel{(n)}{H}\!\!{}^\alpha_{\przerwpod\beta\gamma
\delta\epsilon}
q^{\beta\gamma}\nawias{x}q^{\delta\epsilon}
\nawias{x},
}
whose $n$th-order parameters satisfy the conditions
\refer{TozJacNielXI}-\refer{EqOC} exactly:
\eq{\label{Eq:JacDlaCn}
\left[\stackrel{(n)}C\!\!{}_\beta,~\!\stackrel{(n)}C\!\!{}_\gamma\right]
=\stackrel{(n)}C\!\!{}^\alpha_{\przerwpod\beta\gamma}
\stackrel{(n)}C\!\!{}_\alpha,
}
etc. The relation \refer{eq:Inplusjed} implies then that \eq{\label{Eq:StaleCnplus1}
\stackrel{(n+1)}{C}\!\!{}^\alpha_{\przerwpod\beta\gamma}
=\stackrel{(n)}{C}\!\!{}^\alpha_{\przerwpod\beta\gamma}
+\hbar^{n+1}{c}^\alpha_{\przerwpod\beta\gamma}
+\mathcal{O}\nawias{\hbar^{n+2}}
={e_{{}_R}}^\alpha_{\phantom{\alpha}\beta\gamma}
+\mathcal{O}\nawias{\hbar},
}
while \refer{eq:LinZJoper} requires
\eq{\label{eq:LinCcond}
\left[e_{{}_R\beta},~c_\gamma\right]-\left[e_{{}_R\gamma},~c_\beta\right]
=e_{{}_R\alpha}c^\alpha_{\przerwpod\beta\gamma}
+c_{\alpha}{e_{{}_R}}^\alpha_{\przerwpod\beta\gamma}.
}
Taking into account identity \refer{eq:SingIden}, which gives
$\indgd{c}{\alpha_0}{\beta\gamma}=0$, and performing some manipulations
on \refer{eq:LinCcond} (see e.g. \cite{BBBC}) one finds
\eq{\label{Eq:MaleCPelna}
\indgd{c}{\eta}{\beta\gamma}=
\indgd{z}{\eta}{\epsilon}\indgd{{e_{{}_R}}}{\epsilon}{\beta\gamma}-
\indgd{z}{\epsilon}{\gamma}\indgd{{e_{{}_R}}}{\eta}{\beta\epsilon}-
\indgd{z}{\epsilon}{\beta}\indgd{{e_{{}_R}}}{\eta}{\epsilon\gamma},
}
where\footnote{$\alpha_1$, $\beta_1$, etc. denote non-abelian indices. }
\eq{\label{eq:maleZ}
\indgd{z}{\eta}{\beta}\equiv\indgd{{e_{{}_R}}}{\eta}{{\gamma_1}{\delta_1}}
\mathfrak{K}_{{}_R}^{\delta_1 \alpha_1}\indgd{c}{\gamma_1}{\alpha_1\beta},
}
with $\mathfrak{K}_{{}_R}^{\delta_1 \alpha_1}$ being the inverse of the
Killing form
${\mathfrak{K}_{{}_R}}_{\delta_1 \alpha_1}\equiv\mathrm{tr}
\macierz{{e_{{}_R}}_{\delta_1}{e_{{}_R}}_{\alpha_1}}$.
Let us define
\eq{\nonumber
\stackrel{(n+1)}{Z}\!\!\indgd{}{\eta}{\beta}\equiv\indgd{\delta}
{\eta}{\beta}+\hbar^{n+1}\indgd{z}{\eta}{\beta},
}
and
\eq{\label{Eq:Cnplus1pelne0}
\stackrel{(n+1)}{C}\!\!{}^{\kappa}_{\przerwpod\beta\gamma}\equiv
\macierz{\stackrel{(n+1)}{Z}}^\kappa_{\przerw\delta}
\stackrel{(n)}{C}\!\!{}^{\delta}_{\przerwpod\alpha \epsilon}
\left[\stackrel{(n+1)}{Z}\!\!\!{}^{-1}\right]^\alpha_{\przerw\beta}
\left[\stackrel{(n+1)}{Z}\!\!\!{}^{-1}\right]^\epsilon_{\przerw\gamma}
}
It is clear that
$\stackrel{(n+1)}{C}\!\!{}^{\kappa}_{\przerwpod\beta\gamma}$
are consistent with \refer{Eq:StaleCnplus1} and obey the Jacobi identity
\refer{TozJacNielXI} exactly. Moreover \refer{Eq:Cnplus1pelne0} yields
\eq{\label{Eq:Cnplus1pelne}
\stackrel{(n+1)}{C}\!\!{}^{\kappa}_{\przerwpod\beta\gamma}\equiv
\macierz{\stackrel{(n+1)}{Z_C}}^\kappa_{\przerw\delta}
{e_{{}_R}}^{\delta}_{\przerwpod\alpha \epsilon}
\left[\stackrel{(n+1)}{Z_C}\!\!\!\!\!{}^{-1}\right]^\alpha_{\przerw\beta}
\left[\stackrel{(n+1)}{Z_C}\!\!\!\!\!{}^{-1}\right]^\epsilon_{\przerw\gamma},
}
where
\eq{
\stackrel{(n+1)}{Z_C}=\stackrel{(n+1)}{Z}\stackrel{(n)}{Z_C},\qquad
\stackrel{(0)}{Z_C}=\stackrel{(0)}{Z}=1.
}
The equation \refer{Eq:Cnplus1pelne} is a counterpart of the relation \refer{eq:ZeIZt}
for structure constants ${C}^{\kappa}_{\przerwpod\beta\gamma}$.\\

We now check stability of relation \refer{EqOA} under radiative corrections.
Considerations similar to ones leading to \refer{eq:thetaeta} suggest the
inductive hypothesis
\eq{\label{eq:gp}
\stackrel{(n)}{\hat{g}}=\macierz{\mathrm{d}\!\!\stackrel{(n)}{Z_C}}
\stackrel{(n)}{Z_C}\!\!\!{}^{-1}
+\stackrel{(n)}{\hat{p}}\!\!\!{}^{\sigma_1}\stackrel{(n)}{C}
\!\!\!{}_{\sigma_1},
}
which ensures \refer{EqOA} and holds at the tree level with
$\stackrel{(0)}{\hat{p}}\!\!\!{}^{\sigma_1}=0$. At the $(n+1)$th-order we have
\eq{\label{eq:ExpOfG}
\stackrel{(n+1)}{\hat{g}}=\stackrel{(n)}{\hat{g}}+~\hbar^{n+1}{\hat{G}}
+\mathcal{O}\nawias{\hbar^{n+2}}=\mathcal{O}\nawias{\hbar},
}
and linear constraints \refer{eq:LinZJoper} read
\eq{\label{eq:LinGcond}
\mathrm{d}c_{\beta}=\left[\hat{G},~{e_{{}_R}}_\beta\right]
-{e_{{}_R}}_\alpha \indgd{\hat{G}}{\alpha}{\beta}.
}
Comparing the above equation with the derivative of \refer{Eq:MaleCPelna}
and defining
$\hat{\mathcal{E}}=\hat{G}-\mathrm{d}{z}$
one gets
\eq{\label{eq:LinEcond}
\left[\hat{\mathcal{E}},~{e_{{}_R}}_\gamma\right]
={e_{{}_R}}_\beta \indgd{\hat{\mathcal{E}}}{\beta}{\gamma}.
}
Since ${e_{{}_R}}_{\beta_1}$ are linearly independent, equation
\refer{eq:LinEcond} requires
\eq{\label{LieEcondInd1}
\indgd{\hat{\mathcal{E}}}{\beta_1}{\gamma_0}=0.
}
For non-abelian indices $\beta$ and $\gamma$, \refer{eq:LinEcond} yields
\eq{\label{ESol}
\hat{\mathcal{E}}^\beta_{\phantom{\beta} \gamma}
={e_{{}_R}}^\beta_{\phantom{\beta} \sigma_1 \gamma}
\mathfrak{K}_{{}_R}^{\sigma_1 \delta_1}
\mathrm{tr}\{{e_{{}_R}}_{\delta_1}\hat{\mathcal{E}}\}.
}
Finally, identity \refer{eq:SingIden} requires
$\indgd{\hat{G}}{\beta_0}{\epsilon}=0$, and thus (see \refer{eq:maleZ})
\eq{\label{LieEcondInd2}
\indgd{\hat{\mathcal{E}}}{\beta_0}{\epsilon}=0.
}
Equations \refer{LieEcondInd1} and \refer{LieEcondInd2} show that
\refer{ESol} holds for arbitrary indices $\beta$ and $\gamma$. Defining
\eq{
\stackrel{(n+1)}{\hat{p}}\!\!\!\!\!\!{}^{\sigma_1}\equiv
\stackrel{(n)}{\hat{p}}\!\!\!{}^{\sigma_1}+
\hbar^{n+1}\mathfrak{K}_{{}_R}^{\sigma_1 \delta_1}\mathrm{tr}
\left\{{e_{{}_R}}_{\delta_1}\nawias{\hat{G}-\mathrm{d}{z}}\right\},
}
and
\eq{
\stackrel{(n+1)}{\hat{g}}\equiv\macierz{\mathrm{d}\!\!\!
\stackrel{(n+1)}{Z_C}}\stackrel{(n+1)}{Z_C}\!\!\!{}^{-1}
+\stackrel{(n+1)}{\hat{p}}\!\!\!\!\!\!{}^{\sigma_1}
\stackrel{(n+1)}{C}\!\!\!\!\!\!{}_{\sigma_1},
}
it is easy to convince oneself that the expansion \refer{eq:ExpOfG} is
correct. Thus, equation \refer{EqOA} has been established.\\

Formula \refer{EqOB} holds for $n$th-order parameters provided that
(see \refer{eq:Psieta})
\eq{\label{eq:Fp}
\stackrel{(n)}{\hat{H}}\!\!{}^{\sigma}=
-\mathrm{d}\!\!\!\stackrel{(n)}{\hat{p}}\!\!\!{}^{\sigma}+\!\!
\stackrel{(n)}{\hat{g}}\!\!\!\indgd{}{\sigma}{\delta}\wedge\!\!
\stackrel{(n)}{\hat{p}}\!\!\!{}^{\delta}
-\frac{1}{2}\!\!\stackrel{(n)}{C}\!\!\indgd{}{\sigma}{\kappa\lambda}
\!\!\stackrel{(n)}{\hat{p}}\!\!\!{}^{\kappa}\wedge\!\!\stackrel{(n)}
{\hat{p}}\!\!\!{}^{\lambda},
}
with $\stackrel{(n)}{\hat{p}}\!\!\!{}^{\sigma_0}\equiv 0$. The relevant
part of \refer{eq:Inplusjed} has the form
\eq{\label{eq:ExpOfF}
\stackrel{(n+1)}{\hat{H}}=\stackrel{(n)}{\hat{H}}+~\hbar^{n+1}{\hat{\mathbf{h}}}
+\mathcal{O}\nawias{\hbar^{n+2}}=\mathcal{O}\nawias{\hbar},
}
while \refer{eq:LinZJoper} gives restrictions on the counterterm $\hat{\mathbf{h}}$
\eq{
\hat{\mathbf{h}}^{\alpha_1}{e_{{}_R}}_{\alpha_1}=-d\hat{G}.
}
We have
\eq{
\hbar^{n+1}\hat{G}=\hbar^{n+1}\nawias{\mathrm{d}z+\hat{\mathcal{E}}}=
\hbar^{n+1}\mathrm{d}z+{e_{{}_R}}_{\alpha_1}
\{
\stackrel{(n+1)}{\hat{p}}\!\!\!\!\!\!{}^{\alpha_1}-
\stackrel{(n)}{\hat{p}}\!\!\!{}^{\alpha_1}
\},
}
hence
\eq{
\hbar^{n+1}\hat{\mathbf{h}}^{\alpha_1}{e_{{}_R}}_{\alpha_1}=
\mathrm{d}\!\!\!\stackrel{(n)}{\hat{p}}\!\!\!{}^{\alpha_1}-
\mathrm{d}\!\!\!\!\!\!\stackrel{(n+1)}{\hat{p}}\!\!\!\!\!\!{}^{\alpha_1}
}
thus the formula \refer{eq:Fp} can be extended to the next order without
violating \refer{eq:ExpOfF}. Finally, the `Bianchi identity' \refer{EqOC}
is automatically satisfied if equations \refer{Eq:Cnplus1pelne},
\refer{eq:gp} and \refer{eq:Fp} hold.

\end{section}

\end{document}